\documentclass[aps,prx,amsmath,superscriptaddress,twocolumn,nofootinbib, longbibliography
]{revtex4-1}

\usepackage{gpkmac}

\newcommand{\YL}[1]{{\color{black} #1}}
\newcommand{\MF}[1]{{\color{black} #1}}
\newcommand{\st}[1]{}
\newcommand{\XC}[1]{{\color{black} #1}}
\newcommand{\ZCY}[1]{{\color{black} #1}}

\usepackage{graphicx}
\usepackage{amsmath}
\usepackage{amsfonts}
\usepackage{amssymb}
\usepackage{braket}
\usepackage{multirow}
\usepackage{color}
\usepackage{mathrsfs}
\usepackage{mathtools}
\usepackage{float}
\usepackage[textsize=small]{todonotes}
\usepackage{subfigure}
\usepackage[bottom]{footmisc}
\usepackage{footnote}

 \AtBeginDocument{%
     \newwrite\bibnotes
     \def\bibnotesext{Notes.bib}
     \immediate\openout\bibnotes=\jobname\bibnotesext
     \immediate\write\bibnotes{@CONTROL{REVTEX41Control}}
     \immediate\write\bibnotes{@CONTROL{%
     apsrev41Control,author="08",editor="1",pages="1",title="0",year="1"}}
      \if@filesw
      \immediate\write\@auxout{\string\citation{apsrev41Control}}%
     \fi
 }%

\begin{document}

\title{Entanglement phase transitions in random stabilizer tensor networks}

\author{Zhi-Cheng Yang}
\affiliation{Joint Quantum Institute, University of Maryland, College Park, Maryland 20742, USA}
\affiliation{Joint Center for Quantum Information and Computer Science, University of Maryland, College Park, Maryland 20742, USA}
\author{Yaodong Li}
\affiliation{Department of Physics, University of California, Santa Barbara, CA 93106, USA}
\author{Matthew P. A. Fisher}
\affiliation{Department of Physics, University of California, Santa Barbara, CA 93106, USA}
\author{Xiao Chen}
\affiliation{Department of Physics, Boston College, Chestnut Hill, MA 02467, USA}

\date{\today}

\begin{abstract}

We explore a class of random tensor network models with ``stabilizer'' local tensors which we name Random Stabilizer Tensor Networks (RSTNs).
For RSTNs defined on a two-dimensional square lattice, we perform extensive numerical studies of entanglement phase transitions 
between volume-law and area-law entangled phases of the one-dimensional boundary states. 
These transitions occur when either (a) the bond dimension $D$ of the constituent tensors is varied, or (b) the tensor network is subject to 
random breaking of bulk bonds, implemented by forced measurements.
In the absence of broken bonds, we find that the RSTN supports a volume-law entangled boundary state with 
bond dimension $D\geq3$ where $D$ is a prime number, and an area-law entangled boundary state for $D=2$.
Upon breaking bonds at random in the bulk with probability $p$, there exists a critical measurement rate $p_c$ for each $D\geq 3$ above which the boundary state becomes area-law entangled.
To explore the conformal invariance at these entanglement transitions for different prime $D$, we consider tensor networks on a finite rectangular geometry with a variety of boundary conditions, and extract universal operator scaling dimensions via extensive numerical calculations of the entanglement entropy, mutual information and mutual negativity at their respective critical points.
Our results at large $D$ approach known universal data of percolation conformal field theory, while showing clear discrepancies at smaller $D$, suggesting a distinct entanglement transition universality class for each prime $D$. We further study universal entanglement properties in the volume-law phase and demonstrate quantitative agreement with the recently proposed description in terms of a directed polymer in a random environment.

\end{abstract}

\maketitle

{
\hypersetup{linktocpage, linkcolor=blue}
\tableofcontents
}

\section{Introduction}

Recent progress in understanding quantum many-body systems out of equilibrium has extended our characterizations of phases of matter and phase transitions beyond the symmetry and topology of ground states. It turns out that a new class of phase transitions is instead characterized by a sharp change in the entanglement properties of the many-body wavefunctions.
Highly-excited eigenstates of generic non-integrable quantum Hamiltonians are typically thermal, and the entanglement of a subsystem scales extensively with the volume of the subsystem~\cite{d2016quantum, PhysRevA.43.2046, PhysRevE.50.888, PhysRevLett.71.1291}. On the other hand, ground states of gapped local Hamiltonians are typically area-law entangled, with the entanglement scaling only with the surface area of the subsystem~\cite{hastings2007area, RevModPhys.82.277}. An entanglement phase transition in the quantum state thus signals a dramatic change in the behavior of quantum information dynamics as well as the approach to thermalization of the system.

Many-body localized systems provide a prototypical example of such an entanglement phase transition. Thermalization in such systems is impeded by strong quenched disorder, and the excited eigenstates undergo an entanglement transition from a volume-law to an area-law scaling~\cite{PhysRevB.75.155111, PhysRevB.82.174411, nandkishore2015many}. More recently, a new class of entanglement transitions has been uncovered in monitored quantum systems subject to repeated measurements that are sprinkled at random locations in space and time~\cite{PhysRevB.99.224307, nahum2018transition, PhysRevB.98.205136, li2019hybrid, li2020conformal, bao2018spin, jian2018spin,PhysRevLett.125.030505, gullans2019purification, PhysRevB.103.174309, PhysRevB.103.104306}. One concrete physical realization of such systems is hybrid quantum circuits, where both unitary evolution and measurements are present. As the measurement rate $p$ is varied, the system can sustain a stable phase at small $p$ where the steady states of the individual quantum trajectories are volume-law entangled. The resilience of quantum entanglement against local measurements in the volume-law phase is particularly interesting, and this phase in fact can be viewed as a dynamically generated quantum error-correcting code (QECC)~\cite{PhysRevLett.125.030505, gullans2019purification, PhysRevB.103.174309, PhysRevB.103.104306, fidkowski2020forget,gullans2020lowdepth}.
In this language, local measurements can be interpreted as local errors, and the QECC can retain a finite code rate when the error rate $p$ is low. When the error rate exceeds a certain threshold, the system is no longer robust against errors and the long-range quantum entanglement is destroyed. The measurement rate $p$ thus drives an entanglement transition in the steady states of such hybrid quantum circuit models. Remarkably, at the critical point $p_c$, these models exhibit universal properties akin to a two-dimensional Euclidean conformal field theory (CFT). Although the nature of the CFTs in hybrid quantum circuits remains to be fully understood, their existence has been unambiguously demonstrated numerically in hybrid Clifford circuits~\cite{PhysRevB.98.205136, li2019hybrid,  li2020conformal, zabalo2020critical, zabalo2021ceff}, and analytically established in hybrid Haar random circuits in the limit of infinite local Hilbert-space dimensions~\cite{jian2018spin, bao2018spin}.

\begin{figure}[t]
\centering
\includegraphics[width=0.25\textwidth]{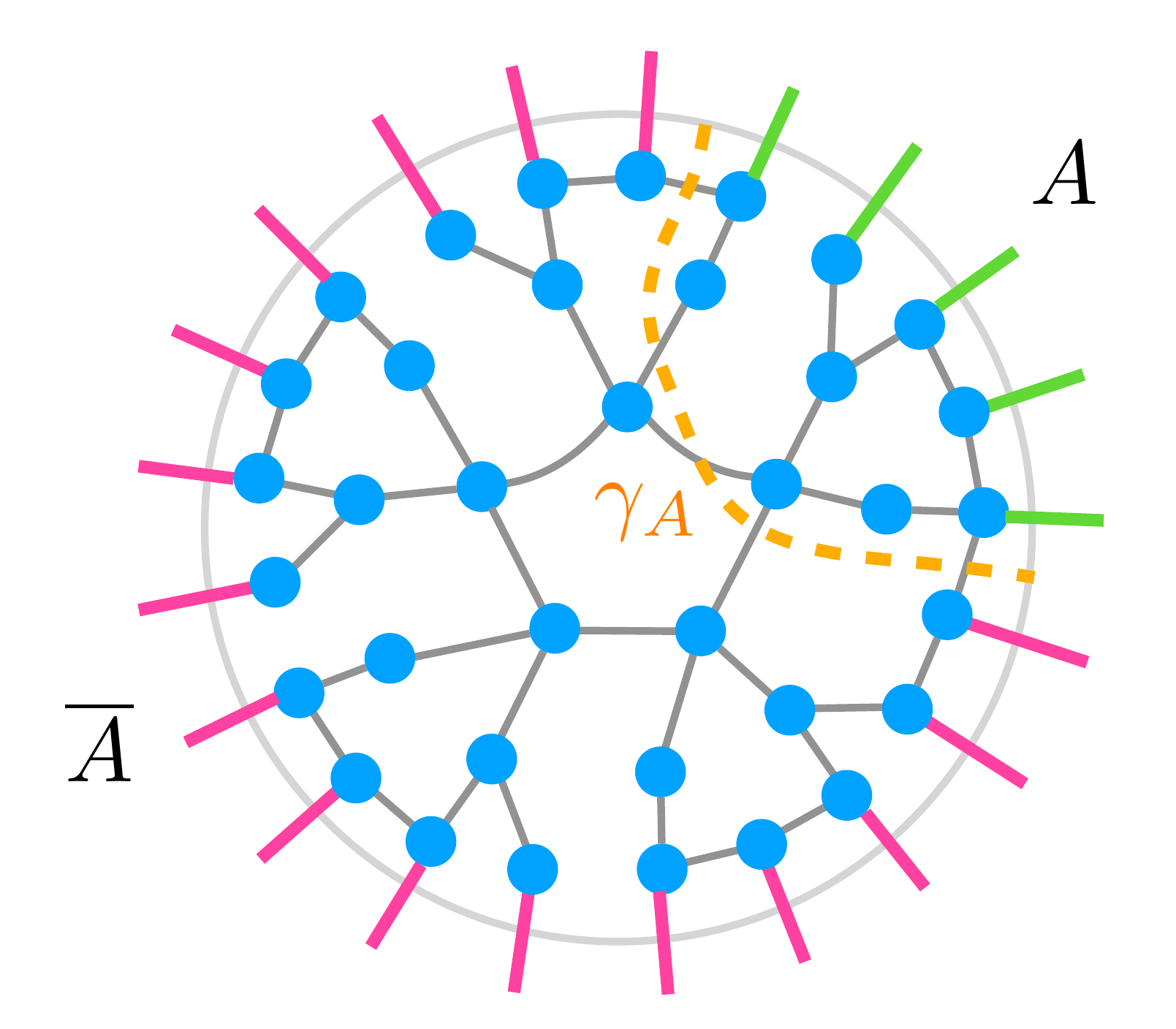}
\caption{A random tensor network defined on a planar graph. The curve $\gamma_A$ denotes a ``minimal cut'' (i.e. a path that cuts through a minimal number of bonds) in the bulk with the endpoints terminating at the boundaries of the subregion $A$.}
\label{fig:minimal_cut}
\end{figure}

Another model that exhibits an entanglement transition is the Random Tensor Network (RTN)~\cite{hayden2016holographic,vasseur2018RTN,Lopez_Piqueres_2020}, that is closely related to hybrid circuits.
A generic tensor network, as depicted in Fig.~\ref{fig:minimal_cut}, represents a one-dimensional quantum state using tensors living in a two-dimensional bulk. The physical degrees of freedom correspond to the uncontracted dangling bonds at the boundary. Such a tensor network state not only provides a useful variational wavefunction for simulating a variety of quantum many-body systems~\cite{perez2006matrix, schollwock2011density, verstraete2008matrix, orus2014practical,Vidal_2008}, but also serves as a conceptual tool for understanding the entanglement properties of many-body wavefunctions ~\cite{hayden2016holographic,jahn2018holography}. For example, consider a random tensor network where each individual tensor is drawn independently at random with a uniform distribution over Haar probability measure. In the limit where the bond dimension $D$ of each tensor becomes infinite, it has been shown that the entanglement of a boundary subregion saturates the upper bound of a minimal cut through the bulk~\cite{hayden2016holographic} (see Fig.~\ref{fig:minimal_cut}). This reproduces the Ryu-Takayanagi (RT) formula in AdS/CFT correspondence~\cite{PhysRevLett.96.181602}, exemplifying the general holographic duality in a tensor network construction~\cite{PhysRevD.86.065007}. For planar graphs such that this minimal cut scales as the length of the boundary subregion, this indicates a maximally entangled boundary state with volume-law scaling. Moving away from the infinite $D$ limit, the entanglement entropy deviates from the RT formula.
Ref.~\cite{vasseur2018RTN} maps the entanglement entropy calculation of a random tensor network state to a classical statistical mechanics model of \YL{a random} magnet, using a replica method. This mapping suggests an entanglement transition as $D$ is varied, corresponding to an ordering transition in the classical spin model. However, the critical properties of the resulting statistical mechanics model in general cannot be solved analytically, nor can it be \MF{easily} accessed numerically.

\begin{table*}[t]
\begin{tabular}{c|c|c|c|c|c|c}
\hline
\multirow{2}{*}{\st{Universal} Data at the critical point} & \multicolumn{4}{c|}{Random stabilizer tensor networks} & \multirow{2}{*}{Hybrid Clifford circuits ($D=2$)} & \multirow{2}{*}{\XC{$S_0$ in Haar} (Percolation)} \\ \cline{2-5}
 & $D=3$    & $D=5$   & $D=23$   & $D=503$   &                           &                              \\ \hline \hline
$p_c$ &  0.188    &  0.354   &   0.476   &   0.499    &  0.160~\cite{li2019hybrid, li2020conformal}  &    0.5                        \\ \hline
$h_{a|b}$  &  0.48 &  0.38   &  0.29 &  0.28 &  0.76~\cite{li2020conformal} &  $\frac{\sqrt{3}}{2\pi} \approx 0.276$~\cite{jiang2019critical}  
                    \\ \hline
$h_{f|f}^{(1)}$  &  0.37  &  0.36   &   0.33   & 0.33  & 0.41~\cite{li2020conformal}  & $\frac{1}{3}\approx 0.33 $                            \\ \hline
$x_{p.b.c}$ &  0.121 & 0.119 & 0.106  &  0.106  & 0.125~\cite{li2020conformal}  &  $\frac{5}{48}\approx 0.104$                                                     \\ \hline
$h_{a|a}^{(1)}$  & 2.0 & 2.0 & 2.0 &  2.0 & 2.0 ~\cite{li2020conformal, sang2020entanglement} &  2.0~\cite{nahum2018transition}                             \\ \hline
$\Delta$ &  2.8  &  2.5   &   2.1   &   2.1    &  3.0 ~\cite{sang2020entanglement}  &        2.0~\cite{sang2020entanglement}                      \\ \hline
\end{tabular}
\caption{A comparison of operator scaling dimensions at the critical point among RSTNs with various bond dimensions $D$ obtained in this work, hybrid Clifford circuits for qubits ($D=2$), and \XC{zeroth R\'enyi entropy $S_0$ in hybrid Haar circuits described by first-passage} percolation CFT~\cite{nahum2018transition, li2020conformal}.
\ZCY{ Notice that the prefactor in front of the logarithmic scaling of the critical entanglement entropy in natural log is given by $2h_{a|b} {\rm ln} D$ for periodic boundary condition and $h_{a|b} {\rm ln}D$ for open boundary condition, as shown in Eq.~(\ref{eq:S_ln}).}
The scaling dimension $h_{f|f}^{(1)}$ is identified with that of the boundary spin operator in percolation: $h_{1,3}=\frac{1}{3}$, and $x_{p.b.c}$ is identified with the bulk spin (magnetization) operator $\Delta_{\sigma}=2h_{\frac{1}{2},0}=\frac{5}{48}$.
\YL{
The scaling dimension $\Delta$ is that of the mutual negativity, and its values for the hybrid random Clifford circuit at $D=2$ and for $S_0$ in the hybrid random Haar circuit are discussed extensively in Ref.~\cite{sang2020entanglement}.}
}
\label{table:scaling}
\end{table*}

In this work, we instead consider a numerically tractable subclass
of tensor networks, 
\YL{namely those with local tensors being random stabilizer states}~\cite{gottesman1997thesis, gottesman1998heisenberg, aaronson2004chp},
which we refer to as Random Stabilizer Tensor Networks (RSTNs), \YL{first introduced in Ref.}~\cite{nezami2016RSTN}. 
Throughout this paper we consider such tensor networks on a two-dimensional square lattice.
The bond dimension of our RSTNs can take the value of a prime power $D = q^n$, by construction~\cite{gottesman1999higherdim}.
In this work, we will focus on the case where $n = 1$, and $D$ itself is a prime number.
By varying the bond dimension $D$, we find that the entanglement entropy of RSTNs satisfies volume law scaling when $D\geq 3$ and has area law scaling at $D=2$, similar to the entanglement transition predicted in RTNs~\cite{vasseur2018RTN}.

We then combine the idea of hybrid quantum circuits and tensor networks, and focus on RSTNs subject to randomly applied single-qudit measurements in the bulk that collapse the dimension of the measured bond to unity, effectively breaking the bonds.
Upon tuning the measurement rate $p$ in the bulk, we observe a continuous entanglement phase transition for each prime $D\geq 3$ separating a volume-law entangled phase at small $p$ and an area-law entangled phase at large $p$. We further extract the universal entanglement properties at the critical points by putting RSTNs on finite rectangles with uniform or mixed boundary conditions, thereby relating the entanglement calculations to universal operator scaling dimensions of the underlying CFT.
Through extensive numerical simulations of RSTNs, we obtain a collection of operator scaling dimensions at each $D$.
\YL{The scaling dimensions for larger values of $D$ (where $D = 23$ and $D = 503$) are close to their counterparts in the CFT of critical first-passage percolation, but clear deviations from percolation for smaller values of $D$ (where $D= 3$ and $D= 5$) are found.
Indeed, the universality classes at different values of $D$ appear to differ from one another, especially clearly when they are distinguishable from those in percolation (i.e. when $D$ is not large).  
}  \MF{This suggests a different universality class for each prime $D$.}

Furthermore, we study universal entanglement properties in the volume-law phase.  We find quantitative agreement with the scaling of 
a directed polymer in a random environment~\cite{ PhysRevLett.54.2708, PhysRevLett.55.2923, PhysRevLett.55.2924}, the same behavior as in the non-thermal volume law phase in hybrid circuits~\cite{li2021entanglement}.

\subsection{Summary of results}

Our main results in this work are summarized in Table~\ref{table:scaling}, where we list the full collection of operator scaling dimensions at the critical point in RSTNs with different $D$, and compare with those in hybrid Clifford circuits obtained in previous works, and with percolation CFT. 

In Sec.~\ref{sec:RNST} we introduce the general setup of RSTNs and methods for numerical simulations. Each individual tensor in a RSTN encodes a random stabilizer state that can be represented in terms of the stabilizers. Using the language of projected entangled pair states (PEPS), tensor contractions correspond to projections onto the Bell-pair state on each bond, which is equivalent to measuring the Pauli operators $\{X_1 X_2, Z_1 Z_2^{-1}\}$ whose outcomes are \textit{forced} to be both +1.

We start by examining the entanglement properties in RSTN states without bulk single-qudit measurements (i.e., only tensor contractions) in Sec.~\ref{sec:no_measurement}. We find numerically that the resulting tensor network state has volume-law entanglement for $D\geq 3$, and area-law entanglement for $D=2$. This is in qualitative agreement with the scenario suggested by a previous statistical mechanics mapping of random (non-stabilizer) tensor networks~\cite{vasseur2018RTN}, \MF{where they argue for the existence of a critical point upon varying the bond dimension $D$, provided $D$ is treated as a continuous variable (as it can be in the spin model mapping).}

In Sec.~\ref{sec:measurement}, we present numerical results for the entanglement phase transitions in RSTNs subject to random breaking of bulk bonds (i.e. randomly applied forced single-qudit measurements in the bulk), where we extract the set of operator scaling dimensions summarized in Table~\ref{table:scaling}.
With the precise definitions of these scaling dimensions in terms of correlation functions given in Sec.~\ref{sec:measurement}, here we first give a brief summary of how each of them is related to a simple physical quantity of interest~\cite{li2020conformal}.
\begin{itemize}
    \item $h_{a|b}$: the entanglement entropy of a subregion $A$ of size $L_A$ on an infinite cylinder or a rectangle with uniform boundary condition [Fig.~\ref{fig:bc}(a)], when the total system size goes to infinity\footnote{\ZCY{Here the entanglement entropy is defined in natural log: $S(\rho)=-{\rm tr}(\rho \ {\rm ln} \rho)$. This differs by a factor of ${\rm ln}D$ from the definition in log base $D$. We explicitly show this factor of ${\rm ln}D$ here, so that $h_{a|b}$ is finite in the large $D$ limit.
    }}:
    \begin{equation}
        S(A) = 2 h_{a|b}\ {\rm ln} D \times {\rm ln} L_A.
        \label{eq:S_ln}
    \end{equation}
    
    \item $h_{a|a}^{(1)}$: the mutual information between two small regions $[z_1,z_2]$ and $[z_3,z_4]$ that are far apart:
    \begin{equation}
        I([z_1,z_2], [z_3,z_4]) \propto \eta^{h_{a|a}^{(1)}}, \quad {\rm as} \ \eta \rightarrow 0,
    \end{equation}
    where $\eta=z_{12}z_{34}/z_{13}z_{24}$ is the cross-ratio;
    
    \item $\Delta$: the mutual negativity between two small regions $[z_1,z_2]$ and $[z_3,z_4]$ that are far apart:
    \begin{equation}
        N([z_1,z_2], [z_3,z_4]) \propto \eta^{\Delta}, \quad {\rm as} \ \eta \rightarrow 0;
    \end{equation}
    
    \item $h_{f|f}^{(1)}$: the entanglement entropy between the top and bottom boundary of a rectangle with free boundary condition on the vertical edges [Fig.~\ref{fig:bc}(c)], when the aspect ratio of the system $\tau \equiv \frac{L_y}{L_x}$ becomes large:
    \begin{equation}
        S \propto {\rm exp} \left(-\pi h_{f|f}^{(1)} \tau \right) ;
    \end{equation}
    
    \item $x_{p.b.c}$: the entanglement entropy between the top and bottom boundary of a cylinder, when the aspect ratio of the system becomes large:
    \begin{equation}
        S \propto {\rm exp} \left(-2\pi x_{p.b.c} \tau \right).
    \end{equation}
\end{itemize}

We see in Table~\ref{table:scaling} that these universal data \MF{agree} 
with the percolation CFT as the prime bond dimension $D$ becomes large.
\YL{
This suggests a 
geometric minimal cut picture in the \MF{$D \rightarrow \infty$ limit}, which we discuss in Sec.~\ref{sec:discussion}.}
However, results at smaller $D$ clearly deviate from percolation.

\YL{We also study RSTNs with reduced randomness, including those with \textit{translationally invariant} tensors but randomly placed measurements, and those with random tensors but \textit{spatially periodic} measurements.
Overall, we find critical properties that are consistent with fully random RSTNs, but generally \XC{with a different $p_c$.} 
}

In Sec.~\ref{sec:kpz} we study universal entanglement properties in the volume-law phase. In particular, we are interested in the sample-to-sample fluctuations of the entanglement entropy:
\begin{equation}
   \delta S(L_A) \equiv \sqrt{\langle S(L_A)^2 \rangle - \langle S(L_A) \rangle^2} \propto (L_A)^{\beta},
\end{equation}
where we find $\beta \approx 0.34$. This agrees with the universal Kardar-Parisi-Zhang (KPZ) scaling $\beta = \frac{1}{3}$ of the free energy fluctuations of a directed polymer in a random medium~\cite{PhysRevLett.54.2708, PhysRevLett.55.2923, PhysRevLett.55.2924, PhysRevLett.56.889}, which was recently proposed to describe the volume-law phase in hybrid quantum circuits~\cite{PhysRevB.103.104306, li2021entanglement}. Drawing a connection to QECC, we also obtain scaling of the (contiguous) code distance~\cite{PhysRevB.103.104306} with system size, and the results are again consistent with KPZ scaling.

\YL{
Finally, we note that upon replacing all the forced measurements (responsible for implementing the bond contractions and for breaking the bonds) by \emph{projective} measurements (see Appendix~\ref{app:prob_measure} for technical details on this distinction), the location of the critical points and the critical exponents do not appear to change.
For this reason, we do not present data on RSTNs with projective measurements.
}

\section{Random stabilizer tensor networks}
\label{sec:RNST}

\begin{figure}[t]
\centering
\includegraphics[width=0.4\textwidth]{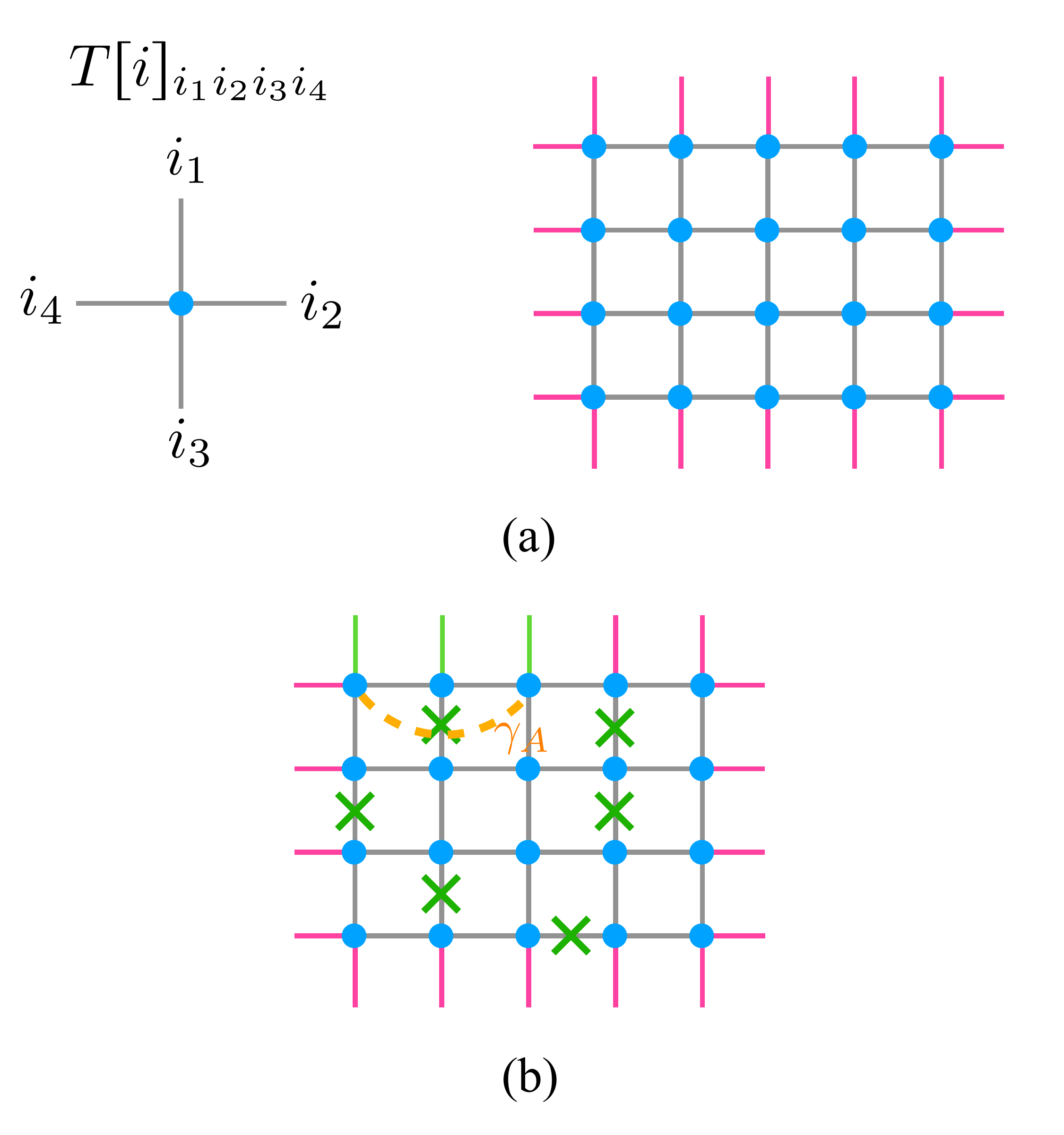}
\caption{(a) A tensor network state defined on a square lattice. The bulk and boundary legs are shown in grey and purple, respectively. A quantum state living on the boundary is constructed by contracting all bulk tensor legs. (b) A tensor network subject to randomly applied 
\YL{breaking}
(green crosses) of the bulk tensor legs. The dotted orange line denotes the minimal cut through the bulk of a boundary region.}
\label{fig:network}
\end{figure}

We first define random tensor network states in general before specializing to stabilizer tensor networks.
Consider an arbitrary graph $G=(V,E)$ where $V$ and $E$ denote the collection of nodes and edges.
With each node $i \in V$ one associates a rank-$l$ tensor $T[i]_{i_1\ldots i_l}$, where $l$ is the degree of the node. Each tensor index runs from 0 to $D-1$: $i_k=0, 1, \dots, D-1$, with $D$ being the ``bond dimension'' of the tensor.
To construct a tensor network state, one further specifies a set of boundary edges $E_\partial$, and obtains a quantum state living on $E_\partial$ by performing tensor contractions (i.e. tracing) over all bulk edges:
\begin{equation}
    |\psi\rangle = \sum_{\{\mu_e\},\ e \in E_\partial}{\rm tTr} \prod_{i \in V} T[i]\ |\{\mu_e\} \rangle,
    \label{eq:tensor_network_state}
\end{equation}
where $|\{\mu_e\}\rangle$ denotes a set of basis states for the dangling boundary legs $e\in E_{\partial}$, and tTr denotes \MF{a tensor contraction over all bulk bonds.} Notice that the state $|\psi\rangle$ is in general unnormalized. For simplicity, in this work we consider RSTNs constructed on a two-dimensional square  lattice ($l=4$), as depicted in Fig.~\ref{fig:network}, where the uncontracted boundary legs are highlighted in purple.

One can alternatively view the tensor network state~(\ref{eq:tensor_network_state}) from the perspective of PEPS, which turns out to be useful within the stabilizer formalism. We define a quantum state for each site $i$
\begin{equation}
    |T[i]\rangle = \sum_{i_1 \ldots i_4 =0}^{D-1} T[i]_{i_1 i_2 i_3 i_4}\ |i_1 i_2 i_3 i_4 \rangle,
    \label{eq:tensor_state}
\end{equation}
where the tensor components are interpreted as wavefunction amplitudes of the virtual degrees of freedom (qudits) in the bulk. On each link of the network $e \in E$, we define a maximally entangled Bell state
\begin{equation}
    |I_e \rangle = \frac{1}{\sqrt{D}}\sum_{i_e=j_e=0}^{D-1} |i_e j_e \rangle,
    \label{eq:bell}
\end{equation}
where sites $i$ and $j$ share the link $e$. Using the PEPS language, the quantum state~(\ref{eq:tensor_network_state}) can be constructed by projecting the virtual degrees of freedom on each bulk bond to the Bell state $|I_e\rangle$
\begin{equation}
    |\psi\rangle = \bigotimes_{i \in V} \bigotimes_{e \in E-E_{\partial}} \langle I_e | T[i]\rangle.
    \label{eq:bell_projection}
\end{equation}
It is easy to see that this projection is equivalent to the tensor contraction in~(\ref{eq:tensor_network_state}).

We \MF{primarily} focus on random stabilizer tensor networks (RSTNs) where the local quantum state~(\ref{eq:tensor_state}) associated with each tensor is a stabilizer state drawn at random. This allows for an efficient representation of the tensor network with prime bond dimensions using the nonbinary stabilizer formalism~\cite{gottesman1999higherdim}, 
which we now briefly review.

Define the following generalization of the Pauli operators to qudits with local Hilbert space dimension $D$:
\begin{equation}
    Z = \sum_{j=0}^{D-1} \omega^j |j \rangle \langle j |, \quad  X = \sum_{j=0}^{D-1} |j \rangle \langle j+1 |,
\end{equation}
where $\omega = e^{i \frac{2\pi}{D}}$. They satisfy $Z^D=X^D=1$ and the commutation relation $XZ = \omega ZX$. For the special case of $D=2$, the above definition coincides with the qubit Pauli matrices.
The \textit{Pauli group} $\mathcal{P}_N$ acting on $N$ qudits consists of all Pauli strings of the form
\begin{equation}
    \omega^r Z_1^{u_1} X_1^{v_1} \otimes Z_2^{u_2} X_2^{v_2} \otimes \cdots \otimes Z_N^{u_N} X_N^{v_N},
\end{equation}
where $r\in \mathbb{Z}_D$ and the $2N$-tuples $({\bm u}, {\bm v})$ are vectors in $\mathbb{Z}_D^{\otimes 2N}$. When $D$ is prime, we may identify $\mb{Z}_D$ with the finite number field with $D$ elements, denoted $\mb{F}_D$.
In this case,
an $N$-qudit \textit{stabilizer state} $|\psi\rangle$ can be defined as the simultaneous eigenstate with eigenvalue $+1$ of an Abelian subgroup of $\mathcal{P}_N$, known as the \textit{stabilizer group} $\mathcal{S}\subseteq \mathcal{P}_N$: $g_i|\psi\rangle = |\psi \rangle$, $\forall g_i \in \mathcal{S}$.
We will mostly work with pure states, for which $\mc{S}$ is generated by $N$ independent, mutually commuting Pauli string operators, and $|\mc{S}| = D^N$.

For a RSTN as shown in Fig.~\ref{fig:network}, each local tensor encodes a four-qudit random stabilizer state that can be constructed as follows. Starting from a reference state $|0000\rangle$ in the computational basis, we randomly draw a unitary from the four-qudit Clifford group $U_i \in \mathcal{C}(D;4)$. The unitary $U_i$ on site $i$ defines a tensor $T[i]$ according to
\begin{equation}
    T[i]_{i_1i_2i_3i_4} = \langle i_1 i_2 i_3 i_4 |U_i|0000\rangle,
\end{equation}
or, equivalently,
\begin{equation}
    |T[i]\rangle \equiv U_i|0000\rangle = \sum_{i_1\ldots i_4=0}^{D-1} T[i]_{i_1 i_2 i_3 i_4} |i_1 i_2 i_3 i_4 \rangle.
\end{equation}
On the full lattice, we have the state
\begin{equation}
    |\Psi(\mathcal{U})\rangle
    = \bigotimes_{i=1}^{|V|}
    \ket{T[i]}
    = \mathcal{U} \bigotimes_{i=1}^{|V|} |0000\rangle_i,
\end{equation}
where $\mathcal{U} = \bigotimes_{i=1}^{|V|} U_i$ is an element of $\mathfrak{C} \coloneqq \mathcal{C}(D;4)^{\otimes |V|}$. Therefore, the (unnormalized) tensor network boundary state is obtained via
\begin{equation}
    \rho(\mathcal{U}) \equiv |\psi\rangle \langle \psi | = 
    {\rm Tr}_{e\in E -  E_{\partial}}
    \left[\mathbb{P}|\Psi(\mathcal{U})\rangle \langle \Psi(\mathcal{U})| \right],
    \label{eq:rho_force}
\end{equation}
where $\mathbb{P}$ denotes projectors onto the Bell state~(\ref{eq:bell}) on all bulk bonds.

\subsection{Forced measurements (tensor contractions) versus projective measurements}

Locally, each $4$-qudit random stabilizer state can be represented by four mutually commuting Pauli string operators.
We now explain how the tensor contractions in~(\ref{eq:tensor_network_state}) can be accounted for in the stabilizer formalism. 
In terms of the PEPS construction, projections onto the Bell state~(\ref{eq:bell_projection}) are equivalent to \textit{forced measurement} of the following two-qudit Pauli operators on the projected bond $e$: 
\begin{equation}
\{ X_{i_e} X_{j_e}=+1, Z_{i_e} Z_{j_e}^{-1}=+1 \}.
\label{eq:bell_force}
\end{equation}
To properly construct the tensor network state corresponding to the definition~(\ref{eq:tensor_network_state}), the Bell-pair measurement must be \textit{forced}.
In practice, one needs to ``post-select" quantum trajectories such that the measurement outcomes of $X_{i_e} X_{j_e}$ and $Z_{i_e} Z_{j_e}^{-1}$ are both equal to +1 on all bonds in the lattice; trajectories with the ``wrong'' measurement outcomes must be rejected.
For finite size systems at finite $D$, there are choices of $\mc{U}$ for which these $+1$ outcomes occur with zero Born probability, corresponding to a case where the state $\rho(\mc{U})$ as defined in (\ref{eq:rho_force}) vanishes.
However, in calculating the ensemble average of any physical quantity, we should only include tensor networks for which $\rho(\mathcal{U})$ does not vanish.

Alternatively, we may replace the forced measurments (i.e. contraction) on each bond by 
\textit{projective measurements}, as usually considered in hybrid quantum circuits, where the measurement outcomes are random and are sampled according to Born's rule, rather than post-selected.

In this work,
we consider both RSTNs with forced measurements (i.e. contraction) as in Eq.~(\ref{eq:rho_force}), and those with all forced measurements replaced by projective measurements.
We denote the state in the latter case as $\rho(\mathcal{U}; {\bm m})$, where ${\bm m} = (m_{e_1}^{ZZ}, m_{e_1}^{XX}, \ldots, m_{e_{|E|}}^{ZZ}, m_{e_{|E|}}^{XX})$ denotes the measurement outcomes of all bonds and labels all admissible quantum trajectories with nonzero probability.
By definition, $\rho(\mathcal{U}) \equiv \rho(\mathcal{U};{\bm m}_0)$, where ${\bm m}_0 = (m_{e_1}^{ZZ}=+1, m_{e_1}^{XX}=+1, \ldots, m_{e_{|E|}}^{ZZ}=+1, m_{e_{|E|}}^{XX}=+1)$. Define the ensemble of $\mathcal{U}$ for which $\rho(\mathcal{U}) \neq 0$:
\begin{equation}
    \mathfrak{C}^> \coloneqq \{ \mathcal{U} \in \mathfrak{C}: \ \rho(\mathcal{U}) \neq 0\}.
\end{equation}
The expectation value of a physical quantity $\mathcal{O}$ in the two tensor network models that we consider is thus defined as:
\begin{itemize}
    \item Forced measurement:
    \begin{equation}
        \langle \mathcal{O} \rangle^{\rm f} = \mathbb{E}_{\mathfrak{C}^>} \mathcal{O}[\rho(\mathcal{U})]; 
    \end{equation}
    \item Projective measurement:
    \begin{equation}
        \langle \mathcal{O} \rangle^{\rm p} = \mathbb{E}_{\mathfrak{C}}\left( \sum_{\bm m} \mathcal{O}[\rho(\mathcal{U}; {\bm m})] \times {\rm Tr}[\rho(\mathcal{U};{\bm m})] \right).
    \end{equation}
\end{itemize}

Clearly, depending on whether the bond measurements are forced or projective, we have different statistical ensembles of RSTNs, and they require different sampling algorithms.
Sampling from the ``projective measurement ensemble'' is straightforward, but some tricks are needed for the  ``forced measurement ensemble'' in order to generate a large number of samples efficiently.
We detail these considerations in Appendix~\ref{app:prob_measure}.

The two ensembles of RSTNs should be viewed as different models, and can have different phase diagrams and/or critical properties, \emph{a priori}.
\YL{See Refs.~\cite{jian2018spin, jian2020fermionRTN, nahum2020alltoall} for relevant discussions on the differences between the physics of forced and projective measurements in similar models.}
In the rest of the paper, we will mainly present numerical results for the forced measurement case, as it is more natural from a conventional tensor network perspective. We will briefly comment on the results for projective measurements in Sec.~\ref{sec:discussion}.

\section{Entanglement scaling of RSTNs on the square lattice}
\label{sec:no_measurement}

The first question regarding the tensor network state defined in Eq.~(\ref{eq:tensor_network_state}) that naturally arises is: how does the entanglement entropy of a boundary region $A$ scale with $|A|$ for a given $D$?
It has been shown that for a random (nonstabilizer) tensor network, the entanglement entropy of a boundary subsystem saturates the ``minimal cut" bound in the limit of $D\rightarrow \infty$: $S(A) = {\rm min}\ |\gamma_A| \times \YL{\ln D} $~\cite{hayden2016holographic}, where we minimize over all paths $\gamma_A$ terminating at the boundary of $A$ (see Fig.~\ref{fig:minimal_cut}).
The same formula holds for RSTNs as well~\cite{nezami2016RSTN}. 
The above expression is akin to the Ryu-Takayanagi formula in AdS/CFT duality~\cite{PhysRevLett.96.181602}.
Indeed, it is possible to reproduce the entanglement scaling of a CFT ground state using a tensor network triangulation of the hyperbolic space~\cite{hayden2016holographic}.
For the regular square lattice considered in our case, this indicates that the boundary region is volume-law entangled in the limit of infinite $D$.
\MF{Note, however, that} \YL{results in Refs.~\cite{hayden2016holographic, nezami2016RSTN} only strictly apply provided that we take $D \to \infty$ while keeping $L$ finite.
Here, we consider the thermodynamic limit instead, taking $L \to \infty$ while keeping $D$ finite.\footnote{\ZCY{For the values of $D$ and $L$ considered in this work, the proof in Refs.~\cite{hayden2016holographic, nezami2016RSTN}, which requires taking $D\gtrsim \exp(L)$, does not directly apply. See Sec.~\ref{sec:discussion} for more discussions on this.}}}

\begin{figure}[t]
\centering
\includegraphics[width=0.43\textwidth]{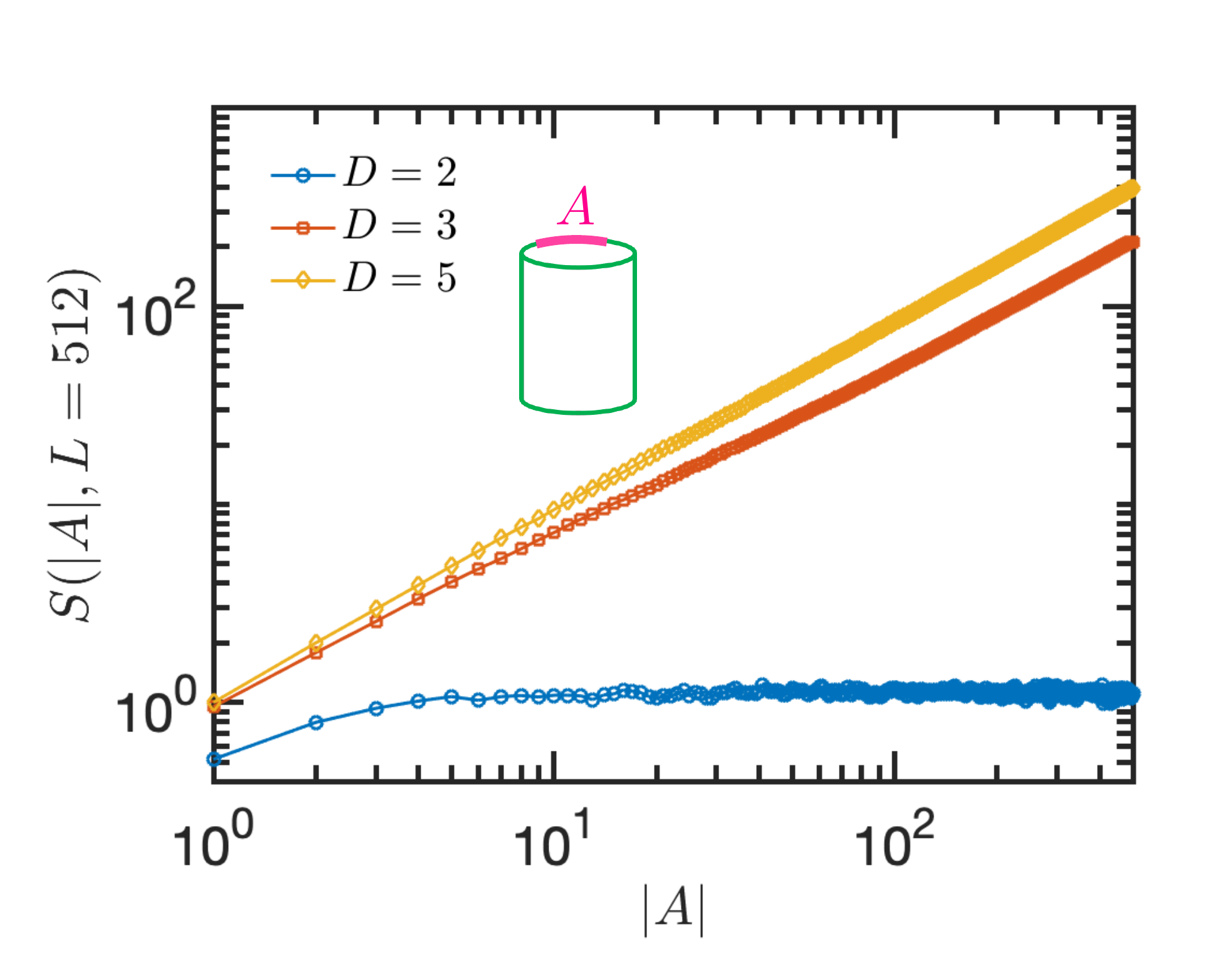}
\caption{Scaling of the entanglement entropy $S(A)$ of a boundary region $A$ for RSTNs with bond dimensions $D=2, \ 3$ and 5. The numerics are performed for tensor networks of size $512 \times 512$, with periodic boundary condition in one direction (i.e. a finite cylinder), as depicted in the inset. The subregion $A$ is chosen as part of the top boundary of the cylinder.}
\label{fig:no_measurement}
\end{figure}

In Fig.~\ref{fig:no_measurement}, we show numerical results on the entanglement entropy of a boundary region for RSTNs with different bond dimensions. The results clearly demonstrate a volume-law entangled boundary state for $D \geq 3$, and an area-law entangled state for $D=2$.
Provided one can define a generalization of the tensor network model that allows for non-integer $D$, this result would suggest that an entanglement transition could occur for bond dimension $2<D<3$.

In fact, in Ref.~\cite{vasseur2018RTN} the authors use a replica method and map the calculation of $S(A)$ in RTNs with arbitrary $D$ to that of the free energy cost of a boundary twist in a classical statistical mechanics model of a magnet, where the bond dimension plays the role of an effective temperature: $\beta J \propto {\rm log}D$.
This statistical mechanics model of magnet has an ordered phase at low temperature $\beta J > (\beta J)_c$, where the free energy cost of a bulk domain wall due to the boundary twist is extensive, and a disordered paramagnetic phase at high temperature $\beta J < (\beta J)_c$, where the domain wall has a free energy cost of order $\mathcal{O}(1)$.
This corresponds to an entanglement transition from a volume-law scaling at $D>D_c$ to an area-law scaling at $D<D_c$. 
However due to the \ZCY{difficulty in taking the replica limit of the resulting statistical mechanics model,}
the nature of the phase transition is not yet fully understood.

Although a statistical mechanics mapping for the RSTN is not available at the moment, \MF{provided one exists and allows a description for continuously variable $D$, we might then} expect that a similar entanglement transition as the bond dimension is varied may be present in RSTNs as well.  This \MF{would} imply the existence of a critical point at a non-integer bond dimension $2<D_c<3$.

\section{Entanglement transitions driven by random breaking of bulk bonds} 
\label{sec:measurement}

\begin{figure*}[t]
\centering
\includegraphics[width=0.95\textwidth]{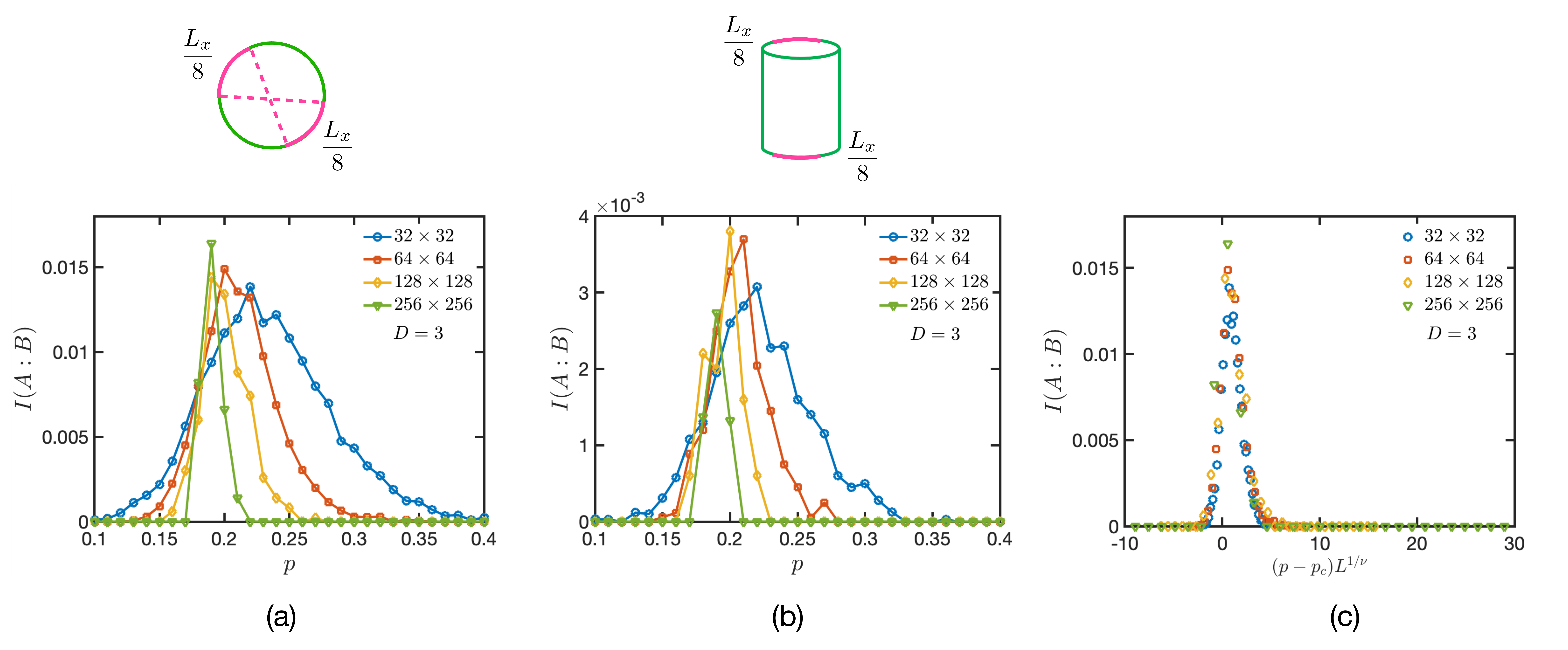}
\caption{Mutual information between two antipodal subregions of size $|A|=|B|=L_x/8$, for RSTNs of size $L_x \times L_y$ and bond dimension $D=3$ on a cylinder (periodic boundary condition along the $x$ direction). (a)Both $A$ and $B$ are placed at the top boundary of the cylinder. (b)$A$ and $B$ are placed separately at the top and bottom boundaries, respectively. (c)Data collapse using the scaling form $I(A:B) = f[(p-p_c)L^{1/\nu}]$, which yields $p_c \approx 0.19$ and $\nu \approx 1.13$.}
\label{fig:MI_d3}
\end{figure*}

We now study the effect of randomly applied bulk bond breakings on the entanglement properties of the boundary region [see Fig.~\ref{fig:network}(c)]. 
In the tensor network state~(\ref{eq:tensor_network_state}), breaking a specific bond $e = (i, j)$ reduces the dimension of that bond to one, such that the tensor components $T[i]_{i_1\ldots i_k \dots i_4}$ are nonzero only for one particular value of $i_k$, say $i_k=0$: $T[i]_{i_1\ldots i_k=0 \ldots i_4}\neq 0$, and $T[i]_{i_1\ldots i_k\neq 0 \dots i_4}=0$.
The tensor on site $j$ is similarly modified by the breaking.
In the stabilizer formalism, this can be implemented by performing a single-qudit Pauli-$Z$ measurement on qudits $i_k$ and $j_k$, whose outcomes are forced to be $+1$:
\begin{equation}
\{Z_{i_k}=+1, Z_{j_k}=+1\},
\label{eq:measure}
\end{equation}
 consistent with tensor contractions.
 
 \ZCY{In RSTNs, the bonds in the bulk are randomly measured/broken with probability $p$. As a function of $p$, the lattice geometry itself undergoes a bond percolation transition at $p_{\rm perc}=0.5$. For $p > p_{\rm perc}$, the minimal path through the bulk that separates the tensor network into two parts (Fig.~\ref{fig:minimal_cut}) has a finite \MF{weight}, i.e., independent of the size of the boundary region. Since the entanglement entropy of a boundary region is upper bounded by the \MF{weight} of this minimal path, it must obey an area-law scaling in this regime. This suggests a possible entanglement phase transition as a function of $p$ with $D\geq 3$. The critical point $p_c$ for such a transition, however, may differ from $p_{\rm perc}$ in general.}

 Below, we shall first demonstrate the existence of measurement-induced entanglement transitions in RSTNs with $D\geq 3$ as diagnosed by the mutual information between two disjoint boundary regions on a cylindrical geometry. \ZCY{In particular, we will show that at finite $D$, the value of the critical point $p_c$ is smaller than $p_{\rm perc}$, indicating that the entanglement phase transition occurs ahead of the geometric phase transition of the underlying lattice structure, as is also the case in hybrid quantum circuits~\cite{nahum2018transition, li2019hybrid, PhysRevB.99.224307, PhysRevB.98.205136}.}
 Then, we further unveil the universal entanglement properties at criticality using the machinery of boundary CFT, where the entanglement entropy is related to correlation functions of boundary scaling operators.
 In particular, we extract universal operator scaling dimensions by computing the entanglement entropy, mutual information and mutual negativity on a finite rectangular geometry subject to different choices of boundary conditions.

\subsection{Mutual information on a cylindrical geometry}

A convenient quantity that signals the existence of an entanglement transition is the 
mutual information between two disjoint boundary regions $A$ and $B$:
\begin{equation}
    I(A:B) = S(A) + S(B) - S(A \cup B).
    \label{eq:MI_def}
\end{equation}
In the thermodynamic limit, the mutual information between two small distant subsystems vanishes in both the volume-law and area-law phases~\cite{nahum2018transition, li2019hybrid}.
In the area-law phase, the boundary region is short-range entangled, and hence the mutual information between two small subregions that are far apart should quickly decay to zero.
In the volume-law phase, quantum information is scrambled across the entire system, thus the mutual information shared between any two small subregions also vanishes.
At criticality, the mutual information is enhanced due to the long-range correlations that decay algebraically in space.

\begin{figure}[tbh]
    \centering
\subfigure[]{
    \includegraphics[width=0.4\textwidth]{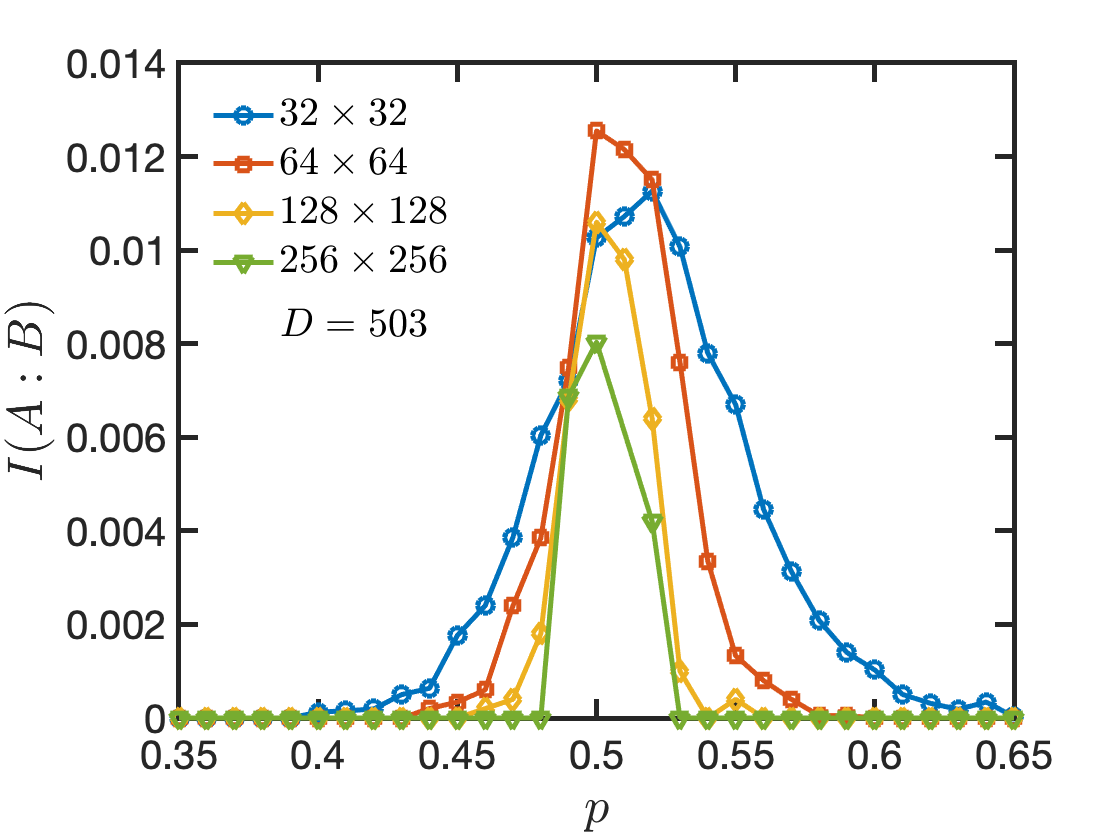}}
\subfigure[]{
    \includegraphics[width=0.4\textwidth]{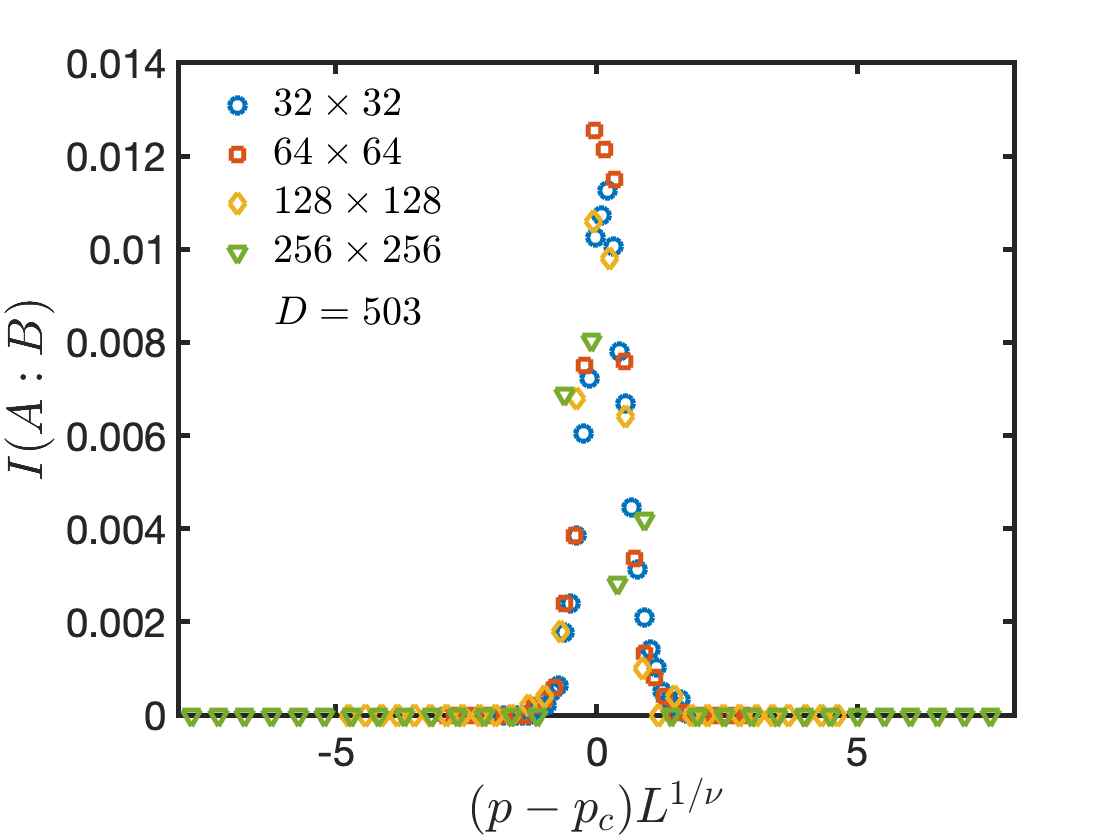}} 
    \caption{Same as Fig.~\ref{fig:MI_d3}, for $D=503$. In (a), we place both subregions $A$ and $B$ at the top boundary. The data collapse in (b) yields $p_c \approx 0.50$ and $\nu \approx 1.4$.
    }
    \label{fig:MI_d503}
\end{figure}

Consider a RSTN of size $L_x \times L_y$ with periodic boundary condition along the $x$ direction. In Fig.~\ref{fig:MI_d3}, we plot the mutual information between two antipodal subregions of size $|A|=|B|=L_x/8$ as a function of the measurement rate $p$, for RSTNs with $D=3$. Since a finite cylinder has a top and a bottom boundary, we consider either placing both $A$ and $B$ at the top boundary [Fig.~\ref{fig:MI_d3}(a)], or separately at the top and bottom boundary [Fig.~\ref{fig:MI_d3}(b)]. We find that the mutual information in both cases peaks at a critical value $p_c \approx 0.19$, and the peak becomes sharper as the system size increases. In Fig.~\ref{fig:MI_d3}(c), we collapse the data for different system sizes using the scaling form $I(A:B) = f[(p-p_c)L^{1/\nu}]$, which yields $\nu \approx 1.13$. In this data collapse, we have assumed that there exists a correlation length $\xi\sim |p-p_c|^{-\nu}$ which diverges at the critical point.

As the bond dimension $D$ increases, we find that the entanglement transition persists, with the critical point $p_c$ shifting upwards towards 0.5 (see Appendix~\ref{sec:MI_additional} for additional numerical results for $D=5$ and $D=23$). In Fig.~\ref{fig:MI_d503}, we show the mutual information versus $p$ for $D=503$. The data collapse yields a critical measurement rate $p_c \approx 0.50$, which is consistent with the critical point $p_{\rm perc}$ of two-dimensional bond percolation on a square lattice. Since in the limit of large $D$, the entanglement entropy of a boundary region of RSTNs  \MF{is expected to be} given by the geometric minimal cut formula $S(A) = {\rm min}\ |\gamma_A| \times {\rm ln}D$, calculation of the entanglement entropy in the presence of random measurement on bulk tensor legs amounts to finding a path with the minimal weight on a lattice where the weight of each individual bond is equal to either 0 with probability $p$, or 1 with probability $1-p$. The latter is known as ``first-passage percolation"~\cite{kesten1986aspects, chayes1986critical}.

The results shown above indicate that there exists a ``breaking/measurement''-induced phase transition in RSTNs for each $D\geq 3$.
In particular, \MF{in the large $D$ limit, the critical exponents approach those of bond percolation}, while for finite $D$ there appear to be distinct critical points for each prime $D$,
\MF{especially clear for small $D$.}
Next, we further probe the universal entanglement properties at these critical points.

\subsection{Finite rectangular geometry and boundary conditions}

\begin{figure}[t]
\centering
\includegraphics[width=0.43\textwidth]{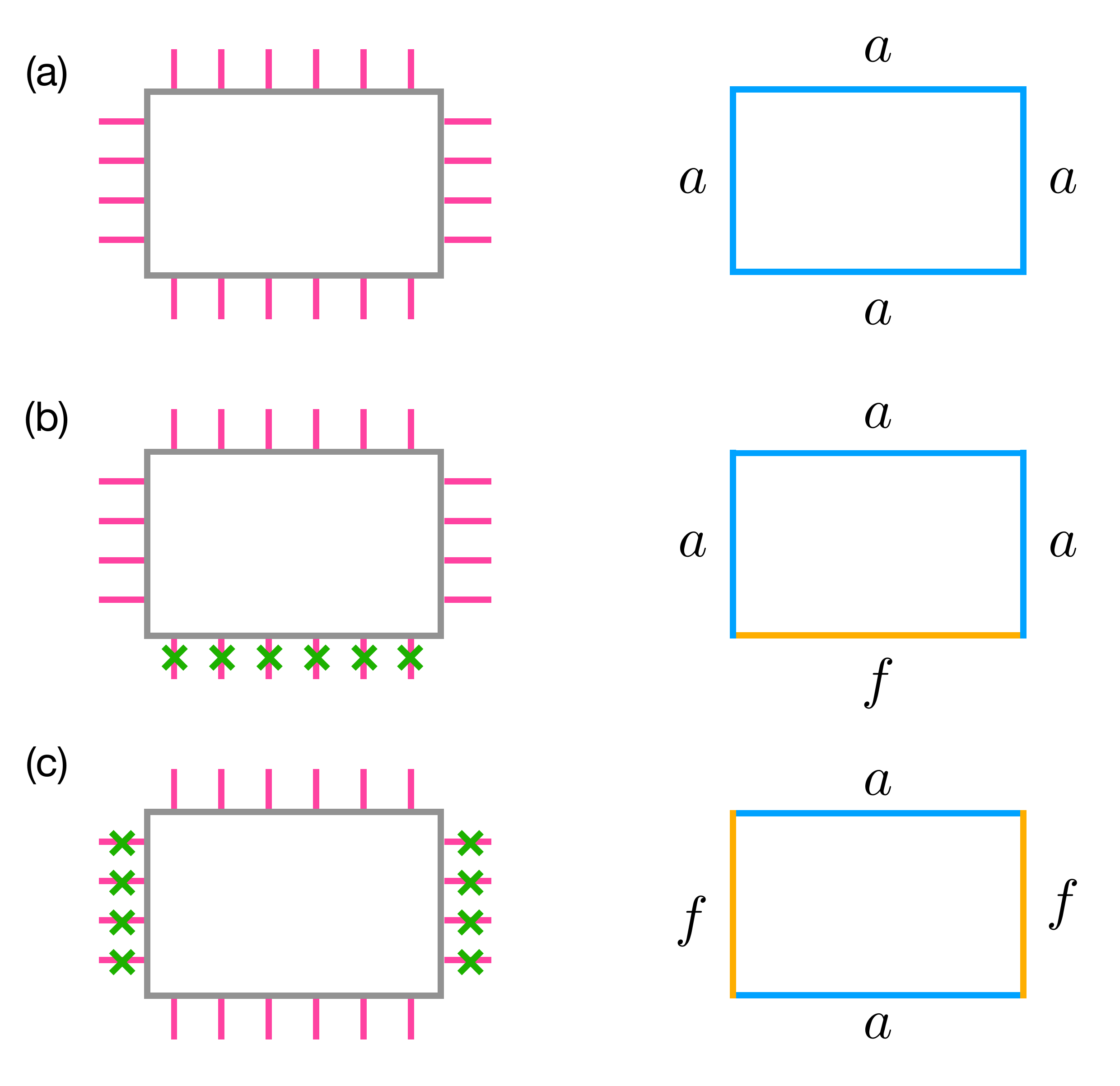}
\caption{Three types of boundary conditions that are considered in this work. A particular edge can either be \textit{free}, when all physical qudits are measured; or \textit{fixed}, in which case they are left uninterrupted by any measurement. We shall label the three cases as: (a) $aaaa$; (b) $afaa$; (c) $fafa$.}
\label{fig:bc}
\end{figure}

To explore the nature of the entanglement criticality, it is more convenient to consider tensor networks on a finite rectangle, with physical qudits on four edges~\cite{li2020conformal}. Boundary conditions need to be specified on each edge in order to define the system. There are two natural choices of boundary conditions in a tensor network with open boundaries: \textit{free}, which means that the physical qudits at the boundary are all measured according to Eq.~(\ref{eq:measure}); and \textit{fixed}, in which case they are left uninterrupted by any measurement. In Fig.~\ref{fig:bc}, we show three types of boundary conditions that we shall focus on in this work, where we label free and fixed boundary conditions by $f$ and $a$, respectively.

We postulate that the tensor network at criticality is described by a two-dimensional CFT, such that these two microscopic boundary conditions become two distinct \textit{conformal boundary conditions} at long distances. At the corner of the rectangle where two different types of conformal boundary conditions meet, a \textit{boundary condition changing} (bcc) operator is inserted~\cite{cardy1984conformal, cardy2004boundary}, which we denote as, e.g., $\phi_{a|f}$. 
We assume that these bcc operators transform as primary fields under conformal transformations~\cite{li2020conformal}.

Having introduced the boundary conditions in Fig.~\ref{fig:bc} and the notion of bcc operators, we are now in position to explain how the entanglement entropy of a subregion on the boundary can be related to correlation functions of the appropriate bcc operators in CFT. As shown in Refs.~\cite{bao2018spin, jian2018spin, vasseur2018RTN, li2020conformal,hayden2016holographic}, the entanglement entropy of a subregion $A$ in random hybrid circuits and random tensor networks can be written as the free energy cost associated with a boundary condition twist in region $A$:
\begin{equation}
    S(A) = - {\rm ln} \frac{Z(A)}{Z_{\rm bg}},
    \label{eq:EE}
\end{equation}
where $Z(A)$ is the partition function of the system with a different boundary condition (twist) imposed in region $A$, and $Z_{\rm bg}$ is the partition function of the background system in the absence of a twist. Eq.~(\ref{eq:EE}) has been derived analytically for Haar random circuits~\cite{bao2018spin, jian2018spin} and random tensor networks~\cite{vasseur2018RTN} using a replica method, and demonstrated numerically for hybrid Clifford circuits~\cite{li2020conformal} where an analytical derivation is currently absent. We thus conjecture that Eq.~(\ref{eq:EE}) holds for RSTNs as well. This conjecture will be substantiated in the next few subsections by extensive numerical simulations. But before that, let us first illustrate how Eq.~(\ref{eq:EE}) is applied in practice using a tensor network with boundary condition $aaaa$ [Fig.~\ref{fig:bc}(a)] as an example.

\begin{figure}[t]
\centering
\includegraphics[width=0.4\textwidth]{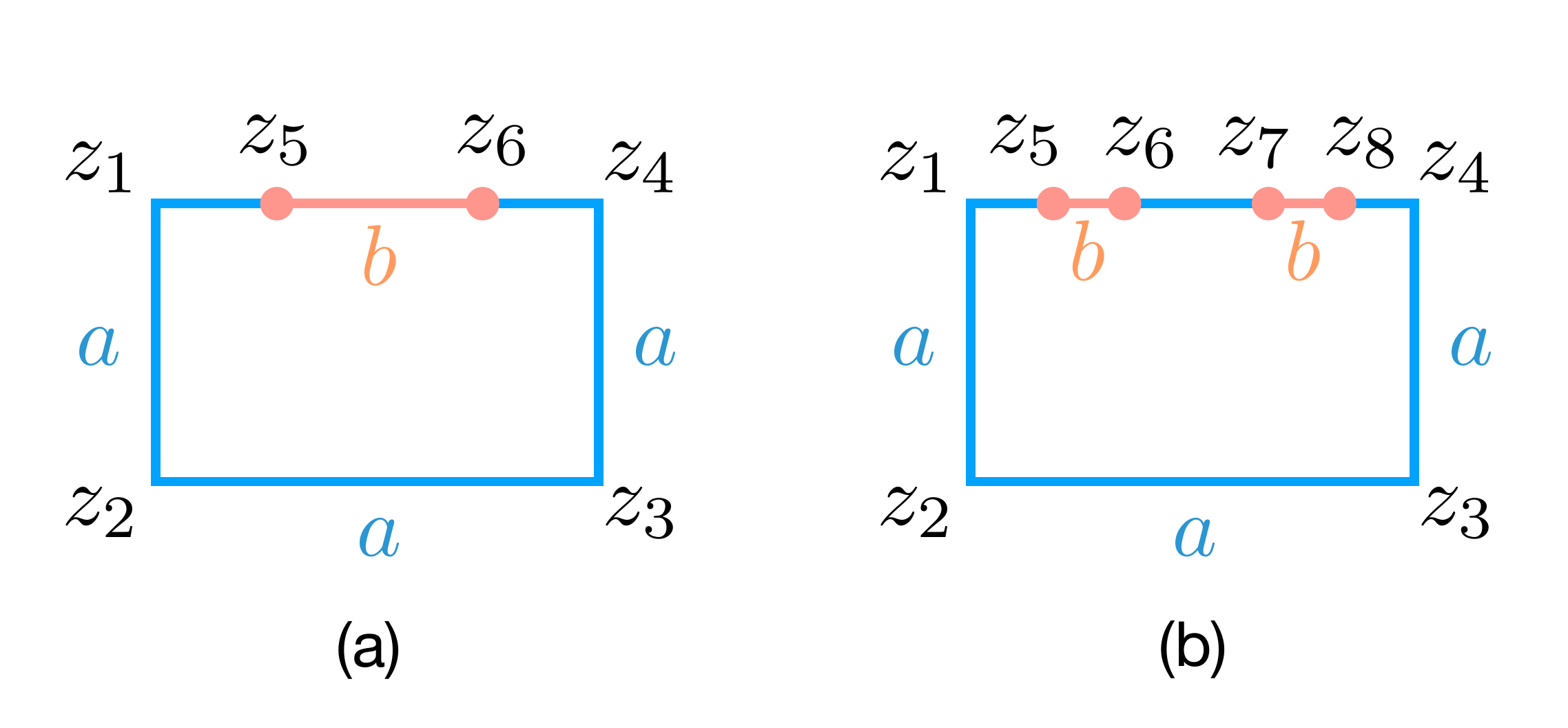}
\caption{Illustration of computing (a) $S(A)$ for a subregion $A=[z_5,z_6]$; (b) mutual information $I([z_5,z_6],[z_7,z_8])$ and mutual negativity $N([z_5,z_6],[z_7,z_8])$ between two subregions on a rectangle with boundary condition $aaaa$.}
\label{fig:aaaa_illus}
\end{figure}

Suppose we would like to compute the entanglement entropy of a subregion $A=[z_5, z_6]$ residing on the top edge of the system, as illustrated in Fig.~\ref{fig:aaaa_illus}(a). According to Eq.~(\ref{eq:EE}), the entanglement entropy is given by the free energy cost when a different type of boundary condition (denoted as $b$) is applied in region $A$. We take $b$ to be of the same nature but distinct from $a$, which means that the scaling dimensions of the bcc operators $\phi_{f|a}$ and $\phi_{f|b}$ are the same. This is motivated by explicit replica calculations, where $a$ and $b$ simply correspond to fixing the boundary regions $A$ and $\overline{A}$ to different states (taking values in permutation groups) of the replicated theory~\cite{bao2018spin, jian2018spin, vasseur2018RTN}. This change of boundary condition within region $A$ can be accounted for by the insertion of bcc operators at the endpoints of $A$. In CFT, the ratio of the partition functions in Eq.~(\ref{eq:EE}) is given by the correlation function of these bcc operators\footnote{We remark that in a boundary CFT, only the holomorphic part of the scaling fields appears~\cite{cardy1984conformal}.}:
\begin{equation}
    {\rm exp}[-S(A)] = \frac{Z(A)}{Z_{\rm bg}} = \langle \phi_{a|b}(z_5) \phi_{b|a}(z_6)\rangle.
    \label{eq:EE_aaaa}
\end{equation}
Our task then is to evaluate this correlation function. 
Since a finite rectangle is simply connected, the Riemann mapping theorem then guarantees that it can be conformally mapped to the lower half plane.
The correlation functions of bcc operators at the boundary of the lower half plane have simple forms, which allows us to extract the universal scaling dimensions of these operators characteristic of the underlying CFT.
Such a conformal transformation is 
a Schwarz-Christoffel mapping~\cite{driscoll2002schwarz}.
We give the explicit form of the Schwarz-Christoffel mapping in Appendix~\ref{sec:schwarz}; see also Ref.~\cite{li2020conformal}.
For now, we simply point out that this mapping depends crucially on the aspect ratio $\tau = \frac{L_y}{L_x}$ of the system, and denote this mapping as $w(z)$, where $w$ lives in the lower half plane. Using the transformation rules of the correlation functions in a CFT under a conformal mapping, we obtain the following expression for the entanglement entropy (see Appendix~\ref{sec:schwarz}):
\begin{equation}
    S(A) = - h_{a|b}\ {\rm ln}D \times {\rm ln}\left[ \frac{\left( \frac{\partial{\omega}}{\partial{z}}\right)_{z_5} \left( \frac{\partial{\omega}}{\partial{z}}\right)_{z_6}}{w_{56}^2}\right]  + {\rm const.},
    \label{eq:EE_aaaa_2}
\end{equation}
where $w_{ij} = w_i - w_j$.
Therefore, one can extract the scaling dimension $h_{a|b} \times \ln D$ by computing $S(A)$ of rectangular RSTNs with boundary condition $aaaa$ at the critical point.

\begin{figure*}[t]
    \centering
\subfigure[]{
    \includegraphics[width=0.35\textwidth]{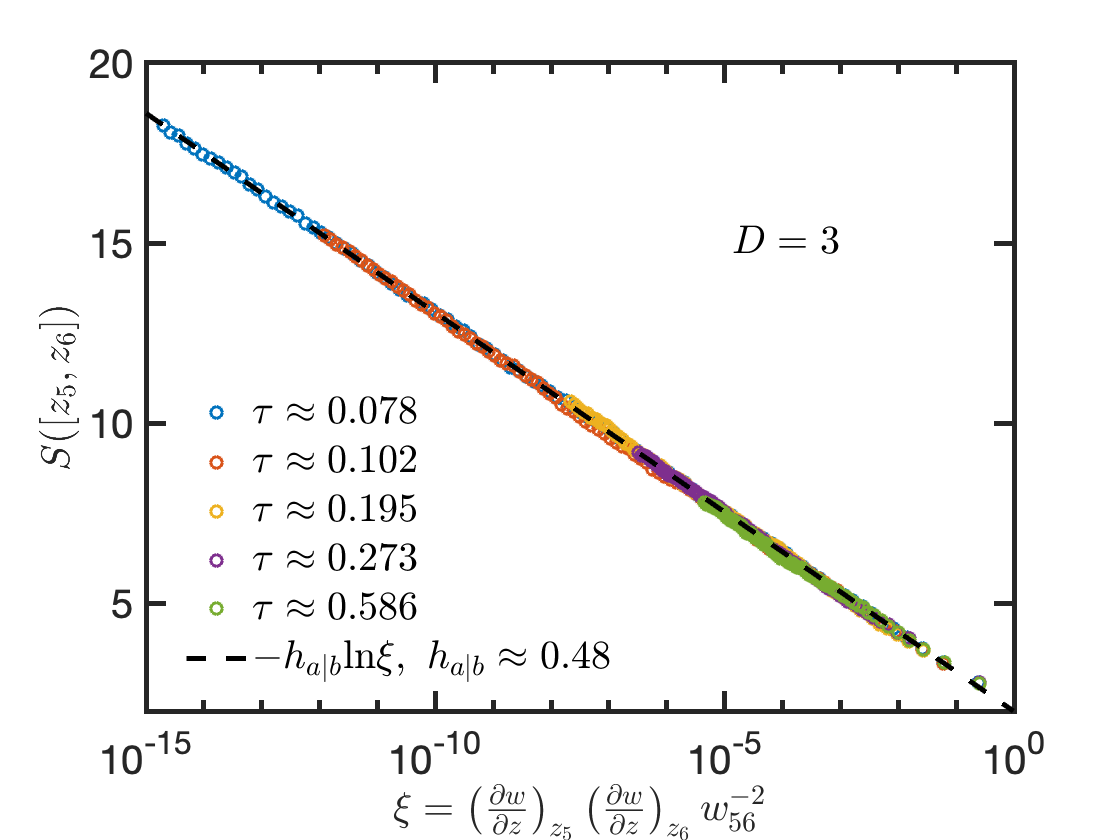}}
\subfigure[]{
    \includegraphics[width=0.35\textwidth]{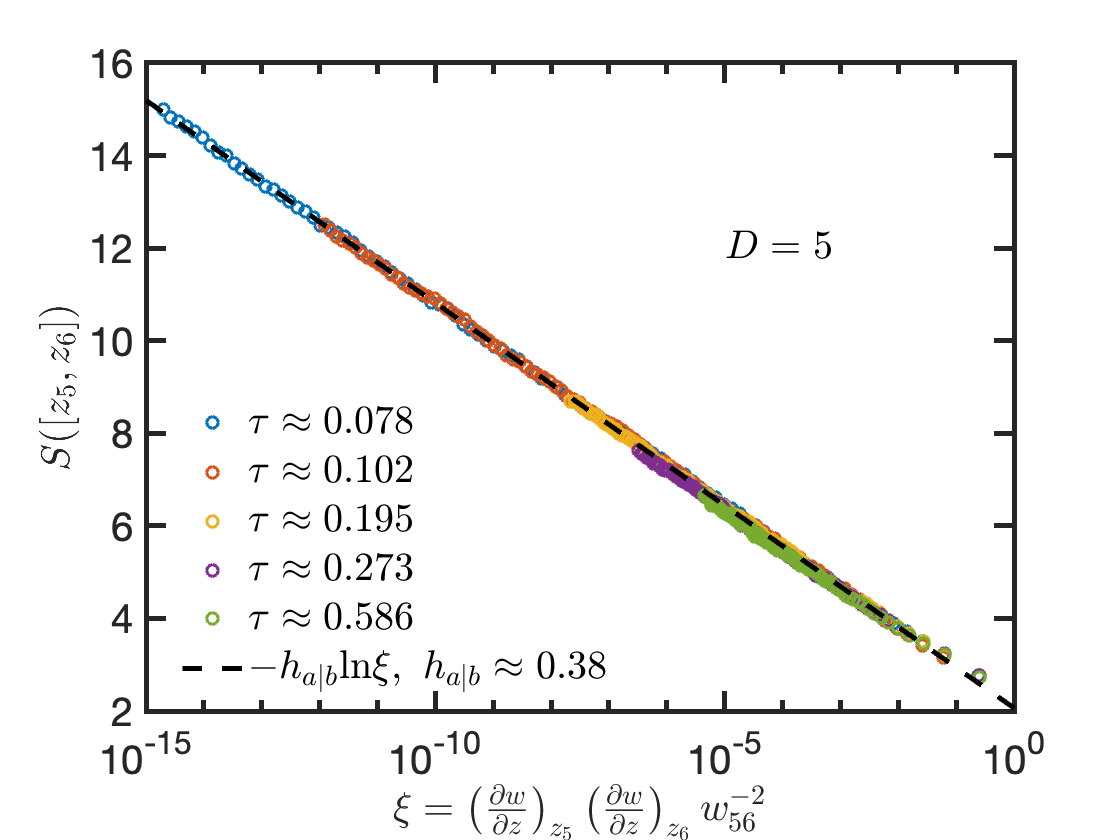}} 
\subfigure[]{
    \includegraphics[width=0.35\textwidth]{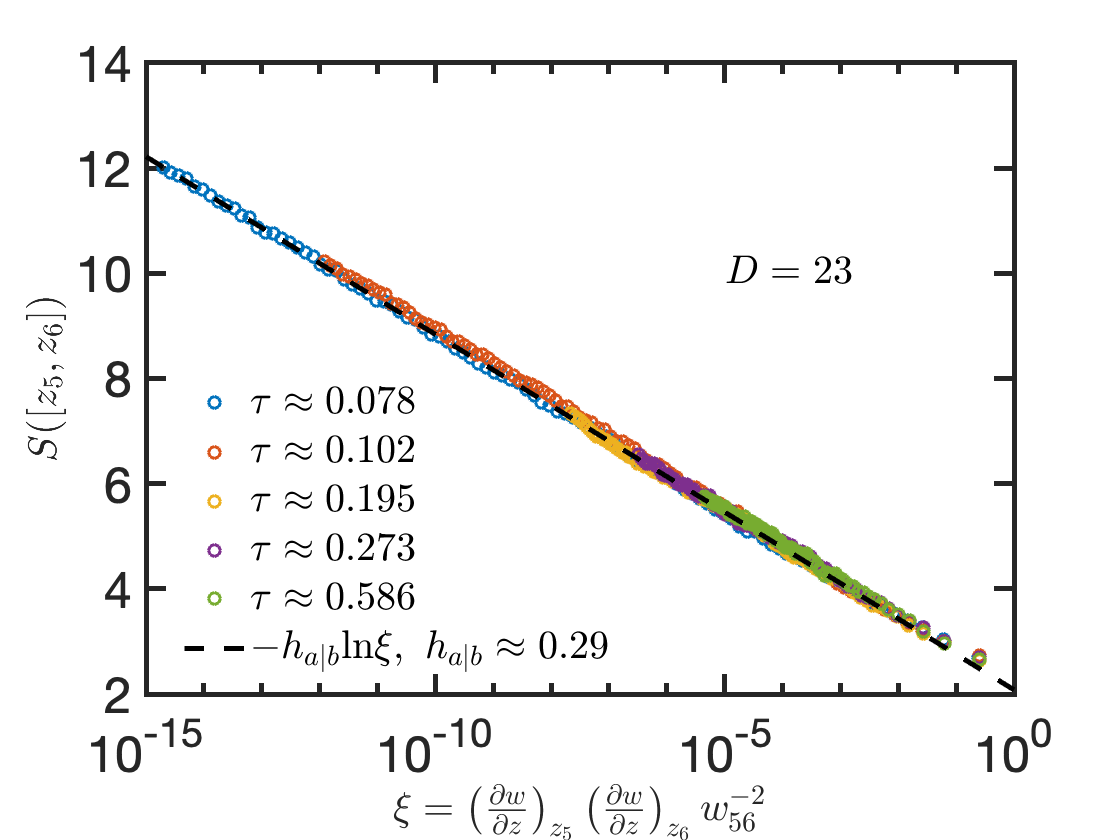}}
\subfigure[]{
    \includegraphics[width=0.35\textwidth]{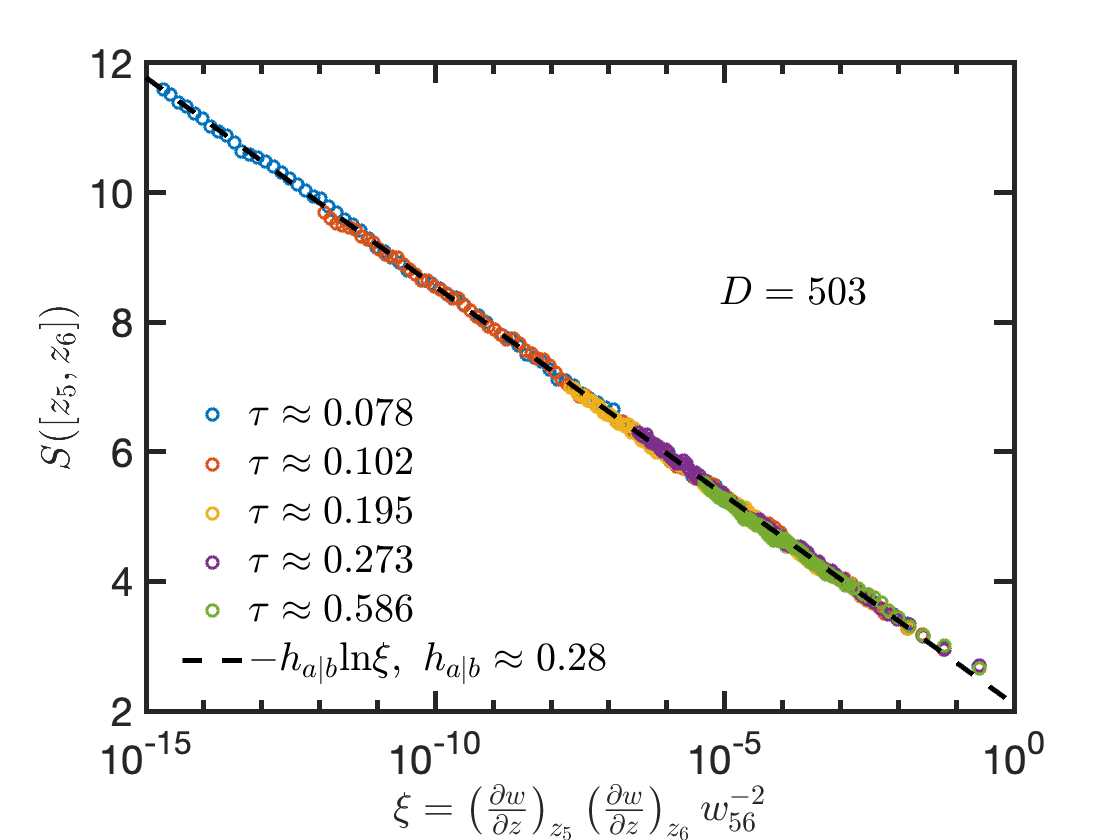}} 
    \caption{Boundary condition $aaaa$: entanglement entropy of a boundary subregion $A=[z_5,z_6]$ [Fig.~\ref{fig:aaaa_illus}(a)] plotted against $\xi$ at the critical point for RSTNs with bond dimensions (a) $D=3$; (b) $D=5$; (c) $D=23$; (d) $D=503$. We fix $L_x$ and consider tensor networks with different aspect ratios $\tau$. The critical points are: (a) $p_c=0.188$; (b) $p_c=0.354$; (c) $p_c=0.476$; and (d) $p_c=0.499$. \ZCY{In each plot, we divide $S$ by ${\rm ln}D$ so as to extract the exponent $h_{a|b}$ defined in Eq.~(\ref{eq:EE_aaaa_2}).} The extracted scaling dimensions of operator $\phi_{a|b}$ in each case are: (a) $h_{a|b}\approx 0.48$; (b) $h_{a|b}\approx 0.38$; (c) $h_{a|b}\approx 0.29$; and (d) $h_{a|b}\approx 0.28$. The result in (d) should be compared with predictions from first passage percolation: $h_{a|b} = \sqrt{3}/(2\pi) \approx 0.276$.
    }
    \label{fig:aaaa}
\end{figure*}

\begin{figure*}[t]
    \centering
\subfigure[]{
    \includegraphics[width=0.35\textwidth]{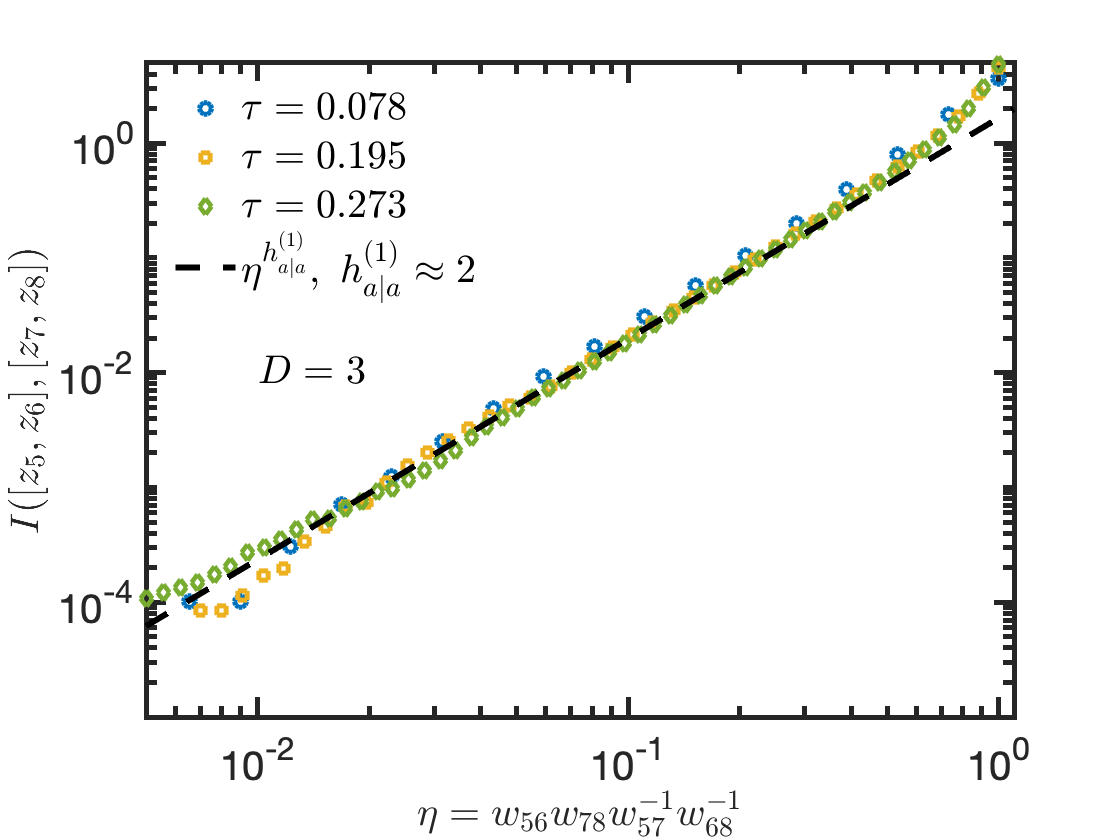}}
\subfigure[]{
    \includegraphics[width=0.35\textwidth]{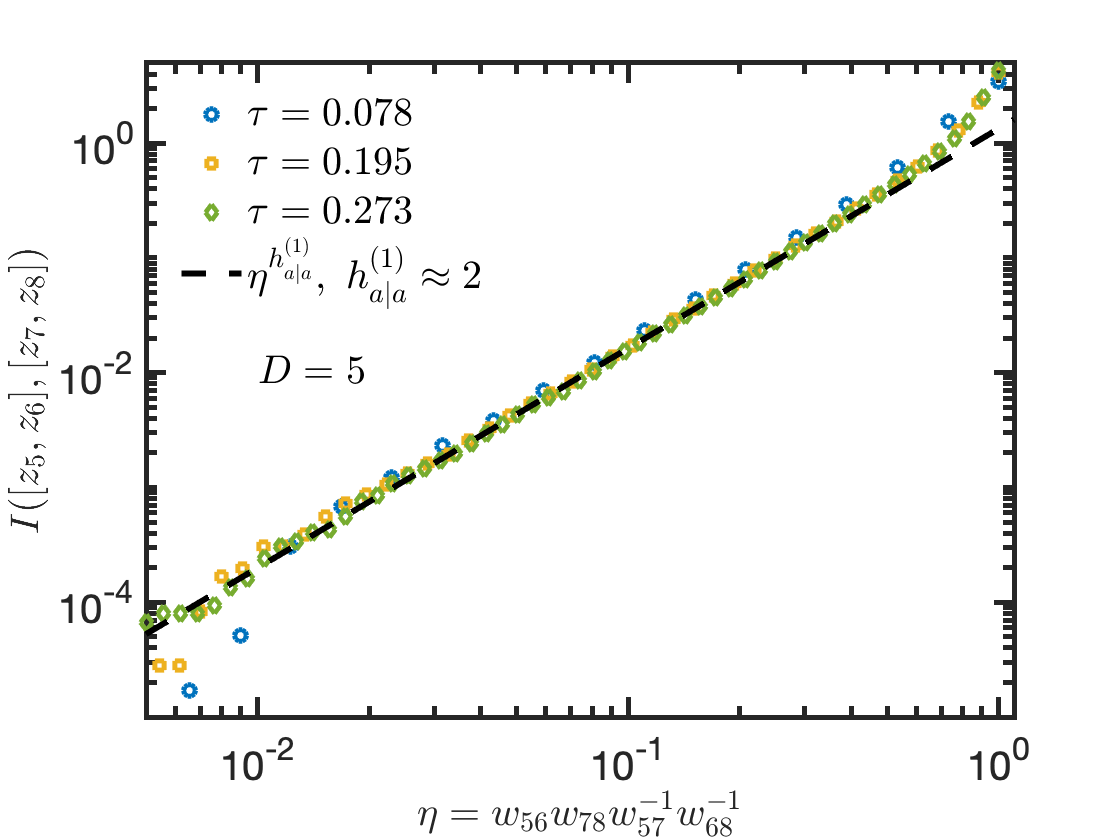}} 
\subfigure[]{
    \includegraphics[width=0.35\textwidth]{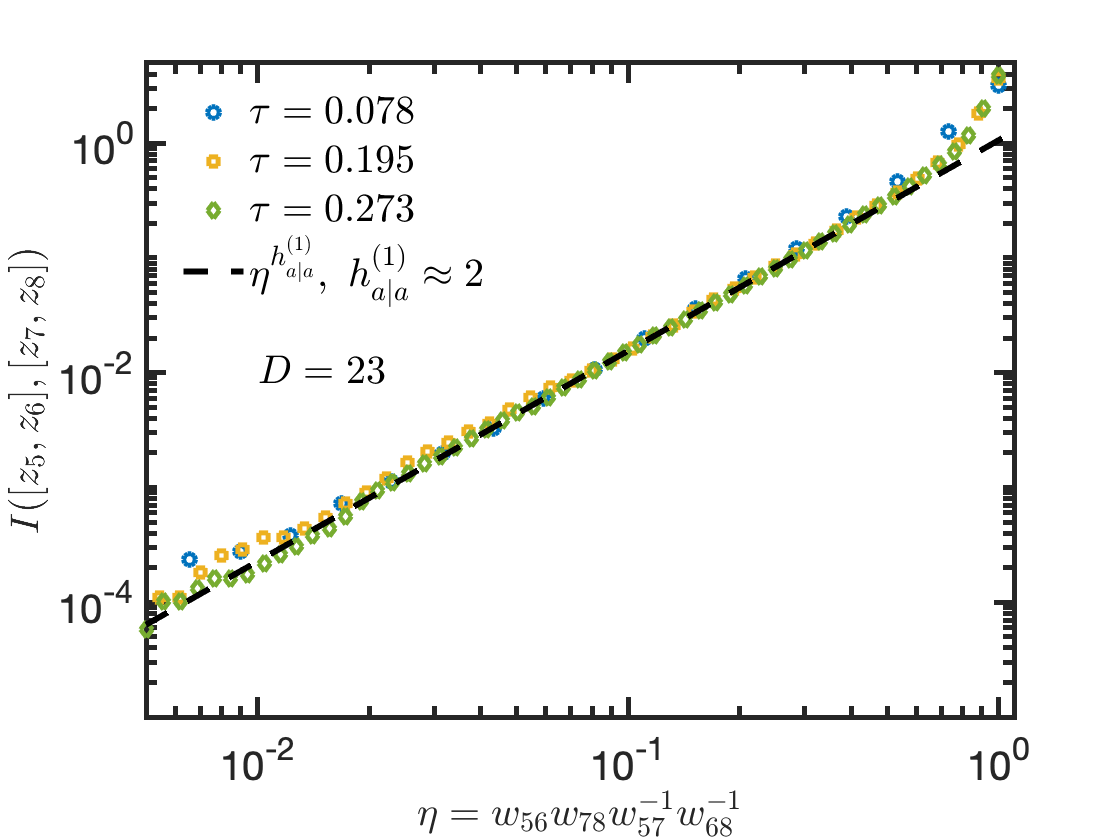}}
\subfigure[]{
    \includegraphics[width=0.35\textwidth]{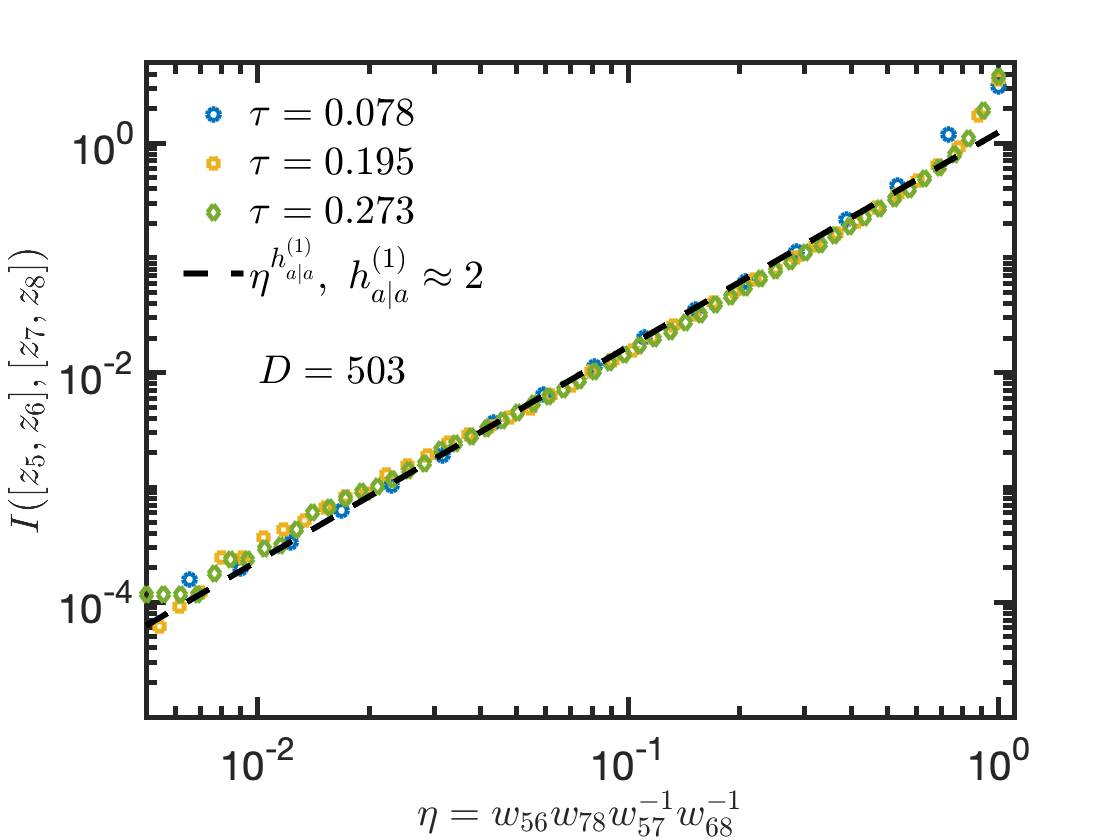}} 
    \caption{ Boundary condition $aaaa$: mutual information between two boundary subregions $[z_5,z_6]$ and $[z_7,z_8]$ [Fig.~\ref{fig:aaaa_illus}(b)] at the critical point for RSTNs with bond dimensions (a) $D=3$; (b) $D=5$; (c) $D=23$; (d) $D=503$. We fix $L_x$ and consider tensor networks with different aspect ratios $\tau$. The exponent for $I(\eta)$ as $\eta \rightarrow 0$ is found to be $h_{a|a}^{(1)}\approx 2$ for all $D$.
    }
    \label{fig:aaaa_MI}
\end{figure*}

\subsection{Boundary condition $aaaa$: scaling dimension $h_{a|b} \times \ln D$, $h_{a|a}^{(1)}$, $\Delta$, mutual information, and mutual negativity}

In this subsection, we present numerical results on RSTNs with boundary condition $aaaa$. We start from the entanglement entropy of a boundary subregion $A$ as depicted in Fig.~\ref{fig:aaaa_illus}(a), which has the form of Eq.~(\ref{eq:EE_aaaa_2}) at the critical point.

In Fig.~\ref{fig:aaaa}, we plot $S(A)$ against $\xi = \left(\frac{\partial{\omega}}{\partial{z}}\right)_{z_5}\left(\frac{\partial{\omega}}{\partial{z}}\right)_{z_6} \omega_{56}^{-2}$, for RSTNs with bond dimensions $D=3,\ 5,\ 23$, and 503 at the critical point. For each bond dimension $D$, we consider tensor networks with $L_x=256$ and different aspect ratios $\tau$. For each choice of the aspect ratio, we vary the endpoints of subregion $A$, thereby computing $S(A)$ as a function of $\xi$. Eq.~(\ref{eq:EE_aaaa_2}) predicts that $S(A)$ at the critical point is only a function of $\xi$, and hence the data for different aspect ratios and subregion endpoints should all collapse onto a single curve. We find that this is indeed the case, as shown in Fig.~\ref{fig:aaaa}. For each $D$, we determine the accurate critical point $p_c$ by searching for the best data collapse in the vicinity of the $p_c$ estimated from the mutual information calculations. This $p_c$ is then fixed for all boundary conditions considered hereafter. We also extract the scaling dimension $h_{a|b} \times \ln D$ by fitting the data according to Eq.~(\ref{eq:EE_aaaa_2}). The scaling dimension $h_{a|b} \times \ln D$ is different for each bond dimension $D$, as shown in Fig.~\ref{fig:aaaa}. 

\MF{Guided by the} minimal cut picture, at large $D$ \MF{we anticipate that} $h_{a|b}$ \MF{might} approach predictions from first passage percolation. More precisely, \MF{for first passage percolation} $S(A)$ corresponds to the minimal weight of a path connecting the two endpoints at the boundary. Exact results of this minimal weight in critical first passage percolation are known~\cite{nahum2018transition, nahum2019majorana, jiang2019critical}.
To see this possible connection more explicitly, consider Eq.~(\ref{eq:EE_aaaa_2}) in the limit of $\tau \rightarrow \infty$ and then taking $L_x \rightarrow \infty$. In this limit, Eq.~(\ref{eq:EE_aaaa_2}) becomes:
\begin{equation}
    S(A) \approx 2h_{a|b}\ {\rm ln}D \times {\rm ln} z_{56}.
\end{equation}
Since the entire edge of the system shares a uniform boundary condition, this can be viewed as a first passage percolation problem in a system with periodic boundary condition, in which case the universal prefactor is given by $2h_{a|b} = \sqrt{3}/\pi$, or $h_{a|b}=\sqrt{3}/(2\pi) \approx 0.276$~\cite{nahum2018transition, nahum2019majorana, jiang2019critical}.
In Fig.~\ref{fig:aaaa}(d) with $D=503$, the numerically obtained value $h_{a|b}\approx 0.28$ is indeed very close to that of critical percolation, \MF{confirming our expectation}. On the other hand, the scaling dimensions at smaller $D$ are distinct from percolation, signaling the failure of a simple geometric minimal cut picture and \MF{suggesting different universality classes for each prime $D$.}

\begin{figure*}[t]
    \centering
\subfigure[]{
    \includegraphics[width=0.35\textwidth]{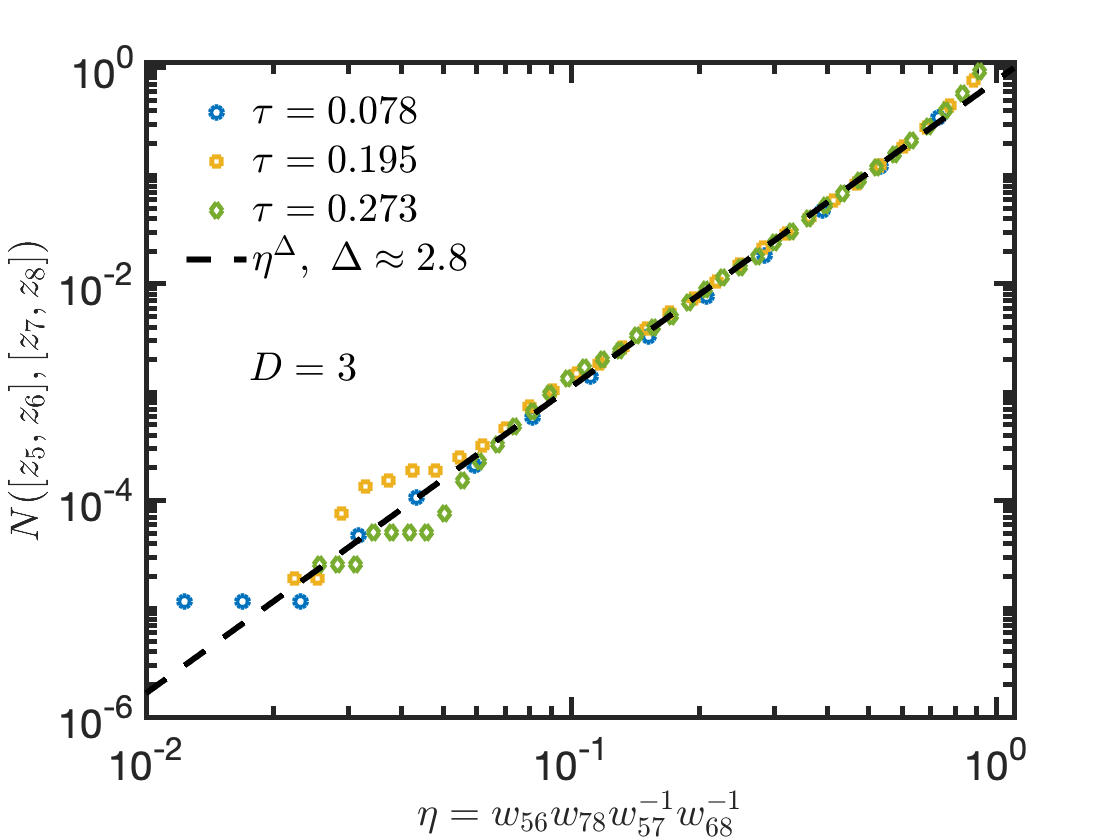}}
\subfigure[]{
    \includegraphics[width=0.35\textwidth]{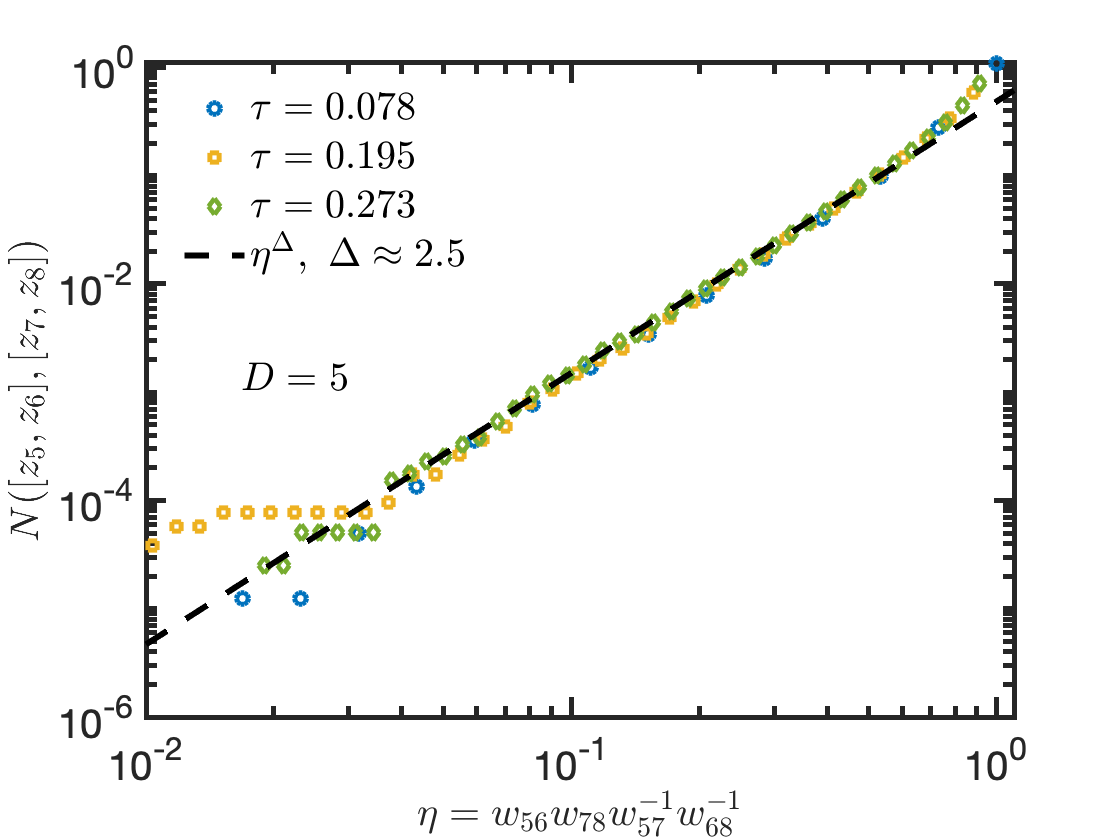}} 
\subfigure[]{
    \includegraphics[width=0.35\textwidth]{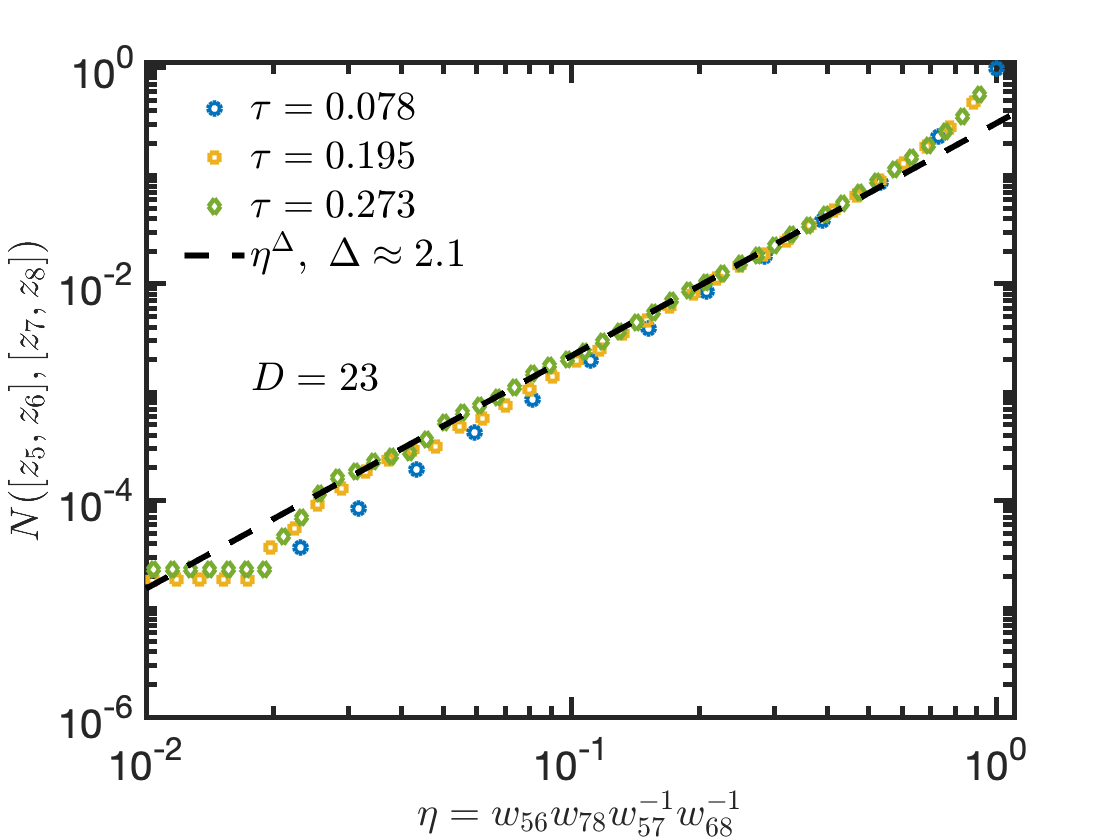}}
\subfigure[]{
    \includegraphics[width=0.35\textwidth]{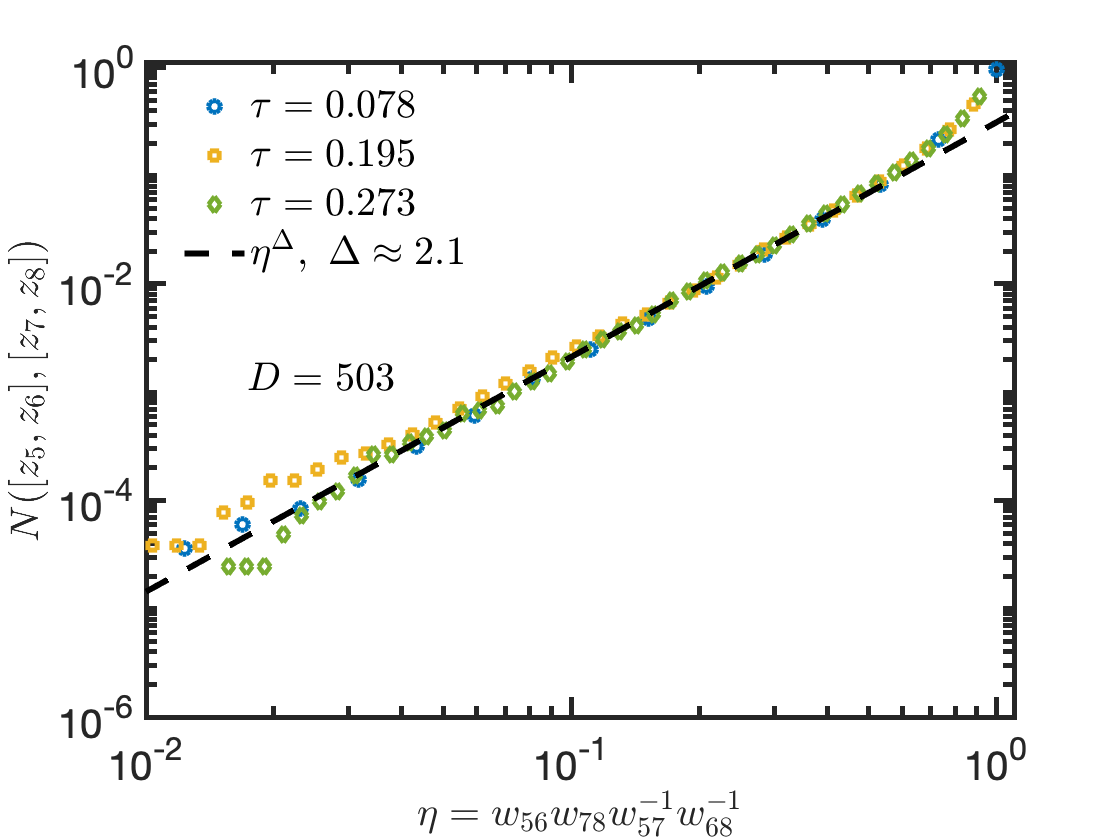}} 
    \caption{ Boundary condition $aaaa$: mutual negativity between two boundary subregions $[z_5,z_6]$ and $[z_7,z_8]$ [Fig.~\ref{fig:aaaa_illus}(b)] at the critical point. We fix $L_x$ and consider tensor networks with different aspect ratios $\tau$. The exponent for $N(\eta)$ as $\eta \rightarrow 0$ gradually decreases as $D$ increases, and approaches $\Delta = 2$ at large $D$.
    }
    \label{fig:aaaa_MN}
\end{figure*}

Next, we turn to the critical behavior of mutual information, which gives us access to another universal scaling dimension. Consider the mutual information between two subregions $A=[z_5,z_6]$ and $B=[z_7,z_8]$ located at the top edge of the rectangle, as illustrated in Fig.~\ref{fig:aaaa_illus}(b). According to the definition~(\ref{eq:MI_def}), the mutual information can be written in terms of the bcc operators as:
\begin{eqnarray}
    &&{\rm exp}\left[-I([z_5,z_6],[z_7,z_8])\right]  \nonumber  \\
    =&& \frac{\langle \phi_{a|b}(z_5) \phi_{b|a}(z_6)\rangle \langle \phi_{a|b}(z_7) \phi_{b|a}(z_8)\rangle}{\langle \phi_{a|b}(z_5) \phi_{b|a}(z_6) \phi_{a|b}(z_7) \phi_{b|a}(z_8)\rangle}.
\end{eqnarray}
Using the general form of four-point correlation functions in a CFT~\cite{BPZ1984}
\begin{equation}
    \langle \phi_{a|b}(z_5) \phi_{b|a}(z_6) \phi_{a|b}(z_7) \phi_{b|a}(z_8)\rangle = F(\eta) (z_{56} z_{78})^{-2\widetilde{h}_{a|b}},
\end{equation}
where $\widetilde{h}_{a|b} \equiv h_{a|b} {\rm ln}D$, and $F(\eta)$ is a function depending solely on the cross ratio
\begin{equation}
    \eta = \frac{z_{56}z_{78}}{z_{57}z_{68}},
\end{equation}
we obtain
\begin{equation}
    I([z_5,z_6],[z_7,z_8]) = {\rm ln} F(\eta),
\end{equation}
i.e., the mutual information is only a function of $\eta$. In Fig.~\ref{fig:aaaa_MI}, we plot the mutual information for RSTNs with different bond dimensions as a function of $\eta$, where we find that the curves for tensor networks with different aspect ratios indeed fall on top of one another. Moreover, for small $\eta$, the mutual information obeys a power law, as indicated by the straight line in a log-log plot in Fig.~\ref{fig:aaaa_MI}. To understand this behavior, notice that $\eta \rightarrow 0$ corresponds to taking $z_5 \rightarrow z_6$ and $z_7 \rightarrow z_8$, in which limit one can invoke the following OPE:
\begin{eqnarray}
    &&\phi_{a|b}(z_1)\phi_{b|a}(z_2) \nonumber  \\
    &\sim& z_{12}^{-2h_{a|b}}\left(\mathbb{1}_{a|a} + C_{a|b|a}^{(1)} z_{12}^{h^{(1)}_{a|a}} \phi^{(1)}_{a|a}(z_1) + \ldots \right)
    \label{eq:OPE}
\end{eqnarray}
where $\phi_{a|a}^{(1)}$ denotes the primary field appearing in the OPE channel with the lowest scaling dimension $h_{a|a}^{(1)}$ other than the identity, and $C^{(1)}_{a|b|a}$ is the OPE expansion coefficient. Using the OPE in Eq.~(\ref{eq:OPE}), we find that the mutual information in the limit $\eta \rightarrow 0$ has the form
\begin{equation}
    I([z_5,z_6],[z_7,z_8]) \approx {\rm ln}(1+\#\eta^{h_{a|a}^{(1)}}) \propto \eta^{h_{a|a}^{(1)}}.
\end{equation}
Therefore, the exponent of $I(\eta)$ at small $\eta$ is given by another operator scaling dimension $h_{a|a}^{(1)}$. Remarkably, in Fig.~\ref{fig:aaaa_MI} we find that the extracted value $h_{a|a}^{(1)} \approx 2$ is robust and remains unchanged for all values of $D$ that we have examined. The same exponent also appears in hybrid Clifford circuits~\cite{li2019hybrid, li2020conformal}, as well as the zeroth R\'enyi entropy~\cite{nahum2018transition} at the critical point, which is described by the first passage percolation problem.

Finally, let us compute another quantity that is closely related to mutual information: the \textit{(logarithmic) mutual negativity} for two boundary subregions $A$ and $B$ defined as~\cite{PhysRevA.65.032314, sang2020entanglement, shi2020entanglement}
\begin{equation}
    N(A,B) \equiv N(\rho_{A \cup B}) = {\rm log}\ ||\rho_{A\cup B}^{\intercal_A}||_1 = {\rm log} \sum_i |\lambda_i|,
    \label{eq:negativity_def}
\end{equation}
where $\intercal_A$ denotes partial transpose on subsystem $A$, $|| \cdot ||_1$ denotes trace norm, and $\lambda_i$ are the eigenvalues of $\rho_{A\cup B}^{\intercal_A}$. To see the physical meaning of this quantity, notice that since ${\rm tr} \rho_{A\cup B}^{\intercal_A}=1$, we can write
\begin{equation}
    ||\rho_{A\cup B}^{\intercal_A}||_1 = 1 + 2\sum_{\lambda_i<0} |\lambda_i|.
\end{equation}
That is, mutual negativity measures the degree of ``negativity" in the partial transpose of the density matrix. Since a non-positive partial transpose indicates that the state is not separable and thus cannot be prepared via local operations and classical communication (LOCC) only, it quantifies the degree of \textit{quantum} correlations in $\rho$. The mutual information, on the other hand, detects \textit{both} quantum \textit{and} classical correlations. It has been recently demonstrated in hybrid quantum circuits that mutual negativity and mutual information exhibit rather different scaling properties at the critical point~\cite{sang2020entanglement, shi2020entanglement}. In particular, for stabilizer states, the mutual negativity receives contribution solely from the bipartite entanglement, whereas the mutual information receives contributions from \text{both} bipartite \textit{and} tripartite GHZ-type entanglement~\cite{sang2020entanglement}.

The mutual negativity for stabilizer states can be computed as follows~\cite{sang2020entanglement, shi2020entanglement}.
Define the stabilizer group for $\rho_{A\cup B}$ as $\mathcal{S}$, and its generating set $\mathcal{G}(\mathcal{S}) = \{g_1, g_2, \ldots, g_{m} \}$, where $m=|\mathcal{G}(\mathcal{S})|$. Define ${\rm proj}_A$ as the projection of $g_i \in \mathcal{G}(\mathcal{S})$ on subsystem $A$, i.e., the Pauli operators in $g_i$ that are supported on $B$ are set to the identity. Define the following $m \times m$ commutation matrix
\begin{equation}
    (K_A)_{ij} = \lambda_{ij}
    \label{eq:commutation}
\end{equation}
where
\begin{equation}
    {\rm proj}_A(g_i) \cdot {\rm proj}_A(g_j) = \omega^{\lambda_{ij}} {\rm proj}_A(g_j) \cdot {\rm proj}_A(g_i),
\end{equation}
with $\omega = e^{2\pi i/D}$.
Then the mutual negativity of $\rho_{A\cup B}$ is given by
\begin{equation}
    N(A,B) = \frac{1}{2} {\rm rank} (K_A).
    \label{eq:negativity}
\end{equation}
We shall prove Eq.~(\ref{eq:negativity}) in Appendix~\ref{sec:proof_negativity}.

In Fig.~\ref{fig:aaaa_MN}, we plot the mutual negativity for RSTNs with different bond dimensions. We find that the mutual negativity exponent $N(\eta) \approx \eta^{\Delta}$ is in general distinct from the mutual information exponent $h_{a|a}^{(1)}$. For $D=3$, we find $\Delta \approx 2.8$, which is consistent with the value $\Delta \approx 3.0$ obtained in hybrid circuit models in Ref.~\cite{sang2020entanglement}. Nonetheless, at large $D$, the amount of tripartite entanglement in RSTNs is scarce~\cite{nezami2016RSTN}.
We thus expect that $h_{a|a}^{(1)}$ and $\Delta$ should agree at large $D$.
In Fig.~\ref{fig:aaaa_MN}(d), we indeed find $\Delta \approx 2$, approaching $h_{a|a}^{(1)}$ at $D=503$.
It was argued in Ref.~\cite{sang2020entanglement} that $\Delta=2$ is a fingerprint of percolation, since this value appears for all occurences of percolation considered there.
Our result in RSTNs is also consistent with this picture.

\begin{figure}[t]
\centering
\includegraphics[width=0.45\textwidth]{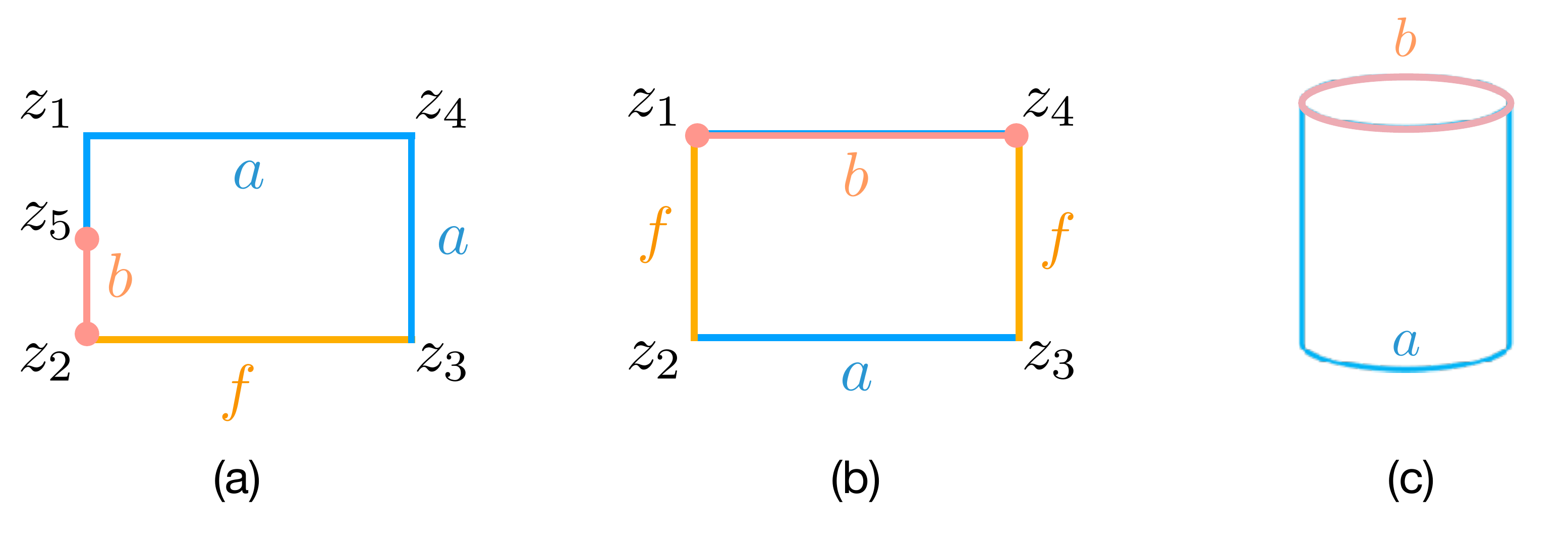}
\caption{Illustration of computing $S(A)$ for a subregion (a) $A=[z_2,z_5]$ with boundary condition $afaa$; (b) $A=[z_1,z_4]$ with boundary condition $fafa$; (c) $A=$ the top edge with periodic boundary condition.}
\label{fig:afaa_illus}
\end{figure}

\subsection{Boundary condition $afaa$: scaling dimension $h_{a|b} \times \ln D$}

We shall now move on to the second type of boundary condition $afaa$, as illustrated in Fig.~\ref{fig:bc}(b). From this boundary condition, we extract the same scaling dimension $h_{a|b} \times \ln D$ from a different set of correlation functions. Therefore, the results in this subsection serve as a cross-check for those obtained in the previous subsection, as well as a consistency check for our general assumption of a CFT description.

\begin{figure*}[t]
    \centering
\subfigure[]{
    \includegraphics[width=0.35\textwidth]{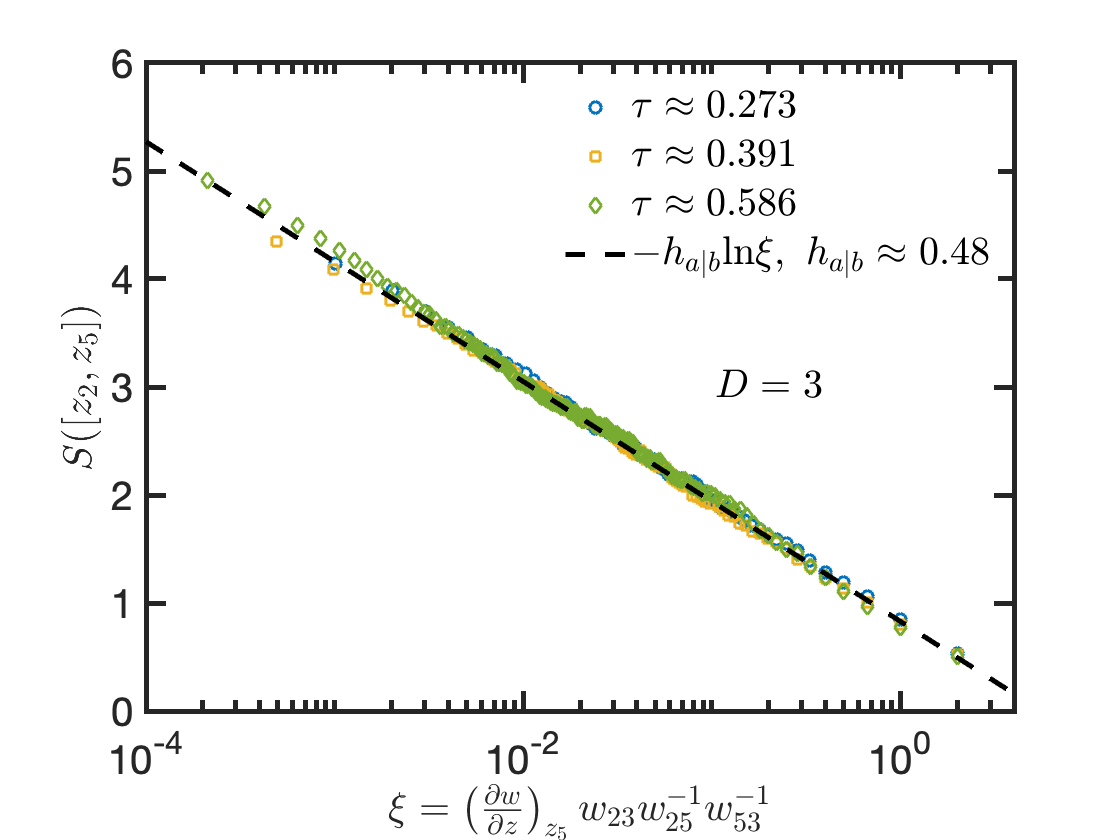}}
\subfigure[]{
    \includegraphics[width=0.35\textwidth]{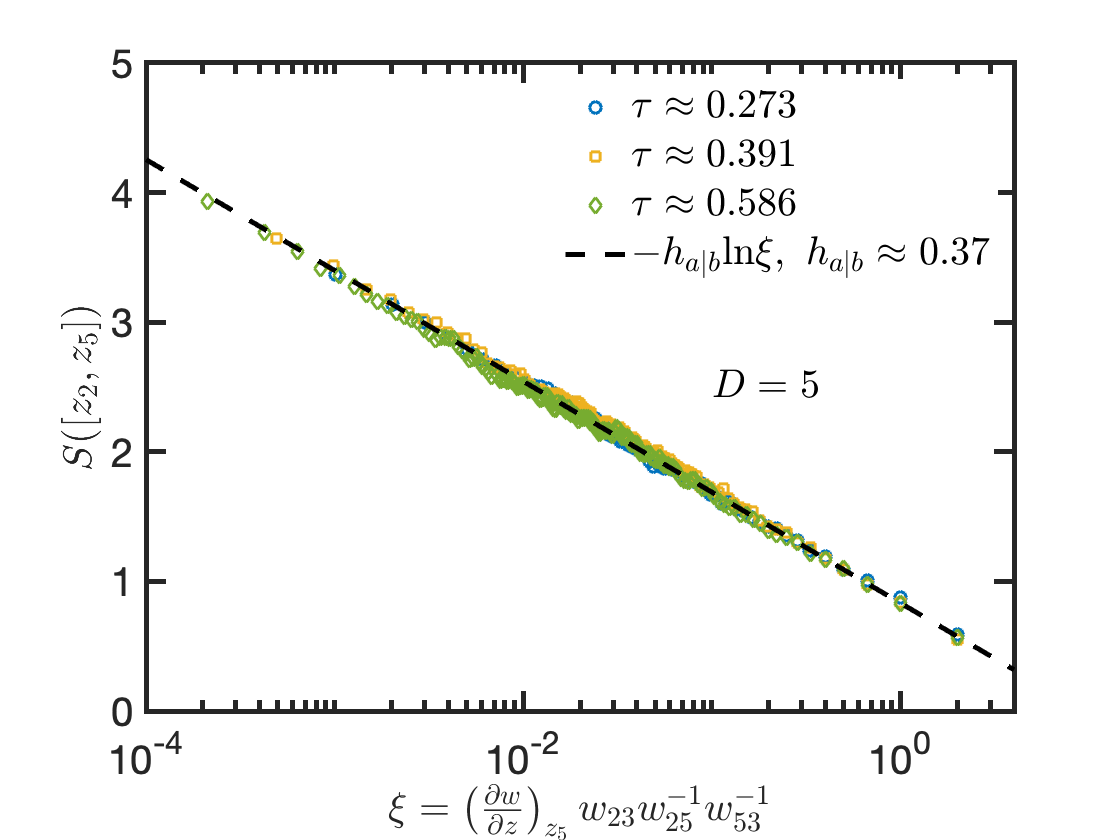}} 
\subfigure[]{
    \includegraphics[width=0.35\textwidth]{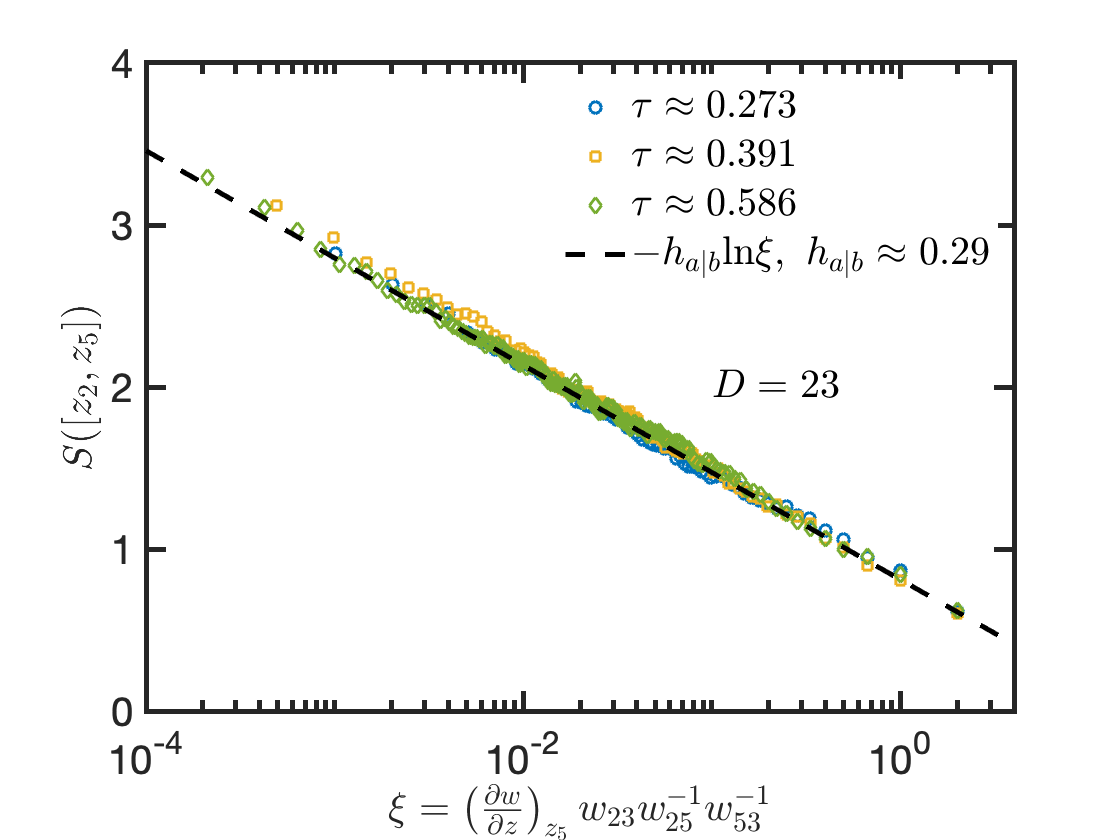}}
\subfigure[]{
    \includegraphics[width=0.35\textwidth]{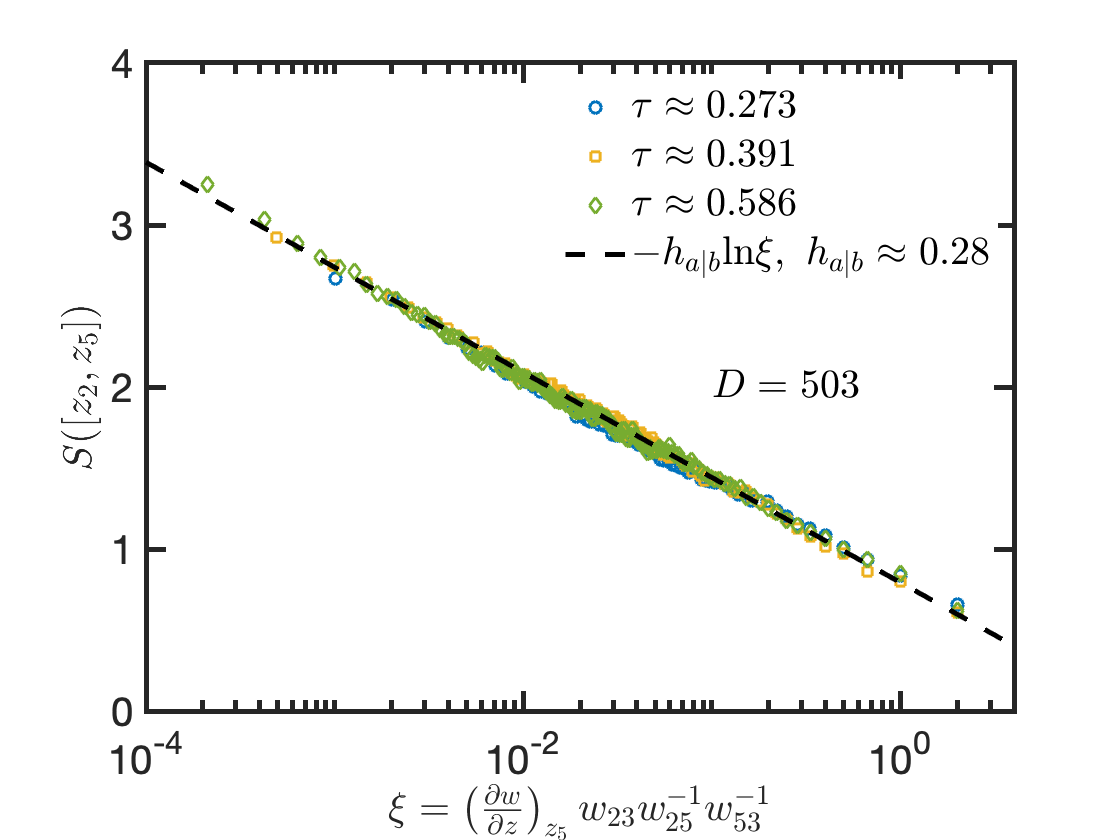}} 
    \caption{Boundary condition $afaa$: entanglement entropy of a boundary subregion $A=[z_2,z_5]$ [Fig.~\ref{fig:afaa_illus}(a)] plotted against $\xi$ at the critical point for RSTNs with bond dimensions (a) $D=3$; (b) $D=5$; (c) $D=23$; (d) $D=503$. We fix $L_x$ and consider tensor networks with different aspect ratios $\tau$. \ZCY{Similarly to Fig.~\ref{fig:aaaa}, we divide $S$ by ${\rm ln}D$ so as to extract the exponent $h_{a|b}$ defined in Eq.~(\ref{eq:EE_afaa}).}
    The extracted scaling dimensions of operator $\phi_{a|b}$ agree with those computed from boundary condition $aaaa$ (Fig.~\ref{fig:aaaa}).
    }
    \label{fig:afaa}
\end{figure*}

\begin{figure}[t]
\centering
\includegraphics[width=0.45\textwidth]{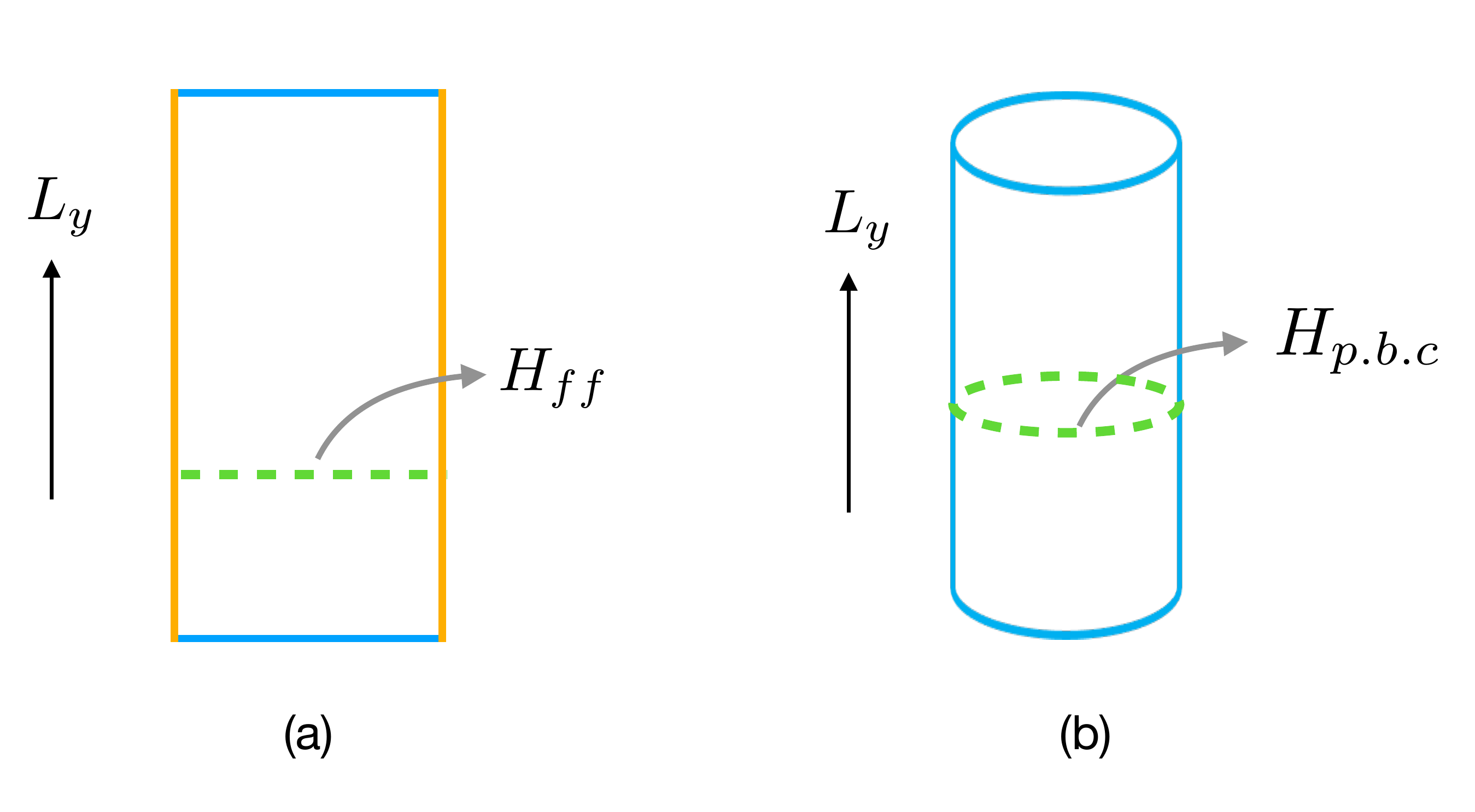}
\caption{Illustration of the transfer matrix formulation considered in (a) boundary condition $fafa$ and (b) periodic boundary condition. The Hamiltonians $H_{ff}$ defined on an open chain with free boundary conditions on both ends and $H_{p.b.c}$ defined on a periodic chain generate translations along the $L_y$ direction in both cases.}
\label{fig:transfer_matrix}
\end{figure}

Consider in this case the entanglement entropy $S(\rho_A)$ of a subregion $A=[z_2,z_5]$ located at the vertical edge of the rectangle, as illustrated in Fig.~\ref{fig:afaa_illus}(a). Since in this case, the background system $Z_{\rm bg}$ no longer has a uniform boundary condition, we have
\begin{eqnarray}
Z_{\rm bg} &=& Z_0 \ \langle \phi_{f|a}(z_2) \phi_{a|f}(z_3) \rangle  \\
Z(A) &=& Z_0 \ \langle \phi_{f|b}(z_2) \phi_{b|a}(z_5) \phi_{a|f}(z_3) \rangle,
\end{eqnarray}
where $Z_0$ is the partition function of the system with free boundary condition on all edges. Thus, we have
\begin{eqnarray}
&&{\rm exp}\left[ -S([z_1,z_5]) \right] = \frac{\langle \phi_{f|b}(z_2) \phi_{b|a}(z_5) \phi_{a|f}(z_3) \rangle}{\langle \phi_{f|a}(z_2) \phi_{a|f}(z_3) \rangle}   \nonumber  \\
&& = \left( \frac{\partial{\omega}}{\partial{z}}\right)_{z_5}^{\widetilde{h}_{a|b}} \frac{\langle \phi_{f|b}(\omega_2) \phi_{b|a}(\omega_5) \phi_{a|f}(\omega_3) \rangle}{\langle \phi_{f|a}(\omega_2) \phi_{a|f}(\omega_3) \rangle}   \nonumber \\
&& \propto \left( \frac{\partial{\omega}}{\partial{z}}\right)_{z_5}^{\widetilde{h}_{a|b}} \left( \frac{\omega_{23}}{\omega_{25} \omega_{53}} \right)^{\widetilde{h}_{a|b}},
\end{eqnarray}
and hence
\begin{equation}
    S([z_1,z_5]) = - h_{a|b}\ {\rm ln}D \times {\rm ln}\left(\frac{\left( \frac{\partial{\omega}}{\partial{z}}\right)_{z_5} \omega_{23}}{\omega_{25}\omega_{53}} \right) + {\rm const.,}
    \label{eq:EE_afaa}
\end{equation}
where we have used the general form of three-point functions in a CFT~\cite{BPZ1984}, \ZCY{and similarly to Eq.~(\ref{eq:EE_aaaa_2}), $\widetilde{h}_{a|b} \equiv h_{a|b} {\rm ln} D$.} In Fig.~\ref{fig:afaa}, we plot the entanglement entropy as a function of $\xi=\left( \frac{\partial{\omega}}{\partial{z}} \right)_{z_5} \omega_{23} \omega_{25}^{-1} \omega_{53}^{-1}$, for RSTNs with different aspect ratios and bond dimensions. Again, we find that the curves for systems with different aspect ratios collapse on top of one another, as predicted by Eq.~(\ref{eq:EE_afaa}), and the extracted values of $h_{a|b}$ are indeed consistent with those computed from boundary condition $aaaa$ in Fig.~\ref{fig:aaaa}.

\subsection{Boundary condition $fafa$: scaling dimension $h_{f|f}^{(1)}$}

\begin{figure*}[t]
    \centering
\subfigure[]{
    \includegraphics[width=0.35\textwidth]{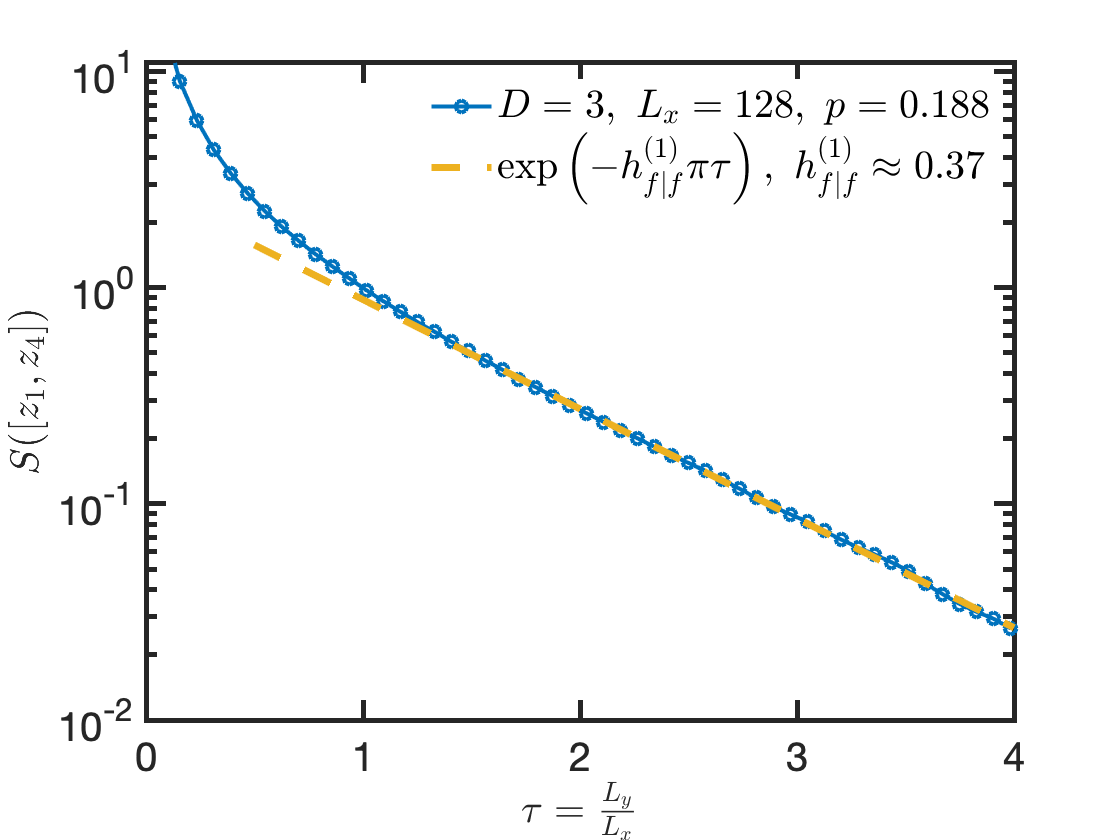}}
\subfigure[]{
    \includegraphics[width=0.35\textwidth]{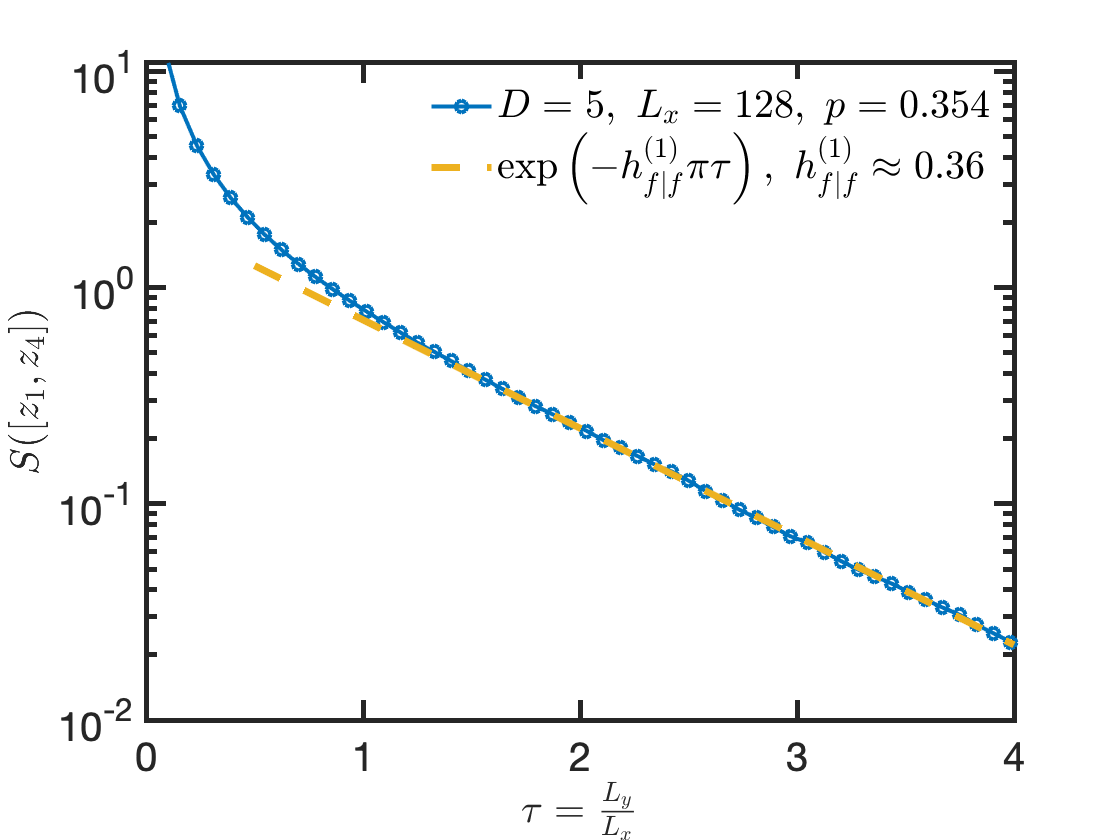}} 
\subfigure[]{
    \includegraphics[width=0.35\textwidth]{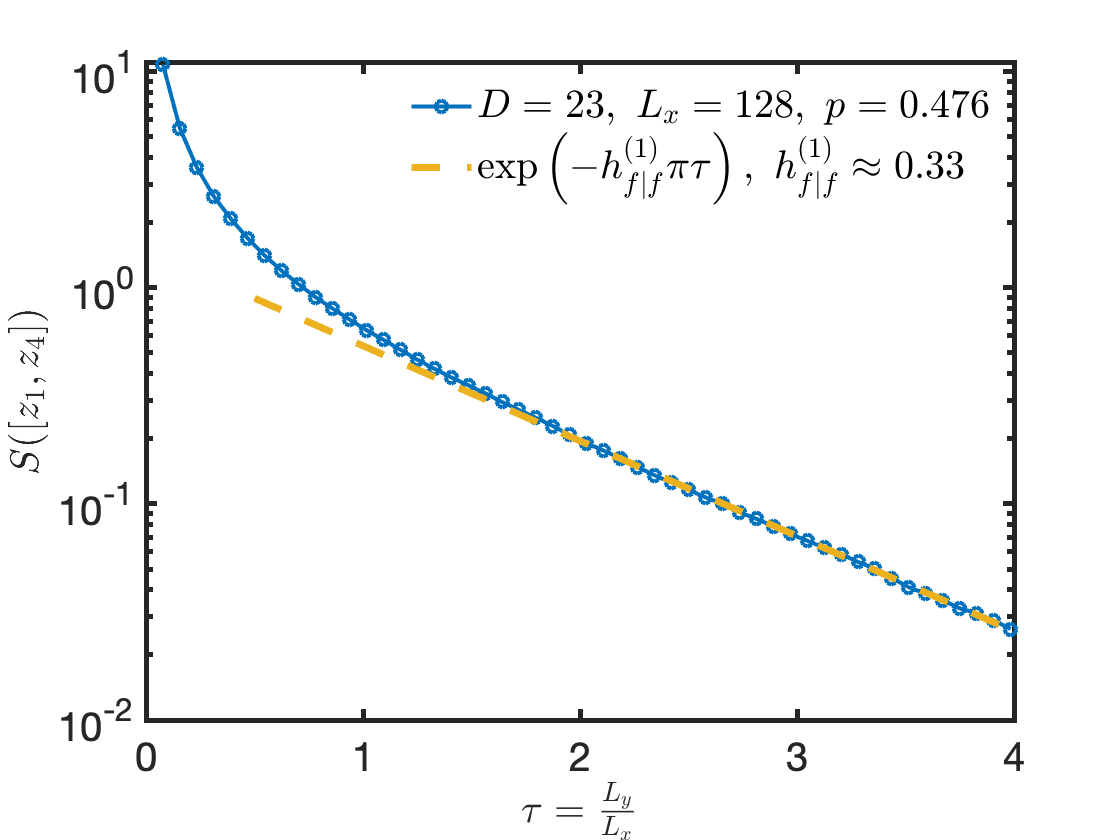}}
\subfigure[]{
    \includegraphics[width=0.35\textwidth]{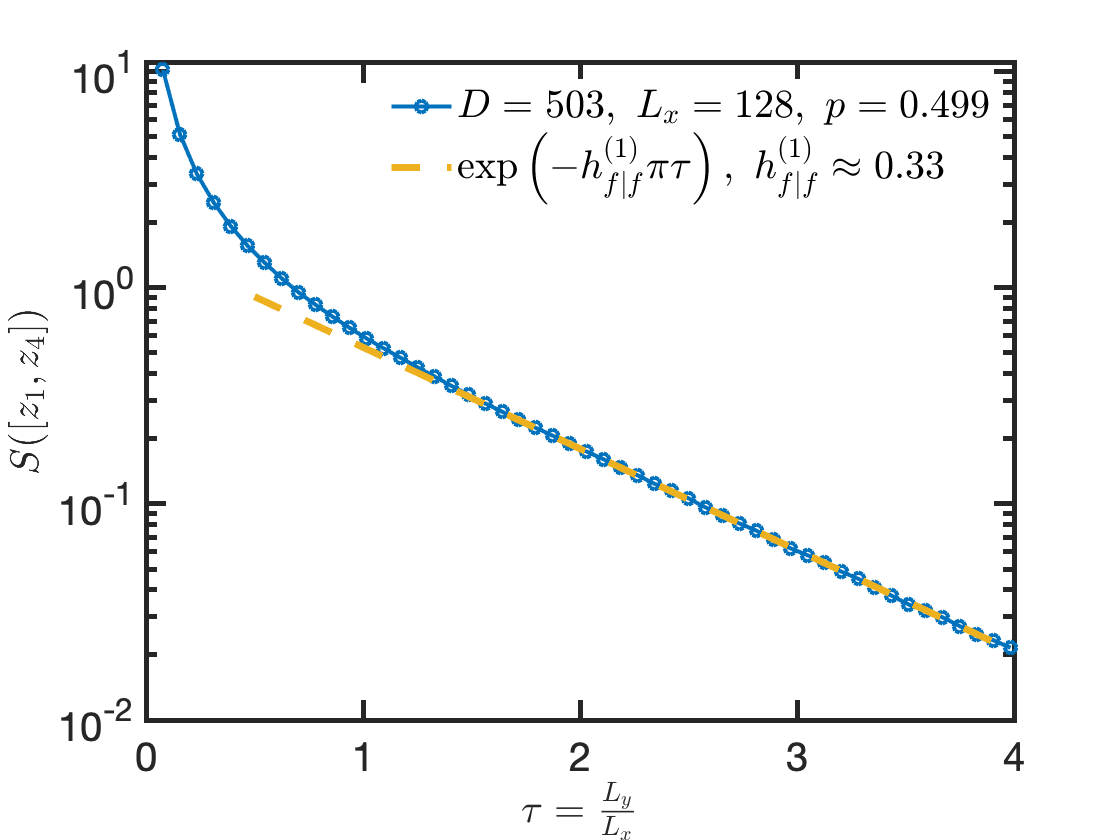}} 
    \caption{Boundary condition $fafa$: entanglement entropy of a boundary subregion $A=[z_1,z_4]$ [Fig.~\ref{fig:afaa_illus}(b)] plotted against the aspect ratio $\tau$ at the critical point for RSTNs with bond dimensions (a) $D=3$; (b) $D=5$; (c) $D=23$; (d) $D=503$. The exponential decay of $S$ at large $\tau$ is controlled by the operator scaling dimension $h_{f|f}^{(1)}$. At large $D$, the value of $h_{f|f}^{(1)}$ agrees with the scaling dimension of a \textit{boundary} spin operator in percolation $h_{1,3}=\frac{1}{3}$.
    }
    \label{fig:fafa}
\end{figure*}

Next, we turn to the third type of boundary condition $fafa$ depicted in Fig.~\ref{fig:bc}(c). \ZCY{This setup is physically interesting in hybrid quantum circuit models, where it was interpreted as the purification dynamics starting from a mixed initial state~\cite{gullans2019purification, gullans2019scalable}.
If we treat the top edge as the system and the bottom edge as the environment, after long time evolutions, the entanglement entropy of the system will decay exponentially to zero and become decoupled from the environment at the critical point~\cite{li2020conformal}. In this subsection, we consider a similar setup in RSTNs. Although there is no temporal direction in RSTNs microscopically, we demonstrate that the entanglement entropy of the top edge also decays exponentially with the aspect ratio $\tau$ of the system, when $\tau$ is large. In particular, we extract the critical exponent associated with this exponential decay and show that it approaches that of percolation in the large $D$ limit.}

Consider the entanglement entropy of the entire top edge $S([z_1,z_4])$. Since now both $Z(A)$ and $Z_{\rm bg}$ involve a four-point correlation function of the bcc operators whose explicit form we do not know, we instead consider the limiting case when $\tau \rightarrow \infty$. Using the transfer matrix formulation, the partition functions can be written as
\begin{eqnarray}
Z(A) &=& \langle b| e^{-H_{ff} \times L_y} | a \rangle, \\
Z_{\rm bg} &=& \langle a | e^{-H_{ff} \times L_y} | a \rangle,
\end{eqnarray}
where we take the spatial direction to be along the $x$ direction, and the imaginary time direction to run along the $y$ direction. $H_{ff}$ is the Hamiltonian of an open chain with free boundary conditions on both ends, as depicted in Fig.~\ref{fig:transfer_matrix}(a). In this language, $|a\rangle$ and $|b\rangle$ become the conformally invariant \textit{boundary states}, corresponding to boundary conditions $a$ and $b$, respectively. In the limit $L_y \rightarrow \infty$, using the spectral decomposition, we have
\begin{eqnarray}
    && e^{-H_{ff}\times L_y} |a \rangle \nonumber \\
    =&& e^{-E_0 L_y}\left( \langle 0|a\rangle \ |0\rangle + e^{-(E_1-E_0)L_y} \langle 1|a\rangle \  |1\rangle + \cdots \right), \nonumber \\
    \label{eq:spectral_decomp}
\end{eqnarray}
where $|0\rangle$, $|1\rangle$ denote the ground state and first excited state of the effective Hamiltonian $H_{ff}$, and $E_0$, $E_1$ are their energies. The Hamiltonian (i.e. generator of infinitesimal translations) on an infinite strip in CFT can be written as~\cite{cardy2004boundary}
\begin{equation}
    H_{\rm strip} = \frac{\pi}{L_x} \hat{L}_0 - \frac{\pi c}{24L_x},
\end{equation}
where $\hat{L}_0$ is the Virasoro generator for dilatation. Specializing to our current situation, the spectrum of $H_{ff}$ thus has the form
\begin{equation}
    E_{i,n} = \frac{\pi (h_{f|f}^{(i)} + n)}{L_x} + E_0,
    \label{eq:spectrum_ff}
\end{equation}
where $n \geq 0$ is an integer, and $h_{f|f}^{(i)}$ denotes the scaling dimensions of all primary boundary operators that can be inserted at a boundary with free boundary condition, and are arranged in ascending order: $h_{f|f}^{(i)} < h_{f|f}^{(i+1)}$. Combining Eqs.~(\ref{eq:spectrum_ff}) and~(\ref{eq:spectral_decomp}), when $\tau\gg 0$, we finally arrive at
\begin{equation}
    S([z_1,z_4]) \propto {\rm exp}\left[-\frac{\pi h_{f|f}^{(1)}}{L_x} L_y \right] = {\rm exp}\left(-\pi h_{f|f}^{(1)}\tau \right),
\end{equation}
where $h_{f|f}^{(1)}$ is the operator with the lowest scaling dimension that can appear at a free boundary. 

In Fig.~\ref{fig:fafa}, we plot $S([z_1,z_4])$ as a function of the aspect ratio $\tau$ for RSTNs with different bond dimensions. The results clearly show the anticipated exponential decay at large $\tau$ for all bond dimensions. The scaling dimension $h_{f|f}^{(1)}$ in general varies for different bond dimensions. \MF{In the large $D$ limit we again anticipate that the critical point might approach a}
percolation CFT.  \MF{In this event}, the boundary operator that can be inserted at a free boundary is the \textit{boundary} spin operator with scaling dimension $h_{1,3} = \frac{1}{3}$~\cite{cardy1984conformal, cardy1992critical}; see also Ref.~\cite{buechler2020projectiveTFIM}.
In Fig.~\ref{fig:fafa}, our numerically extracted value at large $D$: $h_{f|f}^{(1)} \approx 0.33$ indeed agrees with $h_{1,3}$. However, at small $D$, the scaling dimension significantly deviates from percolation CFT.

\begin{figure*}[t]
    \centering
\subfigure[]{
    \includegraphics[width=0.35\textwidth]{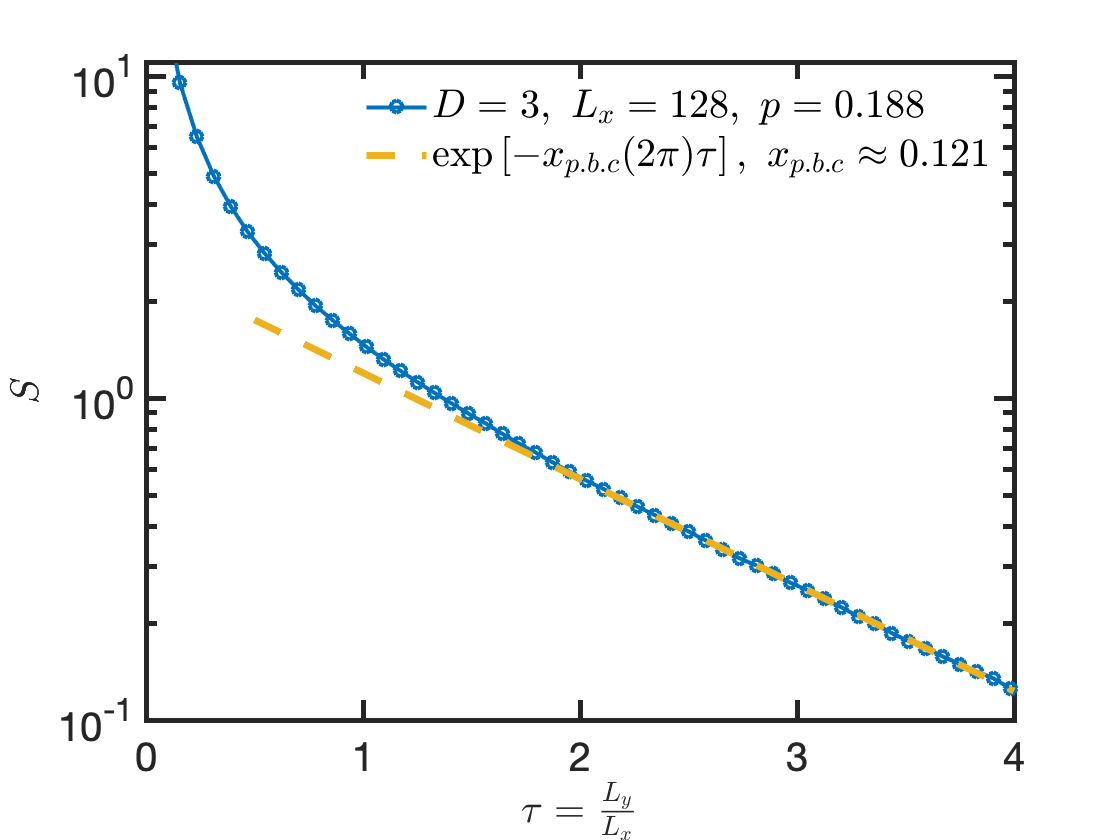}}
\subfigure[]{
    \includegraphics[width=0.35\textwidth]{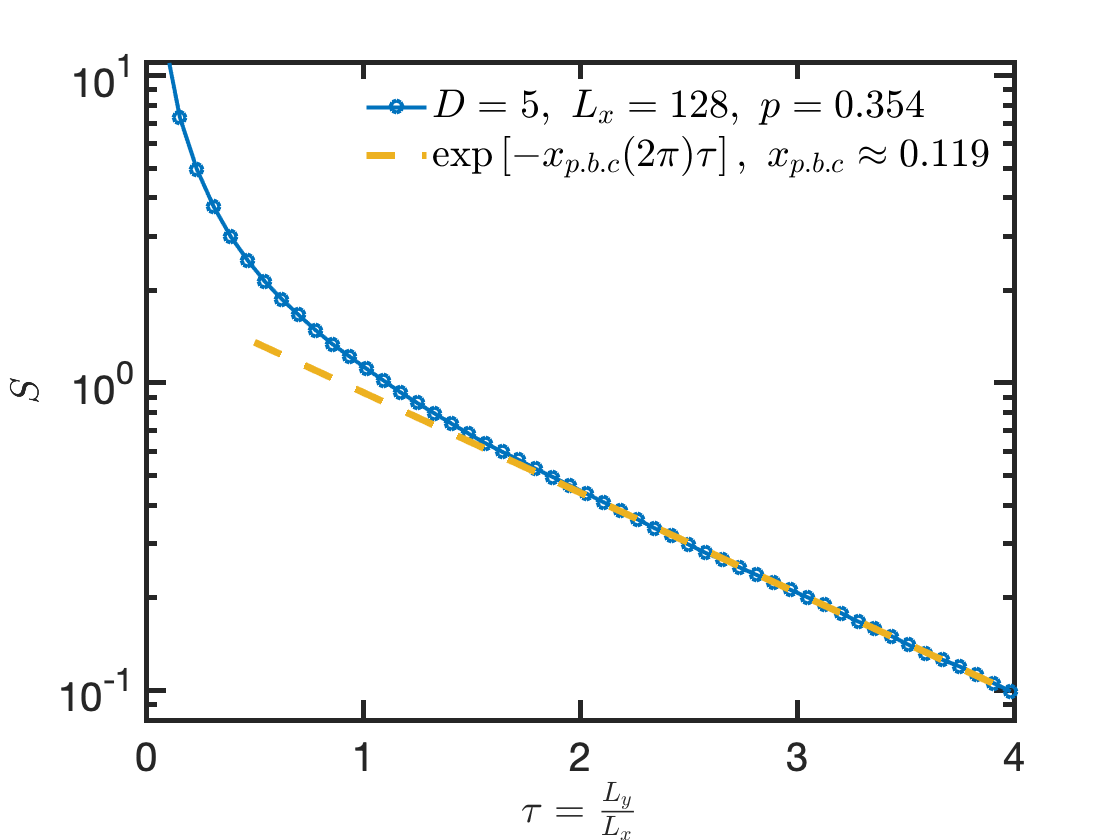}} 
\subfigure[]{
    \includegraphics[width=0.35\textwidth]{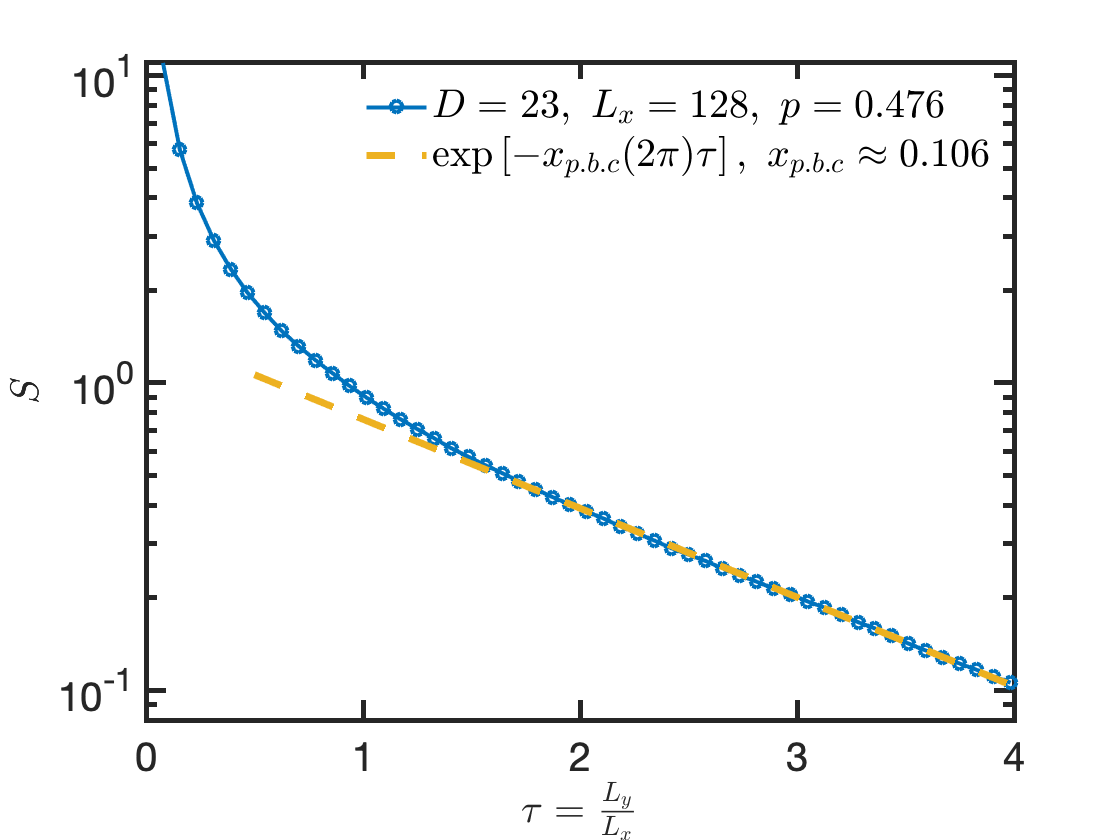}}
\subfigure[]{
    \includegraphics[width=0.35\textwidth]{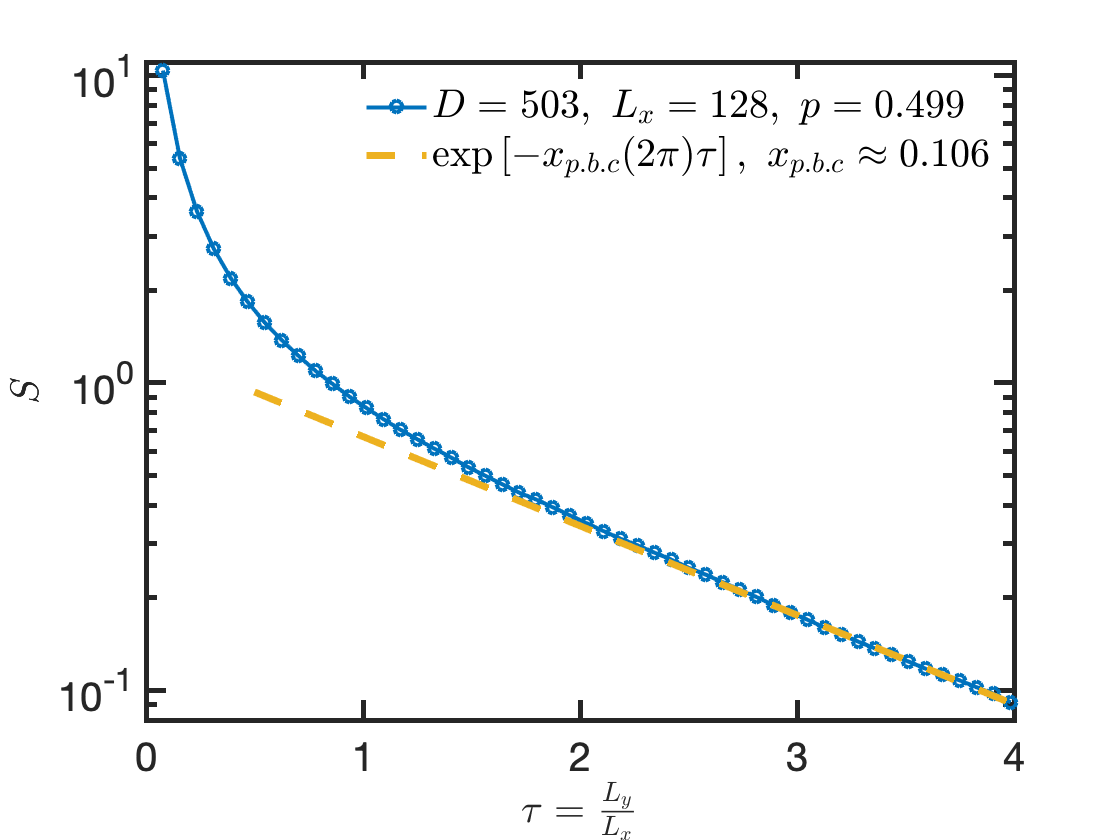}} 
    \caption{Periodic boundary condition: entanglement entropy between the top and bottom edges of the cylinder [Fig.~\ref{fig:afaa_illus}(c)] plotted against the aspect ratio $\tau$ at the critical point for RSTNs with bond dimensions (a) $D=3$; (b) $D=5$; (c) $D=23$; (d) $D=503$. The exponential decay of $S$ at large $\tau$ is controlled by the bulk operator scaling dimension $x_{p.b.c}$. At large $D$, the value of $x_{p.b.c}$ agrees with the scaling dimension of a \textit{bulk} spin (magnetization) operator in percolation $\Delta_{\sigma}=2h_{\frac{1}{2},0}=\frac{5}{48} \approx 0.104$.
    }
    \label{fig:xpbc}
\end{figure*}

\subsection{Periodic boundary condition: scaling dimension $x_{p.b.c}$}

Finally, we go back to the cylindrical geometry with periodic boundary condition along the $x$ direction. However, this time we will analyze the entanglement entropy using the transfer matrix formulation outlined in the previous subsection. As we will see, this allows us to extract one more scaling dimension of a \textit{bulk} primary field, which we denote as $x_{p.b.c}$~\cite{li2020conformal}.

Consider the entanglement entropy between the top and bottom edges of the cylinder, as illustrated in Fig.~\ref{fig:afaa_illus}(c). The partition function in this case can be written as
\begin{eqnarray}
Z(A) &=& \langle b| e^{-H_{p.b.c} \times L_y} | a \rangle, \\
Z_{\rm bg} &=& \langle a | e^{-H_{p.b.c} \times L_y} | a \rangle,
\end{eqnarray}
where $H_{p.b.c}$ denotes the generator of infinitesimal translations along the cylinder [see Fig.~\ref{fig:transfer_matrix}(b)]. One can then apply the same spectral decomposition and obtain Eq.~(\ref{eq:spectral_decomp}). The Hamiltonian $H_{p.b.c}$ with periodic boundary condition, however, has a different spectrum than $H_{ff}$:
\begin{equation}
    H_{p.b.c} = \frac{2\pi}{L_x}( \hat{L}_0 +\hat{\overline{L}}_0) - \frac{\pi c}{6L_x},
\end{equation}
where $\hat{\overline{L}}_0$ is the antiholomorphic part of the Virasoro generator. Therefore, the spectrum of $H_{p.b.c}$ is given by
\begin{equation}
    E_{i,n} = \frac{2\pi(\Delta_i + n)}{L_x} + E_0,
\end{equation}
where $\Delta_i = h_i + \overline{h}_i$ are arranged in ascending order $\Delta_i < \Delta_{i+1}$. The excitation gap $E_1-E_0$ is thus determined by the bulk primary field with the lowest scaling dimension $x_{p.b.c}\equiv \Delta_1$. We finally obtain
\begin{equation}
    S \propto {\rm exp}\left[-\frac{2\pi x_{p.b.c}}{L_x} L_y \right] = {\rm exp}\left(-2\pi x_{p.b.c} \tau \right).
\end{equation}

In Fig.~\ref{fig:xpbc}, we plot the entanglement entropy between the top and bottom edges of the cylinder as a function of $\tau$. We again observe an exponential decay of $S$ at large $\tau$, as predicted by our analysis based on general features of a CFT. The scaling dimension $x_{p.b.c}$ also varies as $D$ changes. In particular, the bulk primary field with the lowest scaling dimension (other than the identity) in percolation is the \textit{bulk} spin (magnetization) operator with $\Delta_{\sigma} = 2h_{\frac{1}{2},0} =\frac{5}{48} \approx 0.104$. Our numerically extracted value at \MF{large} $D=503$ in Fig.~\ref{fig:xpbc}(d): $x_{p.b.c} \approx 0.106$ is \MF{indeed very close the the percolation value.}
\MF{Once again, for smaller $D$,} the scaling dimensions \st{at smaller $D$ in general} deviate from percolation, again suggesting a different universality classes \MF{for each value of (prime) $D$.}

\subsection{Tensor networks with reduced randomness \label{sec:reduced_randomness}}

\begin{figure}[t]
    \centering
\subfigure[]{
    \includegraphics[width=0.4\textwidth]{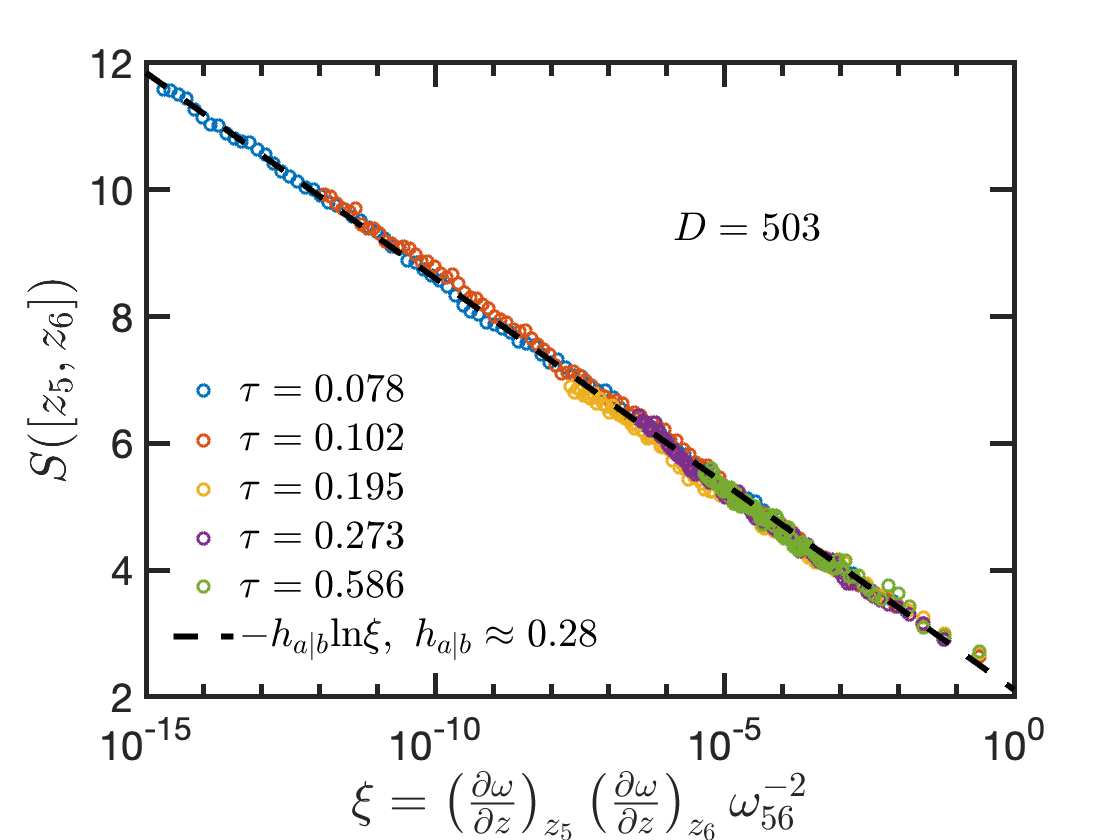}}
\subfigure[]{
    \includegraphics[width=0.4\textwidth]{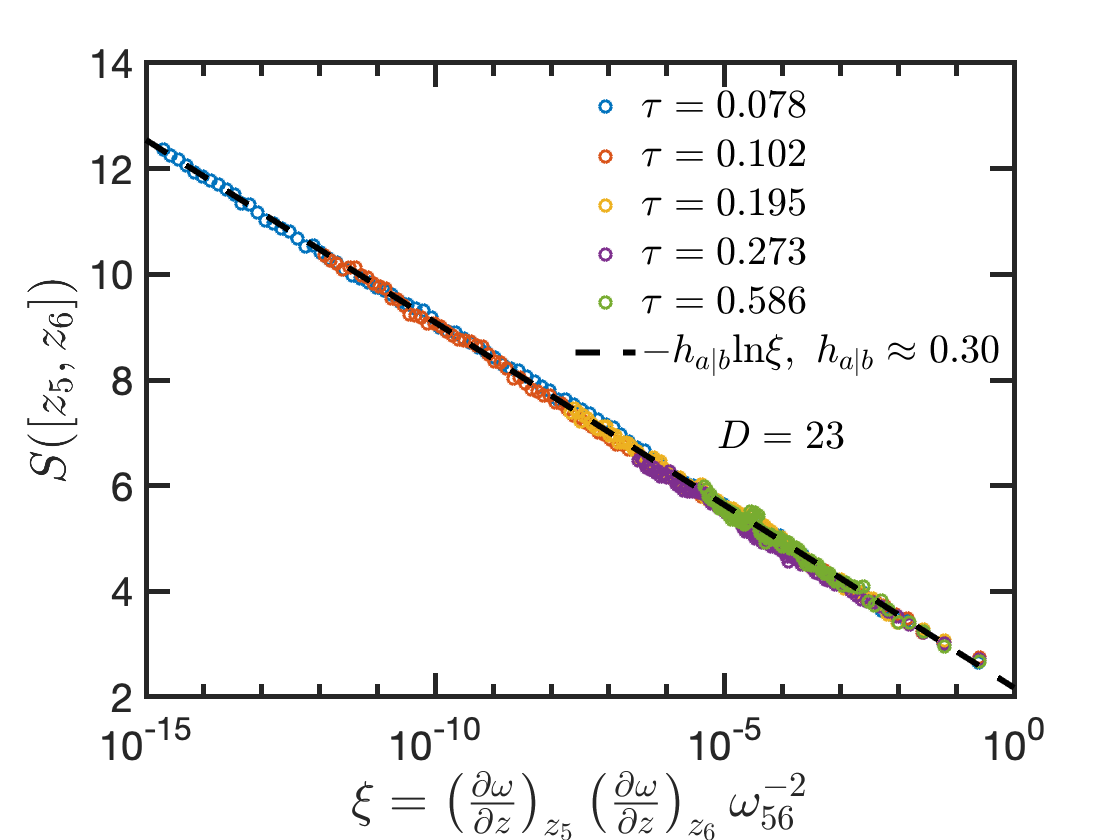}} 
    \caption{Same as Fig.~\ref{fig:aaaa}, but for translationally invariant tensor networks with random measurement. (a) $D=503$, $p_c=0.499$; (b) $D=23$, $p_c=0.490$. The precise locations of $p_c$ may slightly shift compared to random tensor networks. However, the quantity $h_{a|b}$ in both cases are consistent with Fig.~\ref{fig:aaaa}.
    }
    \label{fig:uniform}
\end{figure}

\begin{figure}[t]
    \centering
    \includegraphics[width=0.4\textwidth]{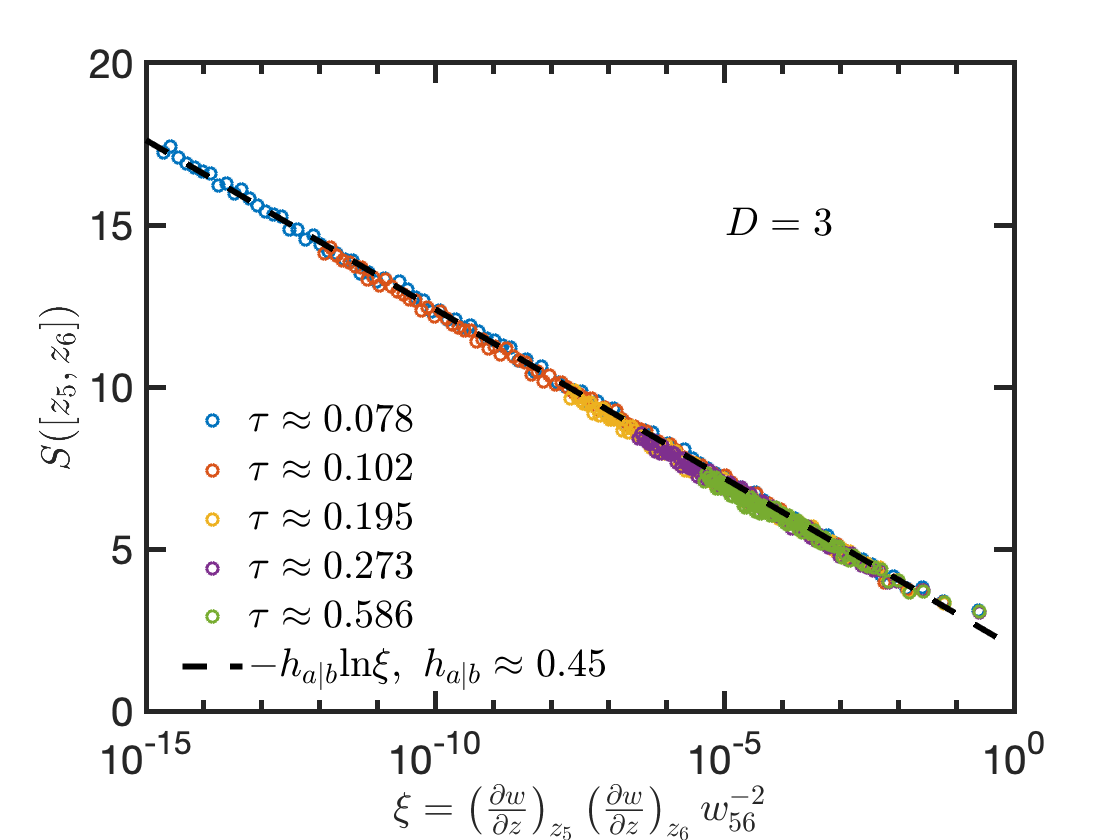}
    \caption{Same as Fig.~\ref{fig:aaaa}, but for random tensor networks with spatially periodic measurement, for $D=3$ and $p_c=0.199$. The quantity $h_{a|b}$ in this case is slightly smaller than that in Fig.~\ref{fig:aaaa}.
    }
    \label{fig:supercell}
\end{figure}

Before closing this section, we discuss the role of randomness in the observed entanglement transitions and criticality. There are two important sources of randomness in RSTNs: randomness in the tensor compositions of the network, and randomness in the measurement locations. In particular, the results shown in this section strongly indicate that \MF{in the large $D$ limit} the entanglement transition in RSTNs approach the two-dimensional bond percolation universality class. In this limit, one then expects that randomness in the measurement locations should be crucial, whereas randomness in the tensor network itself maybe not. On the other hand, the role of each individual source of randomness at small $D$ is not clear.
Below, we shall examine RSTNs with reduced randomness by considering either \textit{translationally invariant} tensor networks with random measurement, or random tensor networks with \textit{spatially periodic} measurement~\cite{li2019hybrid, li2021entanglement}.
In other words, we retain only one type of randomness at a time.

We start from translationally invariant tensor networks with random measurement. We take these tensors to be spatially uniform throughout the system within each realization, while being different \MF{and chosen randomly} for different realizations.
In Fig.~\ref{fig:uniform}(a), we plot the same entanglement entropy as in Fig.~\ref{fig:aaaa} with boundary condition $aaaa$ for $D=503$. We find that the entanglement entropy at the critical point again obeys the CFT prediction in Eq.~(\ref{eq:EE_aaaa_2}), and the quantity $h_{a|b}\approx 0.28$ is also consistent with Fig.~\ref{fig:aaaa}(d) as well as \MF{close to the } percolation \MF{value}. In Fig.~\ref{fig:aaaa}(b), we plot the same quantity for $D=23$, and again we obtain a consistent result with Fig.~\ref{fig:aaaa}(c), although the precise locations of $p_c$ may slightly shift compared to random tensor networks. These results suggest that in the large $D$
limit, randomness in measurement locations can indeed account for the observed criticality, whereas randomness in the tensor compositions is not necessary. On the other hand, for small $D$ (e.g., $D=3$ and $D=5$), we no longer see a sharp signature for a continuous entanglement transition in translationally invariant tensor networks as diagnosed by a peak in the mutual information between two distant subregions, even though the measurement locations are still chosen at random. It is thus unclear whether there is a well-defined notion of volume-law and area-law ``phases" separated by a ``phase transition" in this case.
This indicates that the underlying physics for the entanglement transitions at small $D$ are quite different from that in the large $D$ limit. Indeed, as we have seen, the transitions at small $D$ clearly belong to distinct universality classes from the geometric lattice percolation transition. However, a detailed study of the phase structures for translationally invariant tensor networks at small $D$ is left for future work.

We also consider random tensor networks with spatially periodic measurement. Specifically, denote the coordinate of a bond in the tensor network as $(x,y)$. Then, a single-qudit forced measurement is made on this bond only when~\cite{li2019hybrid}
\begin{equation}
    \lfloor x \sqrt{p} \rfloor < \lfloor (x+1) \sqrt{p} \rfloor, \quad  \lfloor y \sqrt{p} \rfloor < \lfloor (y+1) \sqrt{p} \rfloor,
\end{equation}
where the floor function $\lfloor x \rfloor$ denotes the greatest integer less than or equal to $x$. In this way, there is on average one measurement made each time $x$ or $y$ advances by $1/\sqrt{p}$, so that the measurement rate is $p$.
In this way, the measurement locations form a ``supercell" on the lattice, and are thus spatially periodic. The geometric minimal cut picture, \MF{valid in the large $D$ limit} implies that an entanglement transition should be absent in this case. Indeed, we find that the entanglement transition can only be identified at small $D$ for periodic measurements. As $D$ increases, the critical point shifts toward $p=1$, and the signature for the transition (the peak in the mutual information as in Fig.~\ref{fig:MI_d3}) also becomes weaker. In Fig.~\ref{fig:supercell}, we show $h_{a|b}$ for $D=3$, where we find the value is slightly smaller than that in the case of random measurements.

These results highlight the different roles that the two sources of disorder play in the entanglement criticality at small and large $D$. Randomness in the measurement locations is crucial for the entanglement transition at large $D$, consistent with \MF{expectations from} the minimal cut picture; whereas randomness in the tensor compositions alone is sufficient for a continuous entanglement transition at small $D$.

\section{Entanglement properties in the volume-law phase and KPZ scaling}
\label{sec:kpz}

Having carefully examined the entanglement properties at the critical points, we next turn to the universal entanglement properties in the volume-law phase. The expression~(\ref{eq:EE}) for the entanglement entropy has a physical interpretation in terms of the underlying emergent statistical mechanics model~\cite{bao2018spin, jian2018spin}: it is the free energy cost of a domain wall due to a boundary condition twist in subregion $A$.
Based on this interpretation, it was proposed that most aspects of the entanglement entropy in hybrid circuits in the volume-law phase can be understood by invoking an ``entanglement domain wall'' picture, which behaves as a directed polymer in a random environment (DPRE)~\cite{PhysRevLett.54.2708, PhysRevLett.55.2923, PhysRevLett.55.2924}.
This scenario has recently been confirmed numerically in hybrid Clifford circuits, and analytically demonstrated in hybrid Haar random circuits~\cite{li2021entanglement}.

The DPRE has several characteristic critical exponents, controlled by the Kardar-Parisi-Zhang (KPZ) fixed point~\cite{PhysRevLett.58.2087, PhysRevLett.56.889}.
For example, consider the entanglement entropy of a subregion $A$ belonging to the top edge of the cylinder (see Fig.~\ref{fig:no_measurement}).
The height of the domain wall in the transverse direction scales as $(L_A)^{\zeta}$~\cite{PhysRevLett.54.2708, PhysRevLett.55.2923, PhysRevLett.55.2924}, where $\zeta = 2/3$ is a characteristic ``wandering exponent'' of the DPRE.
For $L_y \gg (L_A)^{2/3}$, the cylinder can be considered as infinitely long from the perspective of the maximal vertical extent of the domain wall. Then, the sample-to-sample fluctuation of the entanglement entropy is given by~\cite{PhysRevLett.54.2708, PhysRevLett.55.2923, PhysRevLett.55.2924}
\begin{equation}
   \delta S(L_A) \equiv \sqrt{\langle S(L_A)^2 \rangle - \langle S(L_A) \rangle^2} \propto (L_A)^{1/3},
\end{equation}
where $\langle \cdot \rangle$ denotes averaging over disorder realizations, and $\beta = 1/3$ is another characteristic ``roughness exponent''.

\begin{figure}[t]
    \centering
\subfigure[]{
    \includegraphics[width=0.4\textwidth]{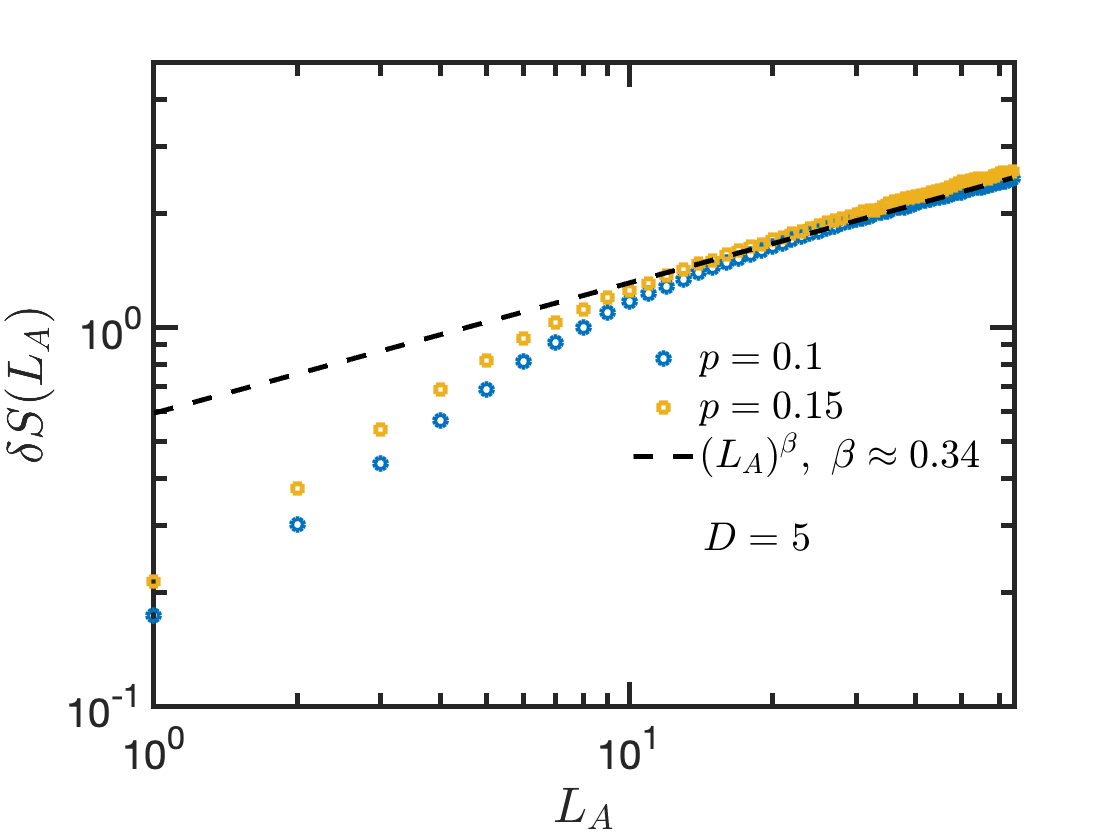}}
\subfigure[]{
    \includegraphics[width=0.4\textwidth]{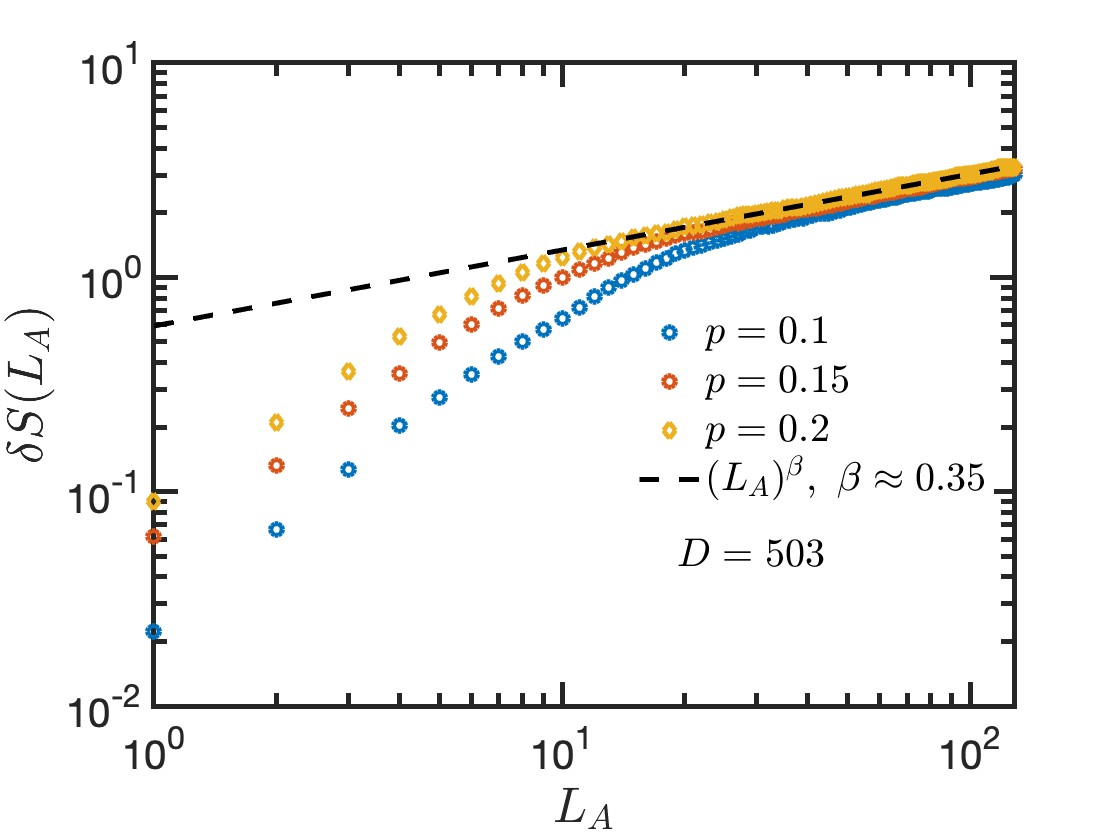}} 
    \caption{Sample-to-sample fluctuations of the entanglement entropy $\delta S(L_A)$ as a function of $L_A$ in the volume-law phase. (a) $D=5$; (b) $D=503$. We find a scaling behavior consistent with KPZ for different measurement rates in both cases.
    }
    \label{fig:EE_fluctuation}
\end{figure}

In Fig.~\ref{fig:EE_fluctuation}, we plot the scaling of $\delta S(L_A)$ with $L_A$ for RSTNs with $D=5$ and $D=503$ in the volume-law phase, respectively.
In both cases, we consider a few different choices of the measurement rates, such that the system is deep in the volume-law phase.
We find, for all cases considered here, a scaling behavior that is consistent with the DPRE.
Our results thus confirm that the universal entanglement properties in RSTNs in the volume-law phase can be quantitatively described in terms of a fluctuating domain wall in a random environment. At large $D$, this result may be thought of as following from the geometric minimal cut in the underlying lattice with random broken bonds, which also scales as the DPRE.
This interpretation does not readily extend to small $D$.

\begin{figure}[t]
    \centering
\subfigure[]{
    \includegraphics[width=0.4\textwidth]{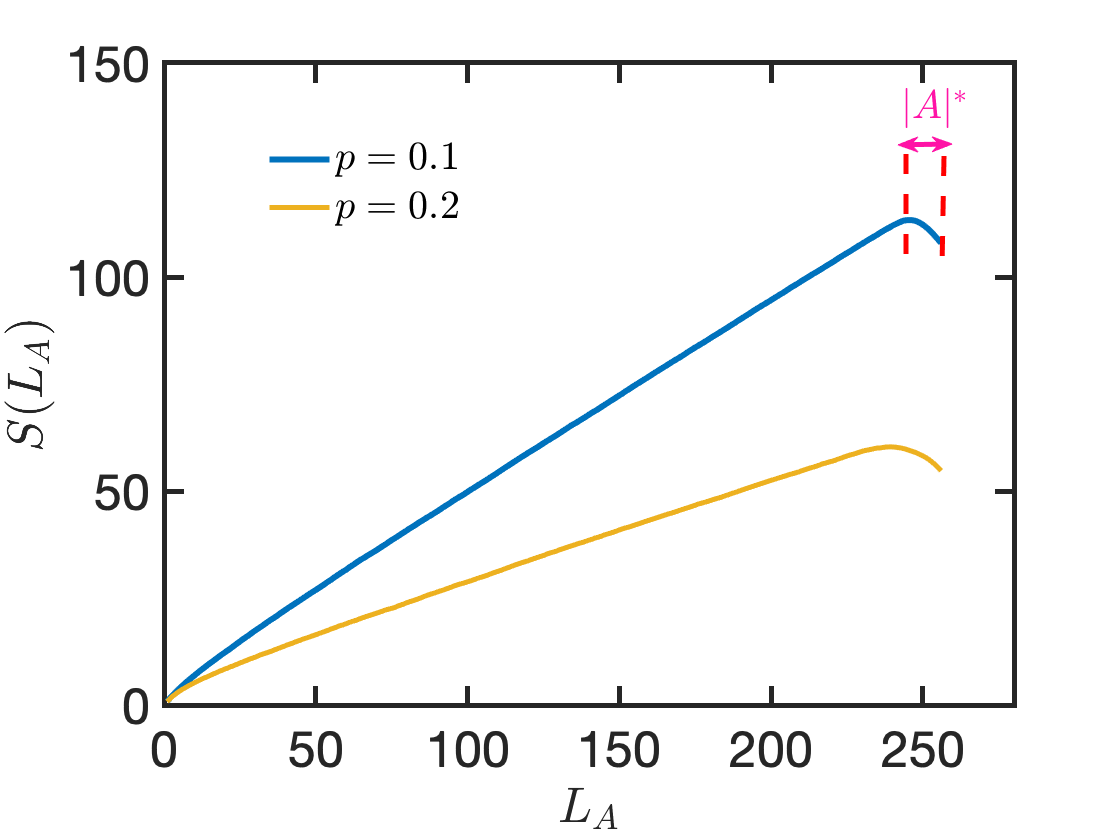}}
\subfigure[]{
    \includegraphics[width=0.4\textwidth]{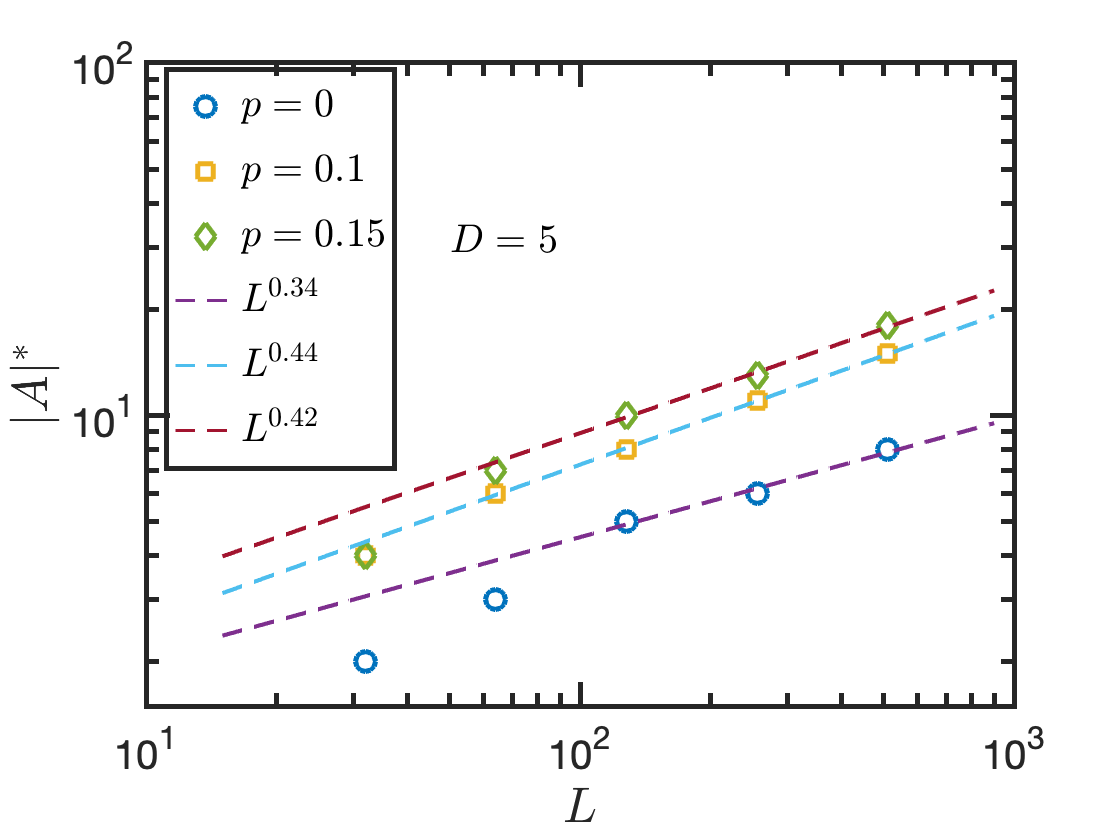}} 
    \caption{Quantum error-correcting code properties of the random stabilizer tensor network states. Numerical results are shown for RSTNs on a cylinder with $D=5$. (a) Non-monotonicity of the subsystem entanglement entropy at the top edge, from which one can define the ``contiguous code distance" denoted as $|A|^*$. (b) Scaling of the contiguous code distance as the system size. The results are qualitatively the same for other choices of $D$.
    }
    \label{fig:EE_code}
\end{figure}

The resilience of the entanglement entropy against measurement in the volume-law phase suggests that the system can be viewed as a dynamically generated quantum error correcting code (QECC)~\cite{PhysRevB.103.104306, PhysRevB.103.174309}.
Specifically, the amount of entanglement entropy can be associated with the number of logical qubits encoded, which decreases to zero as one approaches the transition point. An important characterization of a QECC is the (contiguous) code distance, defined as the minimal length of a logical operator acting (in a contiguous region) within the code space.
As it turns out, the scaling of the contiguous code distance can also be calculated with the domain wall picture,
as a length scale $|A|^*$ where the dominating domain wall configuration contributing to $S(L_A)$ changes~\cite{PhysRevB.103.104306}.
This code distance may be conveniently extracted from a non-monotonic behavior of the subsystem entanglement entropy $S_A$ against $L_A$ in the volume-law phase, starting from a maximally mixed initial state.

In RSTNs on a cylinder, the top (or bottom) edge is naturally in a mixed state, and we indeed observe a similar non-monotonic behavior in the subsystem entanglement entropy in the volume-law phase, as shown in Fig.~\ref{fig:EE_code}(a). The maximum of the entanglement entropy at $L-|A|^*$ corresponds to the point beyond which the mutual information between $\overline{A}$ and the bottom edge is zero: $I_{\overline{A}, {\rm bottom}}\approx 0$, hence measurement in $\overline{A}$ cannot decrease the entanglement entropy $S(L_A)$~\cite{PhysRevB.103.104306}. Therefore, $|A|^*$ can be identified as the contiguous code distance in our system. In Fig.~\ref{fig:EE_code}(b), we plot the scaling of $|A|^*$ with the total system size $L_x$.
Similarly to Ref.~\cite{li2021entanglement}, the exponents we found are slightly larger than the KPZ prediction $\beta = 1/3$, possibly due to the limited system sizes as well as the strong sample-to-sample fluctuations that prevent us from determining $|A|^*$ accurately. We expect that the code distance should also scale as $(L_A)^{1/3}$ for sufficiently large $L_A$.

\section{Discussions and outlook}
\label{sec:discussion}

To summarize, we \ZCY{study} 
two types of entanglement phase transitions in RSTNs.
When tuning the bond dimension $D$ alone, we observe an area law scaling of the boundary entanglement entropy at $D = 2$, and a volume law scaling at prime $D \ge 3$.
This is thus suggestive of a phase transition at $D_c$ between $2$ and $3$, \MF{provided the tensor network model can be appropriately generalized to non-integer bond dimension $D$.}

We observe a second class of phase transitions driven by randomly breaking bonds in the bulk of the RSTN with probability $p$, effectively putting the RSTN on a random lattice.
Here, for all prime $D \ge 3$, we see a continuous phase transition separating a volume law phase of the boundary state at small $p$, and an area law phase at large $p$.
The critical points for \emph{different} values of $D$ are apparently described by \emph{different} \ZCY{CFTs}, 
as summarized in Table~\ref{table:scaling}.
With increasing values of $D$, the critical exponents in the corresponding CFT are approaching their counterparts in critical \emph{first-passage} percolation, suggesting that the critical point is becoming increasingly ``percolation like''.

The last point suggests that a ``minimal cut'' picture is at work as $D \to \infty$.
For example, the entanglement entropy of a single boundary region should differ from the weight of the minimal cut (i.e. the minimal number of bonds a ``domain wall'' must break, see Fig.~\ref{fig:minimal_cut}) by at most an $\mc{O}(1)$ constant:
\begin{align}
    \label{eq:EE_mincut_order_1_correction}
    S(A) = \mathrm{min} |\gamma_A| \times \ln D - \mc{O}(1).
\end{align}
This relation would reproduce the value of $h_{a|b} \approx \frac{\sqrt{3}}{2\pi}$ at large $D$ (Table~\ref{table:scaling}).
Moreover, the entanglement between two disjoint regions, as quantified by the mutual information and the mutual negativity, should also agree with those defined by minimal cuts~\cite{nahum2018transition, sang2020entanglement}, and are supported by the values of $h_{f|f}^{(1)} \approx 1/3$, $h_{a|a}^{(1)} \approx 2$, and $\Delta \approx 2$  at the large $D$ critical points (Table~\ref{table:scaling}).

As we have mentioned earlier in this paper, it is known in RTNs~\cite{hayden2016holographic} and in RSTNs~\cite{nezami2016RSTN} that Eq.~(\ref{eq:EE_mincut_order_1_correction}) holds \emph{exactly} as $D \to \infty$ \textit{if} we keep $L$ finite.
\ZCY{According to Ref.~\cite{nezami2016RSTN},} 
Eq.~(\ref{eq:EE_mincut_order_1_correction}) holds provided that $D$ scales as (at least) $\exp(L)$, \ZCY{such that the deviation from ${\rm min}|\gamma_A| \times {\rm ln}D$ is small}.
In this work, 
while we vary $D$ to take various prime values \ZCY{that} 
are possibly large in themselves, they are nevertheless small compared to $\exp(L)$, thus not within the regime \ZCY{where Eq.~(\ref{eq:EE_mincut_order_1_correction}) can be demonstrated analytically as in} 
Ref.~\cite{nezami2016RSTN}.
For this reason, we \ZCY{are inclined to} 
interpret \ZCY{our results} 
at larger $D$ as reliable estimates of operator dimensions in the underlying ``percolation like'' CFT in the thermodynamic limit $L \to \infty$, rather than \ZCY{artifacts} 
due to finite system sizes.
Thus, a separate justification \YL{(or at least a significant improvement of the previous bound~\cite{nezami2016RSTN})} is needed for the emergence of these percolation like CFTs \ZCY{observed in this work}.

Recently in Ref.~\cite{zabalo2021ceff}, the authors pointed out that for monitored random stabilizer circuits (and RSTNs) with $D = q^n$ where $q$ is a prime number, the underlying statistical mechanics models have different symmetries for different values of $q$ (see Ref.~\cite{li_et_al_2022_largeD} for details).
This explains the series of
universality classes we observe at different values of $D = q$.
It would be interesting to use these results~\cite{zabalo2021ceff} to explain the percolation like exponents at large $D$.

For completeness, we \ZCY{also} briefly summarize the critical properties of monitored random stabilizer circuits for various values of prime $D = q$, \ZCY{for which detailed results will be reported elsewhere}~\cite{li_et_al_2022_largeD}.
Similarly to RSTNs, we find a series of distinct critical points for each value of $D \ge 2$ (since random circuits will always generate volume-law entangled states for any $D\ge 2$), and most of the critical exponents are also approaching critical first-passage percolation with increasing $D$.
However, this series of critical points  \MF{appears to be} distinct from those in the RSTN.
Understanding their difference is left for future work~\cite{li_et_al_2022_largeD}.

We have also considered several variations of the RSTN.
Upon replacing all the forced measurements -- responsible for implementing tensor contraction and bond breaking -- with projective measurements (see Appendix~\ref{app:prob_measure} for a detailed description), we find almost identical phase diagrams and critical exponents (data not displayed).
Reducing the randomness in the RSTN does not seem to alter the critical exponents significantly (see Sec.~\ref{sec:reduced_randomness}), although the location of $p_c$ may \ZCY{change} 
and the critical point may not be easily accessible.
These results suggest the robustness of the critical point against these perturbations.

\XC{There are several \st{other} interesting applications and extensions of \ZCY{the present work}
\st{our work,} 
which we briefly mention.

Firstly, the area law scaling of entanglement entropy observed on the square lattice at $D=2$ is not universal. In principle, \ZCY{one} \st{we} could construct RSTNs \st{with larger $n$ or consider} \ZCY{on lattices or graphs} with a larger coordination number, so that the RSTNs can host a \MF{volume law entangled state.  In this case, upon breaking bonds in the bulk, one could drive an} entanglement transition.
It would then be interesting to explore the critical exponents of these \MF{transitions}.

Secondly, one could introduce \ZCY{physical qudits} \st{qubits} in the bulk which correspond to dangling legs of the bulk tensors. In this setup, the entanglement entropy for the boundary \ZCY{qudits} \st{qubits} \ZCY{may} \st{will} be modified. One could then investigate the bulk-boundary correspondonce and the error correction properties of these RSTNs~\cite{hayden2016holographic,Pastawski_2015, almheiri2014bulklocality, polchinski2015qec}.

\YL{Thirdly, it may be interesting to move away from the fully random RSTN and consider instead those with a restricted class of local tensors, e.g.  less entangled tensors, tensors with a global symmetry, and ``isometric'' tensors for any bipartition of the legs~\cite{Pastawski_2015}.
Previously in the context of hybrid circuits, these ideas have lead to a plethora of ``measurement-protected quantum phases''~\cite{barkeshli2020topological, sang2020protected, ippoliti2020measurementonly, chenxiao2020nonunitary, diehl2020freefermion, ippoliti2020spacetimeduality, bao2021enriched, diehl2021sineGordon, grover2021duality, ippoliti2021fractal}.
}

Lastly, \ZCY{one may be able to construct tensor network states using tensors that encode the Boltzmann weights of a classical statistical mechanics model~\cite{Levin_2007}. It would be interesting to explore the implication of a phase transition of the underlying statistical mechanics model on the entanglement properties of the boundary state.}}

\section*{Acknowledgments}
We acknowledge useful discussions with Tianci Zhou, Andreas Ludwig, and Romain Vasseur. Z.-C.Y. acknowledges financial support from NSF PFCQC program and DoE
ASCR Accelerated Research in Quantum Computing program (award No.
DE-SC0020312).
This work was supported by the Heising-Simons Foundation (Y.L. and M.P.A.F.), and by the Simons Collaboration on Ultra-Quantum Matter which  is  a  grant from the Simons Foundation (651440, M.P.A.F.).
We thank the Aspen Center for Physics where part of this work was carried out, which is supported by National Science Foundation grant PHY-1607611 (M.P.A.F. and X.C.).
The numerical calculations were performed on the Boston University Shared Computing Cluster, which is administered by Boston University Research Computing Services.


\bibliography{reference}


\appendix

\section{Numerical results for the mutual information on a cylinder for $D=5$ and $D=23$}
\label{sec:MI_additional}

\begin{figure}[tbh]
    \centering
\subfigure[]{
    \includegraphics[width=0.4\textwidth]{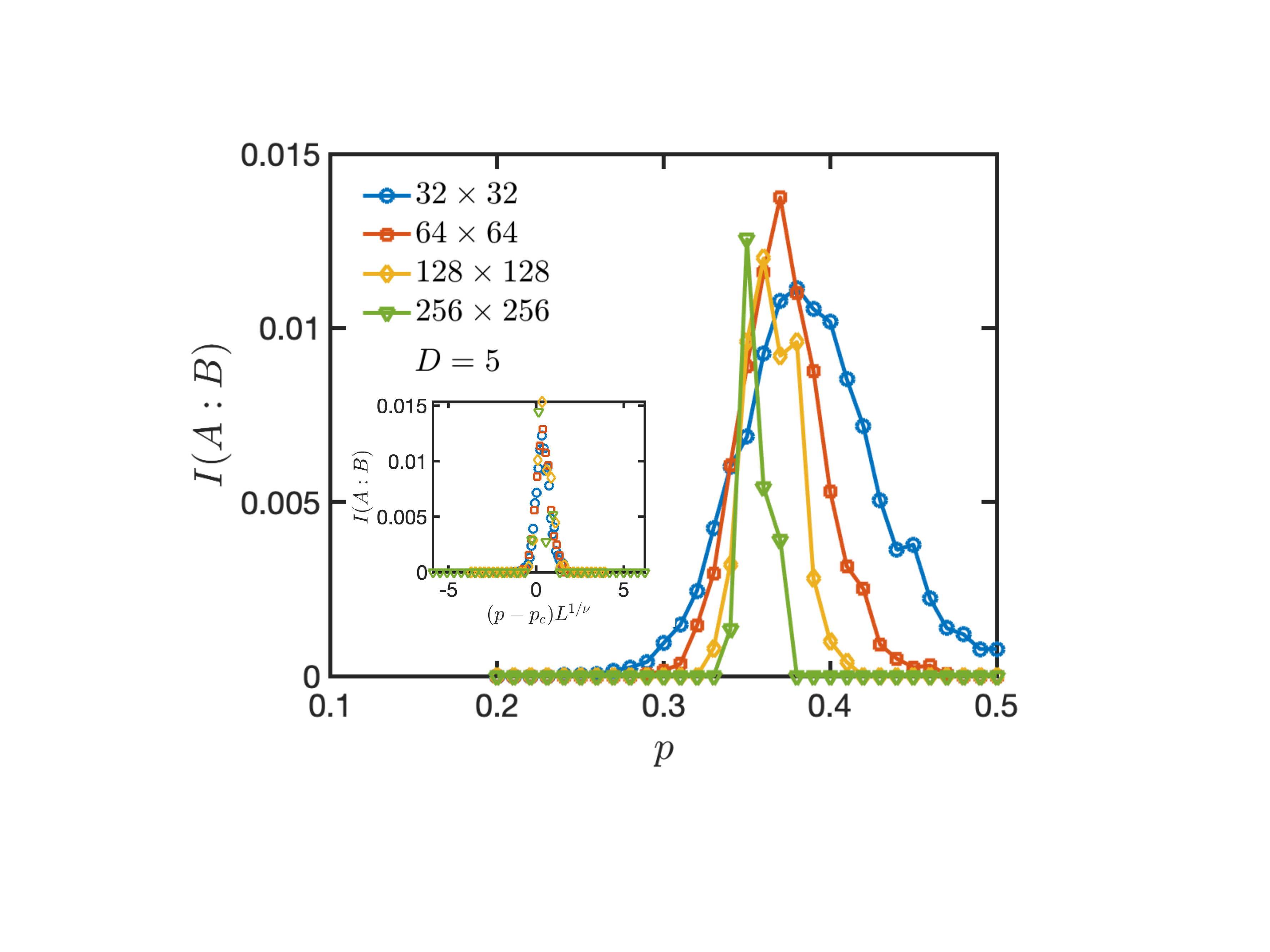}}
\subfigure[]{
    \includegraphics[width=0.4\textwidth]{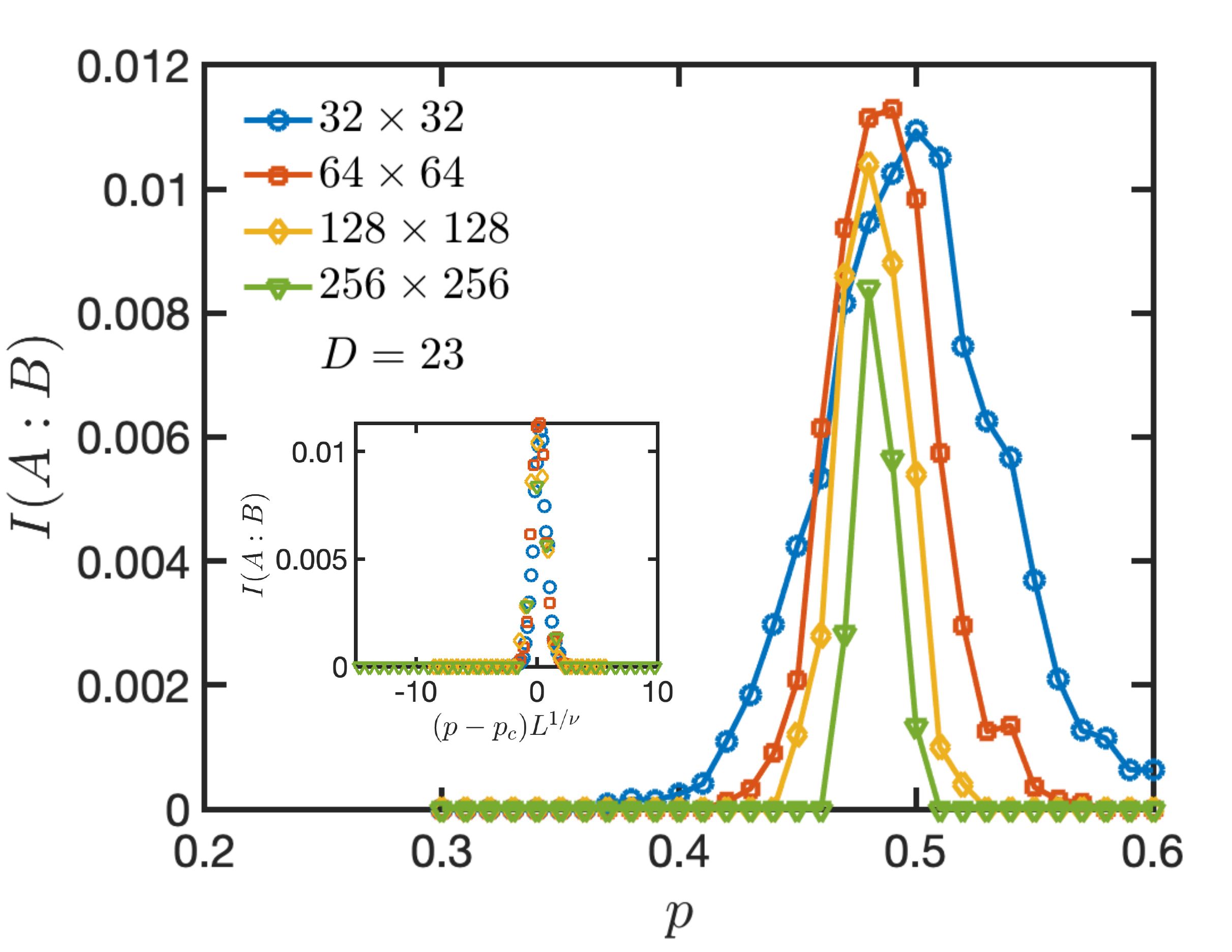}} 
    \caption{Mutual information between two antipodal subregions of size $|A|=|B|=L_x/8$, for RSTNs of size $L_x \times L_y$ on a cylinder (periodic boundary condition along the $x$ direction): (a)$D=5$, (b)$D=23$. Insets: data collapse using the scaling form $I(A:B) = f[(p-p_c)L^{1/\nu}]$, which yields (a) $p_c \approx 0.35$, $\nu \approx 1.5$; and (b) $p_c\approx 0.48$, $\nu \approx 1.26$.
    }
    \label{fig:MI_additional}
\end{figure}

We show additional numerical results for the mutual information of cylindrical random tensor networks with $D=5$ and $D=23$ in Fig.~\ref{fig:MI_additional}. We again see a peak at some critical measurement rate $p_c$ in both cases, indicating the existence of an entanglement transition. Furthermore, the critical value $p_c$ increases towards $0.5$ as $D$ increases.

\section{The Schwarz-Christoffel mapping from a finite rectangle to the lower half plane}
\label{sec:schwarz}

\begin{figure}[t]
\centering
\includegraphics[width=0.5\textwidth]{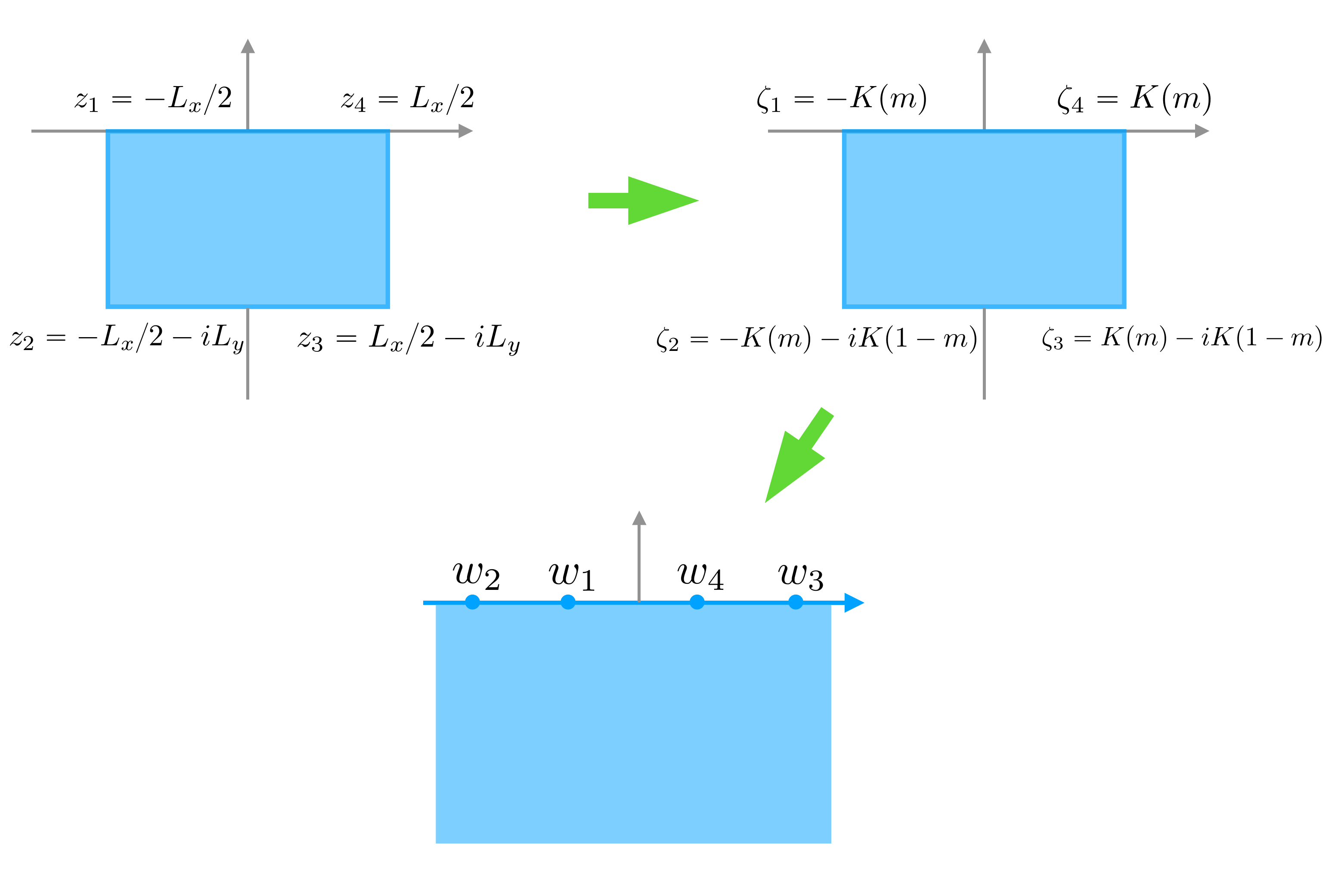}
\caption{The Schwarz-Christoffel mapping that conformally maps the rectangle to the lower half plane. The four corners $z_1,\ldots, z_4$ maps to $w_1=-1$, $w_2=-m^{-1/2}$, $w_3=m^{1/2}$, and $w_4=1$ on the real axis.}
\label{fig:schwarz}
\end{figure}

In this section, we explain in detail the Schwarz-Christoffel mapping that conformally maps a finite rectangle to the lower half plane~\cite{li2020conformal}.
The Schwarz-Christoffel mapping proceeds in two steps, as illustrated in Fig.~\ref{fig:schwarz}: (1) mapping to a ``canonical rectangle" via a scale transformation; (2) mapping from the canonical rectangle to the lower half plane. The first step transforms the original rectangle of size $L_x \times L_y$ to one of size $2K(m) \times K(1-m)$, where $K(m)$ is the complete elliptic integral of the first kind. The parameter $m \in [0,1]$ can be solved numerically by requiring that the aspect ratio remains unchanged:
\begin{equation}
    \tau \equiv \frac{L_y}{L_x} = \frac{K(1-m)}{2K(m)} \in [0,+\infty).
    \label{eq:aspect_ratio}
\end{equation}
Notice that since the tensor network is isotropic along the $x$ and $y$ directions microscopically, the aspect ratio of the system when viewed as a CFT is exactly given by $L_y/L_x$.
Then, one can go from the original complex $z$-plane to the $\zeta$-plane by the rescaling:
\begin{equation}
    \zeta = \frac{2K(m)}{L_x} z.
\end{equation}
The second step is achieved via the Jacobi sn function, which is the inverse of the elliptic integral $K(m)$:
\begin{equation}
    w(\zeta) = {\rm sn}(\zeta, m).
\end{equation}
This maps the boundary of the rectangle to the real axis. In Fig.~\ref{fig:schwarz} we show the image of the four corners of the rectangle under this mapping.

Under a conformal transformation, correlation functions of primary fields in a CFT transform as~\cite{BPZ1984}:
\begin{equation}
    \langle \phi_1(z_1) \ldots \phi_n(z_n) \rangle = \left[ \prod_{i=1}^n \left(\frac{\partial{\omega}}{\partial{z}} \right)_{z_i}^{h_i} \right] \langle \phi_1(\omega_1) \ldots \phi_n(\omega_n) \rangle,
\end{equation}
where $h_i$ is the scaling dimension of the field $\phi_i$. Applying the above transformation to the example of Eq.~(\ref{eq:EE_aaaa}):
\begin{eqnarray}
    \langle \phi_{a|b}(z_5) \phi_{b|a}(z_6) \rangle &=& \left( \frac{\partial{\omega}}{\partial{z}}\right)_{z_5}^{\widetilde{h}_{a|b}} \left( \frac{\partial{\omega}}{\partial{z}}\right)_{z_6}^{\widetilde{h}_{a|b}} \langle \phi_{a|b}(\omega_5) \phi_{b|a}(\omega_6) \rangle   \nonumber \\
    &\propto& \left( \frac{\partial{\omega}}{\partial{z}}\right)_{z_5}^{\widetilde{h}_{a|b}} \left( \frac{\partial{\omega}}{\partial{z}}\right)_{z_6}^{\widetilde{h}_{a|b}} \frac{1}{(w_5-w_6)^{2\widetilde{h}_{a|b}}}.   \nonumber \\
\end{eqnarray}
This leads to Eq.~(\ref{eq:EE_aaaa_2}) in the main text, where we define $\widetilde{h}_{a|b} \equiv h_{a|b} {\rm ln}D$, so as to show explicitly the factor of ${\rm ln}D$ when $D$ is varied, and $h_{a|b}$ is finite as $D$ becomes large.
In particular, since $w(z) = {\rm sn}\left[\frac{2K(m)}{L_x}z,m\right]$, we obtain
\begin{equation}
    \frac{\partial{\omega}}{\partial{z}} = \frac{2K(m)}{L_x} \ {\rm cn}\left[\frac{2K(m)}{L_x}z,m\right] \ {\rm dn}\left[\frac{2K(m)}{L_x}z,m\right],
    \label{eq:sn_deriv}
\end{equation}
where ${\rm cn}(\zeta,m)$ and ${\rm dn}(\zeta,m)$ are the other two Jacobi elliptic functions.

\section{Proof of the negativity formula Eq.~(\ref{eq:negativity})}
\label{sec:proof_negativity}

We now prove Eq.~(\ref{eq:negativity}) in the main text, which provides a simple method for computing the mutual negativity for stabilizer states. The proof follows directly from the method in Ref.~\cite{sang2020entanglement}.
Let us first define the stabilizer subgroup $\mathcal{S}$ supported on subsystem $A\cup B$ and its generators:
\begin{equation}
    \mathcal{G}(\mathcal{S}) = \{ g_1, g_2, \ldots, g_m \},
\end{equation}
where $m = {\rm dim} \ \mathcal{S}$. The density matrix on $A \cup B$ can be represented in terms of the group elements of $\mathcal{S}$:
\begin{equation}
    \rho_{A\cup B} = \frac{1}{D^{|A\cup B|}} \sum_{g\in \mathcal{S}} g.
\end{equation}
The mutual negativity~(\ref{eq:negativity_def}) is related to the partial transpose of $\rho_{A\cup B}$ on subsystem $A$:
\begin{equation}
    \rho_{A\cup B}^{\intercal_A} = \frac{1}{D^{|A\cup B|}} \sum_{g\in \mathcal{S}} g^{\intercal_A},
\end{equation}
where
\begin{equation}
    g^{\intercal_A} = (g_A)^{\intercal} \otimes g_B.
\end{equation}
In terms of the Pauli strings, $g_A$ takes the form:
\begin{equation}
    g_A = Z_1^{u_1} X_1^{v_1} \otimes Z_2^{u_2}X_2^{v_2} \otimes \cdots \otimes Z_{|A|}^{u_{|A|}} X_{|A|}^{v_{|A|}},
\end{equation}
where we have ignored an unimportant global phase factor.
Using properties of the Pauli matrices under transpose $(Z^{u})^{\intercal}= Z^u$, $(X^v)^{\intercal} = X^{D-v}$, it is straightforward to show that
\begin{equation}
    g_A^{\intercal} = \overline{\omega}^{{\bm u}\cdot {\bm v}} Z_1^{u_1}X_1^{D-v_1}\otimes \cdots \otimes Z_{|A|}^{u_{|A|}} X_{|A|}^{D-v_{|A|}} \equiv x_A(g)\ \overline{g}_A,
\end{equation}
where ${\bm u}\cdot {\bm v} = \sum_{i=1}^{|A|} u_i v_i$, and we have defined shorthand notations $x_A(g)\equiv \overline{\omega}^{{\bm u}\cdot {\bm v}}$ and $\overline{g}_A \equiv g_A({\bm v} \rightarrow D-{\bm v})$. Similarly, we define $\overline{g} \equiv \overline{g}_A\otimes g_B$. Thus, we have an explicit form of $g$ under partial transpose:
\begin{equation}
    g^{\intercal_A} = x_A(g)\ \overline{g}_A \otimes g_B = x_A(g)\ \overline{g}.
\end{equation}
Notice that in the special case $D=2$, the above relations become $Z^\intercal = Z$, $X^\intercal = X$, so that $\overline{g}_A=g_A$, and $x_A(g) = \pm 1$ depending on the parity of the $Y$ operators appearing in $g_A$. This is the case considered in Ref.~\cite{sang2020entanglement}.

Next, we define a function $\theta_A(g,h)$, such that
\begin{equation}
    {\rm proj}_A(g) \cdot {\rm proj}_A(h) = \theta_A(g,h)\ {\rm proj}_A(h) \cdot {\rm proj}_A(g).
\end{equation}
We will also need the commutation matrix $K_A$ defined in Eq.~(\ref{eq:commutation}). Since $K_A$ encodes the commutation relations between each pair of generators of $\mathcal{S}$, the phase $\theta_A$ between an arbitrary pair of $g, h \in \mathcal{S}$ can be calculated using $K_A$. Let us write $g$ in terms of the generators of $\mathcal{S}$:
\begin{equation}
    g = \prod_{i=1}^m g_i^{a_i},
\end{equation}
where the expansion coefficients can be encoded in a vector ${\bm a} \in \mathbb{F}_D^{\otimes m}$ for prime $D$. Similarly, $h$ can also be represented as a vector ${\bm b} \in \mathbb{F}_D^{\otimes m}$.
Therefore, we have
\begin{equation}
    \theta_A(g,h) = \omega^{{\bm a}\cdot K_A {\bm b}}.
\end{equation}
Below, we shall use the vector $({\bm u}_g, {\bm v}_g) \in \mathbb{F}_D^{\otimes 2|A|}$ to denote the Pauli string of $g_A$, and ${\bm a}_g\in \mathbb{F}_D^{\otimes m}$ for the expansion of $g$ in terms of the generators $\{g_1,\ldots, g_m\}$.

With all the ingredients above, we shall now derive a key result for the proof of Eq.~(\ref{eq:negativity}): 
\begin{equation}
\left(\rho_{A\cup B}^{\intercal_A}\right)^2 \propto \left(\rho_{A\cup B}^{\intercal_A}\right)^4.
\end{equation}
We start by computing $(\rho_{A\cup B}^{\intercal_A})^2$:
\begin{eqnarray}
&&(\rho_{A\cup B}^{\intercal_A})^2  \nonumber \\
&=& \frac{1}{D^{2|A\cup B|}}\sum_{g,h \in \mathcal{S}} g^{\intercal_A} \cdot h^{\intercal_A}  \nonumber \\
&=& \frac{1}{D^{2|A\cup B|}}\sum_{g,h \in \mathcal{S}} x_A(g) \ x_A(h) \ \overline{g} \cdot \overline{h}  \nonumber  \\
&=& \frac{1}{D^{2|A\cup B|}}\sum_{g,h \in \mathcal{S}} x_A(g) \ x_A(h)\  \overline{\omega}^{2 {\bm u}_h \cdot {\bm v}_g} \ \overline{g\cdot h}  \nonumber \\
&=& \frac{1}{D^{2|A\cup B|}}\sum_{g,h \in \mathcal{S}} \theta_A(h,g)\ x_A(g\cdot h) \ \overline{g\cdot h}  \nonumber \\
&=& \frac{1}{D^{2|A\cup B|}}\sum_{g, t \in \mathcal{S}} \theta_A(g^{-1}\cdot t,g) \ x_A(t) \ \overline{t}  \nonumber \\
&=& \frac{1}{D^{2|A\cup B|}} \sum_{{\bm a}_t \in \mathbb{F}_D^{\otimes m}} \left( \sum_{{\bm a}_g \in \mathbb{F}_D^{\otimes m}} \overline{\omega}^{{\bm a}_g \cdot K_A {\bm a}_t}\right) x_A(t) \ \overline{t}  \nonumber  \\
&=& \frac{1}{D^{2|A\cup B|}} \sum_{{\bm a}_t \in \mathbb{F}_D^{\otimes m}} D^m \ \delta(K_A {\bm a}_t, {\bm 0})\ x_A(t) \ \overline{t}  \nonumber  \\
&=& \frac{1}{D^{2|A\cup B|-m}}\sum_{{\bm a}_t \in {\rm Ker}(K_A)} x_A(t) \ \overline{t}.
\label{eq:partial_transpose_2}
\end{eqnarray}
The calculations above need some explanations. In the fourth line, we have used the following relation:
\begin{equation}
    \overline{g} \cdot \overline{h} = \overline{\omega}^{2{\bm u}_h\cdot {\bm v}_g} \overline{g\cdot h},
\end{equation}
where $({\bm u}_g, {\bm v}_g)$ and $({\bm u}_h, {\bm v}_h)$ denote the Pauli strings in $g$ and $h$, respectively. In the fifth line, we have used the relation:
\begin{equation}
    x_A(g) \ x_A(h) \ \overline{\omega}^{2{\bm u}_h \cdot {\bm v}_g} = \theta_A(h,g) \ x_A(g\cdot h),
\end{equation}
which can be verified via a direct calculation. Notice that in the special case $D=2$, this reduces to the ``cocycle condition" discussed in Ref.~\cite{sang2020entanglement}. Using the result of Eq.~(\ref{eq:partial_transpose_2}), we can now proceed to compute $(\rho_{A\cup B}^{\intercal_A})^4$:
\begin{eqnarray}
&&(\rho_{A\cup B}^{\intercal_A})^4 \nonumber \\
&=& \frac{1}{D^{4|A\cup B|-2m}} \sum_{{\bm a}_g, {\bm a}_h \in {\rm Ker}(K_A)} x_A(g) \ x_A(h) \ \overline{g} \cdot \overline{h}  \nonumber \\
&=& \frac{1}{D^{4|A\cup B|-2m}} \sum_{{\bm a}_t \in {\rm Ker}(K_A)} \left( \sum_{{\bm a}_g \in {\rm Ker}(K_A)} \overline{\omega}^{{\bm a}_g \cdot K_A {\bm a}_t}\right) x_A(t) \ \overline{t} \nonumber \\
&=& \frac{1}{D^{4|A\cup B|-2m}} |{\rm Ker}(K_A)| \sum_{{\bm a}_t \in {\rm Ker}(K_A)} x_A(t) \ \overline{t}   \nonumber  \\
&=& \frac{|{\rm Ker}(K_A)|}{D^{2|A\cup B|-m}} \ \left(\rho_{A\cup B}^{\intercal_A} \right)^2.
\end{eqnarray}
This immediately implies that
\begin{equation}
    (\rho_{A\cup B}^{\intercal_A})^{2n} = \left( \frac{|{\rm Ker}(K_A)|}{D^{2|A\cup B|-m}} \right)^{n-1} \ (\rho_{A\cup B}^{\intercal_A})^2.
\end{equation}
Finally, using the last expression above, we are ready to prove Eq.~(\ref{eq:negativity}):
\begin{eqnarray}
N(A,B) &=& \lim_{n \rightarrow \frac{1}{2}} {\rm log}\ {\rm tr} (\rho_{A\cup B}^{\intercal_A})^{2n}  \nonumber  \\
&=& \lim_{n \rightarrow \frac{1}{2}}{\rm log} \left[\left( \frac{|{\rm Ker}(K_A)|}{D^{2|A\cup B|-m}} \right)^{n-1} \ {\rm tr} (\rho_{A\cup B}^{\intercal_A})^2 \right]  \nonumber  \\
&=& \lim_{n \rightarrow \frac{1}{2}}{\rm log} \left[\left( \frac{|{\rm Ker}(K_A)|}{D^{2|A\cup B|-m}} \right)^{n-1} \ \frac{1}{D^{|A\cup B|-m}} \right]  \nonumber \\
&=& {\rm log} \left(\frac{D^m}{|{\rm Ker}(K_A)|} \right)^{\frac{1}{2}} \nonumber \\
&=& \frac{1}{2}[m - {\rm dim} \ {\rm Ker}(K_A)]  \nonumber \\
&=& \frac{1}{2} {\rm dim} \ {\rm Im}(K_A)  \nonumber \\
&=& \frac{1}{2} {\rm rank}(K_A).
\end{eqnarray}
This is exactly Eq.~(\ref{eq:negativity}).



\section{Forced measurements, projective measurements, and sampling algorithms \label{app:prob_measure}}

In this Appendix, we explain the numerical method for sampling the random tensor networks, and explain two possible probability distributions that we can assign to these random samples.
As we will see, they effectively define two models we considered in the main text, with ``forced meaurements'' and ``projective measurements'', repsectively.

\subsection{Forced measurements for tensor contraction}

The algorithm for sampling a particular instance of the random stabilizer tensor network with forced measurements is as follows.
\begin{enumerate}
\item
We start with the following computational basis state
\begin{align}
    \ket{\Psi_0} = \prod_{\mathbf{r} \in V} \ket{0000}_\mathbf{r}.
\end{align}
Here, we have taken $l = 4$.

\item
Next, for each site $\mathbf{r}$ we sample a random Clifford unitary from the uniform probability distribution on the 4-qudit Clifford group (denoted $\mathcal{C}(D;4)$); the result is denoted $U_{\mathbf{r}}$.

The unitaries $\{U_{\mathbf{r}}\}$ define a tensor on each site, namely
\begin{align}
    T[\mathbf{r}]_{i_1 i_2 i_3 i_4} = {}_\mathbf{r}\bra{i_1 i_2 i_3 i_4} U_{\mathbf{r}} \ket{0000}_\mathbf{r},
\end{align}
or equivalently
\begin{align}
    U_{\mathbf{r}} \ket{0000}_\mathbf{r} = \sum_{i_{1,2,3,4} = 1}^D
    T[\mathbf{r}]_{i_1 i_2 i_3 i_4}  \ket{i_1 i_2 i_3 i_4}_\mathbf{r}
\end{align}

We then have the following state, 
\begin{align}
    \ket{\Psi(\mathcal{U})} \coloneqq 
    \mathcal{U} \ket{\Psi_0}.  
\end{align}
Here $\mathcal{U} \coloneqq \bigotimes_\mathbf{r} U_\mathbf{r}$ is an element of $\mathfrak{C} \coloneqq \mathcal{C}(D;4) ^{\otimes |V|}$.

\item
Finally, we project pairs of qudits on each bulk bond of the network to a fixed Bell state $\frac{1}{\sqrt{D}} \sum_{i=0}^{D-1} \ket{ii}$, effectively contracting the indices of the tensor network.
After that we trace out qudits in the bulk.
These operations can be summarized as follows,
\begin{align}
\rho(\mathcal{U}) \coloneqq
\mathrm{Tr}_{V_{\rm bulk}}
\left[
\mathbb{P}
\ket{\Psi(\mathcal{U})} \bra{\Psi(\mathcal{U})}
\right],
\end{align}
where
\begin{align}
    & \mathbb{P} \nn
    =&\ \bigotimes_{e \in E}
    \left[ \frac{1}{D} \left(\sum_{i=1}^D \ket{ii} \right) \left(\sum_{i=1}^D \bra{ii} \right) \right]_e \nn
    =&\ \bigotimes_{e = (u,v) \in E}
    \(\frac{1}{D} \sum_{k=0}^{D-1} \( \omega^0 Z_u Z_v^{-1}\)^k \)
    \(\frac{1}{D} \sum_{k=0}^{D-1} \(\omega^0 X_u X_v\)^k \).
\end{align}
That is, the projection operator $\mb{P}$ is realized by ``forcing'' or ``post-selecting'' the trajectory where  all measurement outcomes (that is, of all $Z_u Z_v^{-1}$ and $X_u X_v$ operators on all bulk edges) equal to $\omega^0  = +1$.

With the bulk qudits traced out, $\rho(\mathcal{U})$ is a pure state on the boundary qudits of the tensor network.

\end{enumerate}

As defined, $\rho(\mathcal{U})$ is not necessarily normalized.
Its trace $\tr \rho(\mathcal{U})$ is equal to the probability that the aforementioned ``all $+1$'' trajectory occurs, in the ensemble of all possible measurement trajectories.
In numerics, we can formally perform the projection and subsequently normalize the state, without worrying about physical meanings of the post-selection, except when $\tr \rho(\mathcal{U}) = 0$, which means that the contraction of indices gives a zero state, and we must reject the sample given by $\mc{U}$.
This may be numerically unfavorable, for the construction of a state can be quite time consuming.

When $\tr \rho(\mathcal{U}) > 0$, we denote the normalized density matrix as 
\begin{align}
    \widetilde{\rho(\mathcal{U})} \coloneqq \frac{\rho(\mathcal{U})}{\mathrm{Tr} \rho(\mathcal{U})}, \text{ when } \mathrm{Tr} \rho(\mathcal{U}) > 0.
\end{align}
Thus, we have now the following ensemble of normalized boundary states, each occuring with the same probability.
\begin{align}
    \mathcal{M}^{\rm f} =&\ \left\{ \widetilde{\rho(\mathcal{U})} \ :\   \mathcal{U} \in \mathfrak{C} \text{ and } \mathrm{Tr} \rho(\mathcal{U}) > 0 \right\} \nonumber \\
    \coloneqq &\ \left\{ \widetilde{\rho(\mathcal{U})} \ :\   \mathcal{U} \in \mathfrak{C}^> \right\},
\end{align}
where we defined the following subset of $\mathfrak{C}$ 
\begin{align}
    \mathfrak{C}^> \coloneqq \{ \mathcal{U} \in \mathfrak{C} :
    \mathrm{Tr} \rho(\mathcal{U}) > 0
    \}.
\end{align}
Note that while different elements of $\mathfrak{C}^>$ can in principle give rise to the same $\widetilde{\rho(\mathcal{U})}$, we still treat them differently, so that $|\mathcal{M}^{\rm f}| = | \mathfrak{C}^>|$.

We can now discuss what probability distribution (or ``measure'') we want to assign on $\mathcal{M}^{\rm f}$.
Here, we choose to assign each element of $\mathcal{M}^{\rm f}$ with equal probability
$1/|\mathfrak{C}^>|$.
Formally, it means that the ensemble average of an observable $\mathcal{O}$ should read
\begin{align}
\label{eq:avg_forced_M}
    \avg{\mc{O}}^{\rm f} = 
    \frac{1}{|\mathfrak{C}^>|} 
    \sum_{\mathcal{U} \in \mathfrak{C}^> }
        \mathcal{O}[
            \widetilde{\rho(\mathcal{U})}
        ].
\end{align}
The superscript ``f'' means ``forced measurements'', to be distinguished from the case with projective measurements, below.

To numerically estimate $\avg{\mathcal{O}}^{\rm f}$, we may run the sampling algorithm $N$ times (each run $j$ will give us an instance of unitary $\mathcal{U}_j \in \mathfrak{C}$), and average the results against the runs for which the resultant state $\rho(\mathcal{U}_j)$ does not vanish.
Formally, this reads
\begin{align}
\label{eq:avg_forced_M_estimate}
    \avg{\mc{O}}^{\rm f}_N \coloneqq&\
    \frac{
        \sum_{j=1}^N
        \delta(\mathrm{Tr}[\rho(\mathcal{U}_j)] > 0)
        \cdot
        \mathcal{O}\left[
        \widetilde{\rho(\mathcal{U}_j)}
        \right]
    }
    {
        \sum_{j=1}^N
        \delta(\mathrm{Tr}[\rho(\mathcal{U}_j)] > 0)
    },
\end{align}
for which we expect
\begin{align}
    \lim_{N \to \infty} \avg{\mc{O}}^{\rm f}_N = \avg{\mc{O}}^{\rm f}.
\end{align}

\subsection{Projective measurements}

Instead of forcing the $Z_u Z_v^{-1}$ and $X_u X_v$ measurements on the edges and focusing on a single trajectory so that we manage to contract the indices of an abstract tensor network, in this subsection we instead view the measurements as ``physical'', and consider the ensemble of all possible measurement trajectories weighted by their respective Born probabilities, without rejecting any of them.
This way, we can define a different ensemble of RSTN.

Formally, we now have the following ensemble of normalized boundary states,
\begin{align}
\mc{M}^{\rm p} =&\
\{
    \widetilde{\rho(\mc{U}; \mathbf{m})} \coloneqq
    \frac{
        \rho(\mc{U}; \mathbf{m})
    }
    {
        \tr \rho(\mc{U}; \mathbf{m})
    }: \nn
    &
    \quad
    \mc{U} \in \mf{C}, 
    \mathbf{m} \in \{1, \omega, \ldots \omega^{D-1} \}^{2|E|},
    \text{ and }
    \tr \rho(\mc{U}; \mathbf{m}) > 0
\}.
\end{align}
Here, the state is labelled by not only the unitary $\mc{U}$, but also all \textit{admissible} trajectories (i.e. those with nonzero probability) of the $2|E|$ measurements, $\mathbf{m} \coloneqq (m_{e_1}^{ZZ}, m_{e_1}^{XX}, \ldots,  m_{e_{|E|}}^{ZZ}, m_{e_{|E|}}^{XX})$, where $\tr \rho(\mc{U}; \mathbf{m}) > 0$

The weight we assign to states in $\mc{M}^{\rm p}$ is the Born probability, $\mathrm{Tr} \rho(\mathcal{U}; \mathbf{m})$.
The ensemble average of an observable $\mc{O}$ is thus
\begin{align}
\label{eq:avg_physical_M}
    \avg{\mc{O}}^{\rm p} = 
    \frac{1}{|\mf{C}|}
    \sum_{\mc{U} \in \mf{C} }
    \sum_{\mathbf{m} \in \{1, \omega, \ldots \omega^{D-1} \}^{2|E|}}
    \tr \rho(\mc{U}; \mathbf{m})
    \cdot
        \mc{O}[
            \widetilde{\rho(\mc{U}; \mathbf{m})}
        ].
\end{align}

From now on, we focus on entanglement properties of the boundary stabilizer states (and $\mc{O}$ can be the entropy, mutual information between subregions, entanglement negativities, as we considered in the main text), for which all trajectories $\mathbf{m}$ lead to the same value of $\mathcal{O}$.
The result $\mc{O}[
            \widetilde{\rho(\mc{U}; \mathbf{m})}
        ]$
really only depends on the unitary $\mc{U}$, which we denote as $\mc{O}[
            \widetilde{\rho(\mc{U};
            \bm{\mu})}
        ]$,
where $\bm{\mu}$ can be chosen to be any admissible trajectory.
Thus,
\begin{align}
\label{eq:avg_physical_M_simp}
    \avg{\mc{O}}^{\rm p} =&\
    \frac{1}{|\mf{C}|}
    \sum_{\mc{U} \in \mf{C} }
    \sum_{\mathbf{m} \in \{1, \omega, \ldots \omega^{D-1} \}^{2|E|}}
    \tr \rho(\mc{U}; \mathbf{m})
    \cdot
        \mc{O}[
            \widetilde{\rho(\mc{U};
            \bm{\mu})}
        ]
    \nn
    =&\ 
    \frac{1}{|\mf{C}|}
    \sum_{\mc{U} \in \mf{C} }
        \mc{O}[
            \widetilde{\rho(\mc{U};
            \bm{\mu})}
        ],
\end{align}
where we used the conservation of probability,
\begin{align}
    \sum_{\mathbf{m} \in \{1, \omega, \ldots \omega^{D-1} \}^{2|E|}}
    \tr \rho(\mc{U}; \mathbf{m}) = 1.
\end{align}
We emphasize that the simplification in Eq.~\eqref{eq:avg_physical_M_simp} only occurs for entanglement properties of stabilizer states.
In contrast, Eq.~\eqref{eq:avg_forced_M} is completely general, and remains correct for generic (non-stabilizer states) and any observable $\mc{O}$.

To numerically estimate the ensemble avaraged entanglement properties $\avg{\mathcal{O}}^{\rm p}$, we may run the sampling algorithm $N$ times.
In each run, we can take any trajectory $\bm{\mu}$, and calculate $\mc{O}$ for the resultant state $\widetilde{\rho(\mc{U}; \bm{\mu})}$.
In practice, in running the stabilizer simulation, one can simply choose to not record the measurement outcomes at all; the resultant state is guaranteed to be on an admissible trajectory.\footnote{
The same method can also be applied in the simulation of random Clifford circuits where the results are weighted by the Born probability.}
Thus, we do not need to reject any of the runs.
The estimate is thus
\begin{align}
\label{eq:avg_physical_M_estimate}
    \avg{\mc{O}}^{\rm p}_N \coloneqq&\
    \frac{1}{N}
        \sum_{j=1}^N
        \mathcal{O}\left[
        \widetilde{\rho(\mathcal{U}_j; \bm{\mu}_j)}
        \right]
\end{align}
for which we also expect
\begin{align}
    \lim_{N \to \infty} \avg{\mc{O}}^{\rm p}_N = \avg{\mc{O}}^{\rm p}.
\end{align}

\subsection{Relating the two ensembles}

We have defined $\avg{\mc{O}}^{\rm f}$  and $\avg{\mc{O}}^{\rm p}$, and defined sampling methods for estimating them, in Eqs.~(\ref{eq:avg_forced_M_estimate}, \ref{eq:avg_physical_M_estimate}), respectively.
As we explained above, in estimating $\avg{\mc{O}}^{\rm f}$ with Eq.~\eqref{eq:avg_forced_M_estimate}, some samples must be rejected.
Here, we describe a single sampling method for entanglement properties $\mc{O}$, where both $\avg{\mc{O}}^{\rm f}$  and $\avg{\mc{O}}^{\rm p}$ can be estimated, without the need of rejectign any sample.

We start by comparing the definitions of  $\avg{\mc{O}}^{\rm f}$ [Eq.~\eqref{eq:avg_forced_M}] and  $\avg{\mc{O}}^{\rm p}$ [Eq.~\eqref{eq:avg_physical_M_simp}].
It will be convenient now to explicitly include the trajectory information in Eq.~\eqref{eq:avg_forced_M}, replacing
\begin{align}
    \widetilde{\rho(\mc{U})} \ \to\  \widetilde{\rho(\mc{U}; \mathbf{m}_0)},
\end{align}
where $\mathbf{m}_0$ denotes the trajectory where all measurement results are $+1$ (as we needed for contracting the indices)
\begin{align}
    \mathbf{m}_0 = (m_{e_1}^{ZZ} = +1, m_{e_1}^{XX} = +1, \ldots,  m_{e_{|E|}}^{ZZ} = +1, m_{e_{|E|}}^{XX} = +1).
\end{align}
With this replacement, Eq.~\eqref{eq:avg_forced_M} now reads
\begin{align}
\label{eq:avg_forced_M_m0}
    \avg{\mc{O}}^{\rm f} = 
    \frac{1}{|\mathfrak{C}^>|} 
    \sum_{\mathcal{U} \in \mathfrak{C}^> }
        \mathcal{O}[
            \widetilde{\rho(\mathcal{U}; \mathbf{m}_0)}
        ].
\end{align}

For $\avg{\mc{O}}^{\rm p}$ in Eq.~\eqref{eq:avg_physical_M_simp}, for all $\mc{U} \in \mf{C}^>$, we may also choose $\bm{\mu} = \mathbf{m}_0$, by definition.
Thus,
\begin{align}
    \avg{\mc{O}}^{\rm p} =&\
    \frac{1}{|\mf{C}|}
    \(
        \sum_{\mc{U} \in \mf{C}^> }
        \mc{O}[
            \widetilde{\rho(\mc{U};
            \mathbf{m}_0)}
        ]
        +
        \sum_{\mc{U} \in \mf{C} - \mf{C}^> }
        \mc{O}[
            \widetilde{\rho(\mc{U};
            \bm{\mu})}
        ]
    \).
\end{align}
For $\mc{U} \in \mf{C} - \mf{C}^>$, $\bm{\mu}$ must not be equal to $\mathbf{m}_0$, but otherwise arbitrary as long as $\tr \rho(\mc{U}; \bm{\mu}) > 0$.

It is a property of stabilizer states that $\tr \rho(\mc{U}, \mathbf{m})$ must either be $0$ or of the form $D^{-n_{\rm r}(\mc{U})}$, where $n_{\rm r}(\mc{U})$ is an integer between $0$ and $2|E|$, which is equal to the number measurements whose results are random, and only depends on $\mc{U}$.
For example, when $\tr \rho(\mc{U}, \mathbf{m}) = D^{-2|E|}$ on one trajectory $\mathbf{m}$, all $2|E|$ measurements are random; and in fact we have the same trace for all $D^{2|E|}$ trajectories,
\begin{align}
    & \tr \rho(\mc{U}, \mathbf{m}) = D^{-2|E|} \text{ for any } \mathbf{m} \in \{1, \omega, \ldots \omega^{D-1} \}^{2|E|} 
    \nn
    \Rightarrow&\ 
    \tr \rho(\mc{U}, \mathbf{m}) = D^{-2|E|} \text{ for all } \mathbf{m} \in \{1, \omega, \ldots \omega^{D-1} \}^{2|E|}.
\end{align}
In general, when $\tr \rho(\mc{U}, \mathbf{m}) = D^{-n_{\rm r}(\mc{U})}$, there are $D^{n_{\rm r}(\mc{U})}$ trajectories, and all of them will have the same trace.

For the $2|E|$ mutually commuting operators that has been measured $\{g_1, \ldots, g_{2|E| } \}$, we introduce for each of them a ``destabilizer'', $\{ h_1, \ldots, h_{2|E|} \}$, which are single qudit unitaries that satisfy the following commutation relations
\begin{align}
    g_i h_j =&\ \omega^{\delta_{ij}} h_j g_i, \\
    h_i h_j =&\ h_j h_i.
\end{align}

We now construct a mapping $\sigma$ from $\mf{C}$ to $\mf{C}^>$ as follows.
\begin{enumerate}
\item
If $\mc{U} \in \mf{C}^>$, we simply take $\sigma(\mc{U}) = \mc{U}$.
\item
If $\mc{U} \in \mf{C} - \mf{C}^>$, we must have that $n_{\rm r}(\mc{U}) < 2|E|$, and exactly $n_{\rm d}(\mc{U}) = 2|E| - n_{\rm r}(\mc{U})$ measurements have deterministic results.
Among these $n_{\rm d}(\mc{U})$, at least one of the deterministic results is not equal to $1$.

In this case, we define
\begin{align}
    \sigma(\mc{U}) = \( \prod_{j : g_j \neq +1} (h_j)^{k_j} \) \mc{U} \( \prod_{j : g_j \neq +1} (h_j)^{k_j} \)^\dg.
\end{align}
That is, $\sigma(\mc{U})$ and $\mc{U}$ are related by local Clifford unitaries that do not change the entanglement properties.
Morever, measurements of $g_j$ that are random for $\mc{U}$ will remain random for $\sigma(\mc{U})$, and those that are deterministic remain deterministic for $\sigma(\mc{U})$.
The powers $k_j$ are uniquely fixed 
by the condition that the deterministic results $g_j$ for $\sigma(\mc{U})$ are all $+1$.
Thus, $\sigma(\mc{U}) \in \mf{C}^>$.

\end{enumerate}

Clearly, the mapping $\sigma$ assign a unique image to each $\mc{U} \in \mf{C}$, thus well defined.
Thus, we may partition $\mf{C}$ according to the image under $\sigma$,
\begin{align}
    \mf{C} = \bigcup_{\mc{U} \in \mf{C}^>} \sigma^{-1}(\mc{U}),
\end{align}
where by definition $\sigma^{-1}(\mc{U})$ and $\sigma^{-1}(\mc{U}^\p)$ are disjoint sets for $\mc{U} \neq \mc{U}^\p$.
Moreover, for all $\mc{U} \in \mf{C}^>$, we have.
\begin{align}
    | \sigma^{-1}(\mc{U}) | = D^{n_{\rm d}(\mc{U})} = D^{2|E| - n_{\rm r}(\mc{U})} = D^{2|E|} \tr \rho(\mc{U}; \mathbf{m}_0).
\end{align}

With these, we have
\begin{align}
    |\mf{C}| =&\ 
    \sum_{\mc{U} \in \mf{C}^> }
    | \sigma^{-1}(\mc{U}) |
    =
    \sum_{\mc{U} \in \mf{C}^> }
    D^{n_{\rm d}(\mc{U})}, \\
    |\mf{C}^>| =&\
    \sum_{\mc{U} \in \mf{C}}
    D^{-n_{\rm d}(\mc{U})},
\end{align}
and
\begin{align}
    \avg{\mc{O}}^{\rm p}
    =&\ 
    \frac{1}{|\mf{C}|}
    \sum_{\mc{U} \in \mf{C} }
        \mc{O}[
            \widetilde{\rho(\mc{U};
            \bm{\mu})}
        ] \nn
    =&\
    \frac{1}{|\mf{C}|}
    \sum_{\mc{U} \in \mf{C}^> }
    \mc{O}[
            \widetilde{\rho(\mc{U};
            \mathbf{m}_0)}
        ]
    \cdot
    D^{n_{\rm d}(\mc{U})}
    .
\end{align}
We used again the fact that the mapping $\sigma$ does not affect the value of $\mc{O}$.

These results allow us to use the estimates $\avg{\mc{O}}^{\rm p}_N$ for estimating $\avg{\mc{O}}^{\rm f}$.
Again, we sample $N$ unitaries $\mc{U} \in \mf{C}$, in the same fashion that led to $\avg{\mc{O}}^{\rm p}_N$ Eq.~\eqref{eq:avg_physical_M_estimate}.
Then, we calculate the following modified weighted estimate
\begin{align}
\label{eq:estimate_modified}
    \avg{\mc{O}}^{\rm m}_N \coloneqq
    \frac
    {
        \frac{1}{N}
        \sum_{j=1}^N
        \mathcal{O}\left[
        \widetilde{\rho(\mathcal{U}_j; \bm{\mu}_j)}
        \right]
        \cdot
        D^{-n_{\rm d}(\mc{U}_j)}
    }
    {
        \frac{1}{N}
        \sum_{j=1}^N
        D^{-n_{\rm d}(\mc{U}_j)}
    }.
\end{align}
Both the numerator and the denominator takes the form of a $\avg{\mc{O}}^{\rm p}_N$. 
As we take $N \to \infty$, 
\begin{align}
    &
    \lim_{N \to \infty} \avg{\mc{O}}^{\rm m}_N \nn
    =&\
    \frac
    {
        \lim_{N \to \infty}
        \frac{1}{N}
        \sum_{j=1}^N
        \mathcal{O}\left[
        \widetilde{\rho(\mathcal{U}_j; \bm{\mu}_j)}
        \right]
        \cdot
        D^{-n_{\rm d}(\mc{U}_j)}
    }
    {
        \lim_{N \to \infty}
        \frac{1}{N}
        \sum_{j=1}^N
        D^{-n_{\rm d}(\mc{U}_j)}
    } \nn
    =&\
    \frac
    {
        \avg{
        \mathcal{O}\left[
        \widetilde{\rho(\mathcal{U}; \bm{\mu})}
        \right]
        \cdot
        D^{-n_{\rm d}(\mc{U})}
        }^{\rm p}
    }
    {
        \avg{
        D^{-n_{\rm d}(\mc{U})}
        }^{\rm p}
    } \nn
    =&\
    \frac{
        \frac{1}{|\mf{C}|}
        \sum_{\mc{U} \in \mf{C} }
        \mc{O}[
            \widetilde{\rho(\mc{U};
            \bm{\mu})}
        ]
        \cdot
        D^{-n_{\rm d}(\mc{U})}
    }
    {
        \frac{1}{|\mf{C}|}
        \sum_{\mc{U} \in \mf{C} }
        D^{-n_{\rm d}(\mc{U})}
    }\nn
    =&\
    \frac{
        \sum_{\mc{U} \in \mf{C}^> }
        \mc{O}[
            \widetilde{\rho(\mc{U};
            \mathbf{m}_0)}
        ]
        \cdot
        D^{-n_{\rm d}(\mc{U})}
        D^{+n_{\rm d}(\mc{U})}
    }
    {
        \sum_{\mc{U} \in \mf{C}^> }
        D^{-n_{\rm d}(\mc{U})}
        \cdot
        D^{+n_{\rm d}(\mc{U})}
    }\nn
    =&\
    \frac{
        1
    }
    {
        |\mf{C}^>|
    }
    \sum_{\mc{U} \in \mf{C}^> }
        \mc{O}[
            \widetilde{\rho(\mc{U};
            \mathbf{m}_0)}
    ]
    \nn
    =&\
    \avg{\mc{O}}^{\rm f},
\end{align}
compare Eqs.~(\ref{eq:avg_forced_M}, \ref{eq:avg_forced_M_m0}).

Thus, by assigning the weight $D^{-n_{\rm d}(\mc{U})}$ to each sample, and compute the weighted average, we obtain an estimate of $\avg{\mc{O}}^{\rm f}$.

\subsection{Sampling from the forced measurement ensemble with local rejections}

Directly computing $\avg{\mc{O}}^\mathrm{m}_N$ in the numerics leads to a problem of ``undersampling'', i.e. having an insufficient number of samples.
The denominator of Eq.~\eqref{eq:estimate_modified} $\frac{1}{N} \sum_{j=1}^N D^{-n_{\rm d}(\mc{U}_j)}$ can be thought of as the effective number of samples, and is observed to grow slowly with $N$.
Recall that $n_{\rm d}(\mc{U})$ is the total number of bond contraction measurements whose outcomes are deterministic, thus its typical value grows with the system size, severely suppresing the weight of the samples.

In practice, instead of assigning a weight $D^{-n_{\rm d}(\mc{U}_j)}$ to the entire tensor network at the end, we choose to probablistically reject samples at a local level, so that each tensor network occur with a probability proportional to $D^{-n_{\rm d}(\mc{U}_j)}$, and no adjustments of the weights need to be made at the end.
The detailed procedure is as follows.
\begin{enumerate}
\item
We generate the local tensors at each location $(x,y)$ one by one, in the order of increasing $x$ and $y$.
Recall that the local tensor is generated by sampling a 4-qudit Clifford unitary.
\item
Once the local tensor at $(x,y)$ is generated, we attempt to contract its bonds with $(x-1, y)$ and $(x,y-1)$, so that it becomes a part of the entire tensor network.
\item
The contractions are realized by performing two measurements (of $Z_i Z_j^{-1}$ and of $X_i X_j$) that are forced to have results $+1$.
In the case of both measurements have random outcomes, we \emph{accept} the local tensor.
If exactly one of the measurements have random outcomes, we accept the local tensor with probability $D^{-1}$, by flipping a biased coin.
If both measurements have deterministic outcomes, we accept the local tensor with probability $D^{-2}$.
\item
If a local tensor is not accepted, we say it is rejected, in which case we regenerate the local tensor by sampling a new 4-qudit Clifford unitary, and attempt the two contraction measurements again.
\end{enumerate}

To see that this algorithm is correct, we compare it
with the sampling procedure in Eq.~\eqref{eq:estimate_modified}.
\begin{enumerate}
\item
Instead of assigning weights to different samples as in $\avg{\mc{O}}^{\rm m}_N$ (see Eq.~\eqref{eq:estimate_modified}), here we use probabilistic acceptance, which is completely equivalent.
We can imagine generating each sample as in  Eq.~\eqref{eq:estimate_modified}, but accept with probability $D^{-n_{\rm d}(\mc{U}_j)}$.
Although highly inefficient, in the limit $N \to \infty$ the unweighted average will converge to $\avg{\mc{O}}^{\rm m}_N$ and $\avg{\mc{O}}^{\rm f}$.
\item
Following the previous point, instead of deciding whether or not to accept the entire tensor network at the end, we choose to perform the probabilistic acceptance/rejection at a local level, which only depends on local information.
In other words, we can perform the rejections early, so we do not have to contract all the tensors before we find out this sample needs to be rejected.
\item
Moreover, since each sample has a finite probability $D^{-n_{\rm d}(\mc{U}_j)}$ to be accepted, we never run into a case where it is impossible to find a acceptable local tensor.
Thus, we are guaranteed to get a sample, of weight 1, for each run of the algorithm.
This property is again special to the stabilizer tensor networks.
\end{enumerate}
To generate $N$ samples with this algorithm, the running time is proportional to $N$.
The local rejections lead to a mere constant multiple of overhead.

Although the algorithm above is described only for the forced measurements involved in contraction of bonds, it is straightforward to modify it so that forced measurements responsible for breaking bonds can also be implemented.



\end{document}